\documentclass[11pt]{article}
\usepackage{epsfig,cite}
\usepackage{amsmath}
\usepackage{amssymb}
\usepackage{amsfonts}
\usepackage{latexsym}
\usepackage{wasysym}
\usepackage{url}
\usepackage{hyperref}
\usepackage{bm}
\usepackage{graphics}
\usepackage{multirow}
\usepackage{rotating}

\def\bea{\begin{eqnarray}}
\def\eea{\end{eqnarray}}
\def\beq{\begin{equation}}
\def\eeq{\end{equation}}
\catcode`@=11 
\def\slash#1{\mathord{\mathpalette\c@ncel#1}}
 \def\c@ncel#1#2{\ooalign{$\hfil#1\mkern1mu/\hfil$\crcr$#1#2$}}
\def\lsim{\mathrel{\mathpalette\@versim<}}
\def\gsim{\mathrel{\mathpalette\@versim>}}
 \def\@versim#1#2{\lower0.2ex\vbox{\baselineskip\z@skip\lineskip\z@skip
       \lineskiplimit\z@\ialign{$\m@th#1\hfil##$\crcr#2\crcr\sim\crcr}}}
\catcode`@=12 

\def\smallfrac#1#2{\hbox{${{#1}\over {#2}}$}}

\def\({\left(}
\def\){\right)}
\def\[{\left[}
\def\]{\right]}

\relax

\def    \hepph  #1 {{\tt hep-ph/#1}}
\def    \hepex  #1 {{\tt hep-ex/#1}}

\newcommand{\la}{\left\langle}
\newcommand{\ra}{\right\rangle}
\newcommand{\lc}{\left[}
\newcommand{\rc}{\right]}
\newcommand{\lp}{\left(}
\newcommand{\rp}{\right)}
\newcommand{\be}{\begin{equation}}
\newcommand{\ee}{\end{equation}}

\newcommand{\tot}{\rm tot}
\newcommand{\val}{\rm val}

\newcommand{\tr}{\rm tr}

\begin{document}
\begin{figure}[h]
\epsfig{width=0.32\textwidth,figure=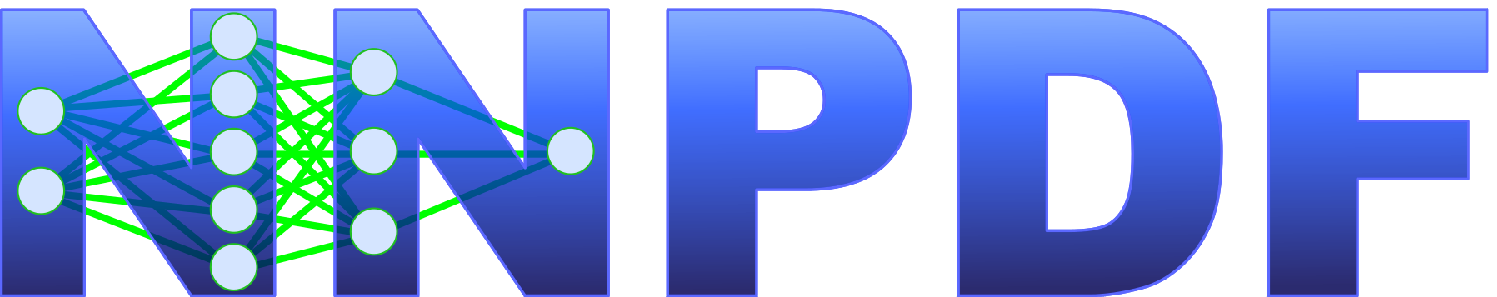}
\end{figure}
\vspace{-2.0cm}

\begin{flushright}
CERN-PH-TH/2012-036 \\
IFUM-987-FT\\
Edinburgh 2013/04 \\
\end{flushright}
\begin{center}
{\Large \bf Unbiased determination of polarized parton distributions \\
  and their uncertainties}
\vspace{0.8cm}

The NNPDF Collaboration: Richard~D.~Ball,$^{1}$ Stefano~Forte,$^2$
Alberto~Guffanti,$^{3}$\\  
Emanuele~R.~Nocera,$^2$  Giovanni~Ridolfi$^4$ and
Juan~Rojo.$^5$

\vspace{1.cm}

{\it ~$^1$ Tait Institute, University of Edinburgh,\\
JCMB, KB, Mayfield Rd, Edinburgh EH9 3JZ, Scotland\\
~$^2$ Dipartimento di Fisica, Universit\`a di Milano and
INFN, Sezione di Milano,\\ Via Celoria 16, I-20133 Milano, Italy\\
~$^3$ The Niels Bohr International Academy and Discovery Center, \\
The Niels Bohr Institute, Blegdamsvej 17, DK-2100 Copenhagen, Denmark\\
~$^4$ Dipartimento di Fisica, Universit\`a di Genova and
INFN, Sezione di Genova,\\ Genova, Italy\\
~$^5$ PH Department, TH Unit, CERN, CH-1211 Geneva 23, Switzerland \\
}

\end{center}

\vspace{0.8cm}

\begin{center}
{\bf \large Abstract:}
\end{center}

We present a determination of a set of polarized parton
distributions (PDFs) of the nucleon, at next-to-leading order, from a global
set of longitudinally polarized
deep-inelastic scattering data: {\tt NNPDFpol1.0}.  The
determination is based on the NNPDF methodology: a Monte Carlo
approach, with neural networks used as unbiased interpolants, previously
applied to the determination of unpolarized parton
distributions, and designed to provide a faithful and statistically
sound representation of PDF uncertainties.  We
present our dataset, its statistical features, and its Monte Carlo
representation. We summarize the technique used to solve the polarized
evolution equations and its benchmarking, and the method used to
compute physical observables.  We review the NNPDF methodology for
parametrization and
fitting of neural networks,  the algorithm used to determine the
optimal fit, and its adaptation to the polarized case. 
We finally present our set of polarized parton
distributions. We discuss its statistical properties, test for its
stability upon various modifications of the fitting procedure, and
compare it to other recent polarized parton sets, and in particular
obtain predictions for polarized first moments of PDFs based on it. We
find that the uncertainties on  the gluon, and to a lesser extent
the strange PDF, were 
substantially underestimated in previous determinations.

\clearpage

\tableofcontents

\clearpage

\section{Introduction}

The interest in the determination of polarized parton distributions
(PDFs) of
the nucleon is largely related to the experimental discovery in the
late 80s that the singlet axial charge of the proton is anomalously
small~\cite{Ashman:1987hv,Ashman:1989ig}, soon followed by the theoretical
realization~\cite{Altarelli:1988nr,Altarelli:1990jp} that the
perturbative behavior of polarized PDFs deviates from parton model
expectations, according to which gluons decouple in the asymptotic limit.
The theoretical interpretation of these results has spawned a
huge literature, while at the same time 
experimental information on polarized PDFs from
deep-inelastic scattering but also from a variety of other processes has
been accumulating over the years
(see e.g.~\cite{deFlorian:2011ia} and references therein).

First studies of the polarized structure of the nucleon were aimed at
an accurate determination of polarized first moments (including
detailed uncertainty
estimates)~\cite{Ball:1995td,Altarelli:1996nm,Altarelli:1998nb}, but
did not attempt a determination of a full PDF set, which was first
proposed in Ref.~\cite{Gehrmann:1995ag}, but without uncertainty
estimation. More recently, polarized PDF sets with uncertainties
have been constructed by
at least four groups (BB~\cite{Bluemlein:2002be,Blumlein:2010rn}, AAC~\cite{Hirai:2008aj},
LSS~\cite{Leader:2006xc,Leader:2010rb} and
DSSV~\cite{deFlorian:2008mr,deFlorian:2009vb}). These PDF sets slightly differ 
in the choice of datasets, the form of  PDF parametrization, and in
several details of the QCD analysis (such as the treatment of 
higher twist corrections), but they are all based on the standard
Hessian methodology for PDF fitting and uncertainty
determination, which has been widely used in the
unpolarized case (see~\cite{Forte:2010dt,Forte:2013wc} and references therein).
This methodology is known~\cite{Forte:2010dt} 
to run into difficulties especially when
information is scarce, because of the intrinsic bias of the Hessian
method based on a fixed parton parametrization. This is
likely to be particularly the
case for polarized PDFs, which rely on data both less
abundant and less accurate than their unpolarized counterparts.

In order to overcome these difficulties, the NNPDF collaboration
has proposed and developed a new methodology for PDF 
determination~\cite{DelDebbio:2004qj,DelDebbio:2007ee,Ball:2008by,Ball:2009qv,Ball:2009mk,
Ball:2010de,Ball:2010gb,Ball:2011mu,Ball:2011uy,Ball:2011eq, Ball:2011gg,Ball:2012cx}.
The NNPDF technique uses a robust set of statistical tools, which include Monte Carlo
methods for error propagation, neural networks for PDF parametrization, and genetic algorithms 
for their training. The NNPDF sets are now routinely used by the Tevatron
and LHC collaborations in their data analysis and for data-theory comparisons.
In this work we extend the application of the NNPDF methodology to the
determination of polarized parton distributions of the nucleon. As we
will see, some PDF uncertainties will turn out to be 
underestimated in existing PDF determinations: in particular those of the polarized gluon distribution, but also those of the strange distribution. 

The outline of this paper is as follows. In Sect.~\ref{sec:expdata} we
present the data set used to determine polarized PDFs, and we review
the relationship between measured asymmetries and structure
functions. In Sect.~\ref{sec:polpdfs} we discuss the parametrization of
polarized PDFs in terms of neural networks, and the construction
of polarized structure functions. Then in Sect.~\ref{sec:minim}
we discuss the minimization strategy. The results for the {\tt NNPDFpol1.0}
polarized partons are presented in Sect.~\ref{sec:results}, and
in Sect.~\ref{sec:phenoimplications} we discuss the phenomenological
implications for the spin content of the proton and the test of the
Bjorken sum
rule. Finally in Sect.~\ref{sec:conclusions} we summarize our results
and outline future developments. Some details
on the benchmarking of polarized PDF evolution are given in the Appendix.


\section{Experimental data}
\label{sec:expdata}

The bulk of the experimental information on (longitudinal) polarized
proton structure
comes from inclusive polarized deep-inelastic scattering with
charged lepton beams. Deep-inelastic scattering with longitudinally
polarized beams and targets allows a determination of the longitudinal
structure function $g_1(x,Q^2)$, which in turn admits a factorized
expression in terms of polarized PDFs. Neutral-current deep-inelastic
scattering does not allow to us to disentangle the contribution of quarks and
antiquarks. Using both proton and neutron (deuteron or ${}^3$He)
targets it is possible to separate the isospin singlet and triplet
quark contributions to structure functions, with the gluon determined
from scaling violations. A weak control on the separation of the
isospin singlet quark contribution into its SU(3) octet and singlet
component is possible using baryon decays to fix the respective
normalization of these contributions, with in principle their different
scale dependence providing some constraint on their shape.

Only charged-current deep-inelastic scattering would
allow for full flavor separation~\cite{Forte:2001ph}: this could be
feasible with 
neutrino beams (such as available at a neutrino
factory~\cite{Mangano:2001mj}), or perhaps very  
high-energy polarized charged lepton beams (such as available at an
electron-ion collider~\cite{Boer:2011fh}). Therefore, current
constraints  on flavor
separation are only provided by semi-inclusive deep-inelastic scattering
data or by polarized hadron collider processes, 
such as  polarized Drell-Yan production in fixed target collisions
and polarized $W$ production at the relativistic Heavy Ion Collider (RHIC). 
Likewise, direct constraints
on the medium and large-$x$ polarized gluon require hadron
and jet production either in fixed target experiments or at RHIC, while
the small-$x$ gluon can only be probed by going to higher energy, such
as at a polarized Electron-Ion Collider.
 
In this paper  we will concentrate on 
inclusive longitudinally polarized DIS data, and thus we will only
determine a subset of PDF combinations. This first polarized  PDF set
based on NNPDF methodology will then be available for inclusion of
other datasets through the reweighting technique of
Refs.~\cite{Ball:2010gb,Ball:2011gg}.

We will first review the experimental observables which we use for the
determination of polarized structure functions, and the information
which various experiments provide on them. Then, we
will summarize the features of the data we use, and finally
the construction and validation of the  Monte Carlo pseudodata sample from
the input experimental data.

\subsection{Experimental observables and longitudinal polarized
  structure functions}
\label{sec:asysf}

Standard perturbative factorization provides predictions for polarized
structure functions $g_1(x,Q^2)$. However,
experiments measure cross section asymmetries,
defined by considering longitudinally polarized leptons
scattering off a hadronic target, polarized either
longitudinally or transversely with respect to the collision axis,
from which the longitudinal ($A_{\parallel}$) and transverse ($A_{\perp}$)
asymmetries are determined as
\be
\label{eq:xsecasy}
 A_{\parallel}=
\frac{d\sigma^{\rightarrow\Rightarrow}-d\sigma^{\rightarrow\Leftarrow}}
{d\sigma^{\rightarrow\Rightarrow}+d\sigma^{\rightarrow\Leftarrow}};\quad
 A_{\perp}=
\frac{d\sigma^{\rightarrow\Uparrow}-d\sigma^{\rightarrow\Downarrow}}
{d\sigma^{\rightarrow\Uparrow}+d\sigma^{\rightarrow\Downarrow}}.\quad
\ee

The hadronic tensor for  polarized, parity conserving deep-inelastic
scattering can be parametrized in terms of four structure
functions: two of them, $F_1(x,Q^2)$ and $F_2(x,Q^2)$, 
characterize spin-averaged deep-inelastic scattering, while
$g_1(x,Q^2)$ and $g_2(x,Q^2)$ appear when both the lepton beam and
the nucleon target are in definite polarization states. 
For the conventional definition of the hadronic tensor in terms
of structure functions, see e.g.~\cite{Ellis:1991qj}.

The two polarized structure functions are related to the
 measurable asymmetries Eq.~(\ref{eq:xsecasy}) by
\begin{align}
\label{g1toA}
g_1(x,Q^2)&=
\frac{F_1(x,Q^2)}{(1+\gamma^2)(1+\eta\zeta)}
\left[
(1+\gamma\zeta)\frac{A_{\parallel}}{D}-(\eta-\gamma)\frac{A_{\perp}}{d}
\right]
\mbox{ ,}
\\
\label{g2toA}
g_2(x,Q^2)&=
\frac{F_1(x,Q^2)}{(1+\gamma^2)(1+\eta\zeta)}
\left[
\left(\frac{\zeta}{\gamma}-1\right)\frac{A_{\parallel}}{D}
+\left(\eta+\frac{1}{\gamma}\right)\frac{A_{\perp}}{d}
\right]
\mbox{ .}
\end{align}
In Eqs.~(\ref{g1toA}-\ref{g2toA}) the dependence on the
nucleon mass $m$ is taken into account through the factor
\be\label{eq:gamdef}
\gamma^2\equiv\frac{4m^2x^2}{Q^2},
\ee
which also appears in the definitions of the other kinematic factors 
in Eqs.~(\ref{g1toA}-\ref{g2toA}):
\begin{align}
\label{eq:ddef}
d&=\frac{D\sqrt{1-y-\gamma^2 y^2/4}}{1-y/2},\\
\label{eq:Ddef}
D&=\frac{1-(1-y)\epsilon}{1+\epsilon R(x,Q^2)},\\
\label{eq:etadef}
\eta&=\frac{\epsilon\gamma y}{1-\epsilon(1-y)},\\
\label{eq:zetadef}
\zeta&=\frac{\gamma(1-y/2)}{1+\gamma^2 y/2},\\
\label{eq:epsilondef}
 \epsilon &= \frac{4(1-y) - \gamma^2 y^2}{2 y^2 + 4 (1-y) + \gamma^2 y^2}.
\end{align}
Here $y$ is the standard lepton scaling variable, given by
\be
\label{eq:ydef}
y=\frac{p\cdot q}{p\cdot k}=\frac{Q^2}{2xmE}
\ee   
in terms of the nucleon, lepton and virtual photon momenta, $p$,  $k$
and $q$, or, 
in the target rest frame, in terms of the energy $E$ of the incoming
lepton beam. 

The unpolarized structure function $F_1$ and unpolarized
structure function ratio $R$ which enter the definition
Eq.~(\ref{g1toA}-\ref{g2toA}) of the asymmetry may be
expressed in terms of $F_2$ and
$F_L$ by
\begin{eqnarray}\label{eq:fonedef}
F_1(x,Q^2)&\equiv&\frac{F_2(x,Q^2)}{2x\left[1+R(x,Q^2)\right]}
\left(1+\gamma^2\right)
\\\label{eq:Rdef}
R(x,Q^2)&\equiv&\frac{F_L(x,Q^2)}{F_2(x,Q^2)-F_L(x,Q^2)}.
\end{eqnarray}

The longitudinal and transverse asymmetries are sometimes expressed in terms
of the virtual photo-absorption asymmetries $A_1$ and $A_2$ according
to
\be
\label{eq:asyrel}
A_\parallel=D(A_1+\eta A_2)
\mbox{ ,}
\qquad\qquad
A_\perp=d(A_2-\zeta A_1),
\ee
where 
\be
\label{eq:gammaasy}
A_1(x,Q^2)
\equiv 
\frac{\sigma^T_{1/2}-\sigma^T_{3/2}}{\sigma^T_{1/2}+\sigma^T_{3/2}}
\mbox{ ,}
\qquad\qquad
A_2(x,Q^2)
\equiv
\frac{2\sigma^{TL}}{\sigma^T_{1/2}+\sigma^T_{3/2}}.
\ee
Recall that  $\sigma^T_{1/2}$ and $\sigma^T_{3/2}$
are  cross sections for the scattering of
virtual transversely polarized photons
(corresponding to longitudinal lepton polarization)
with helicity of the photon-nucleon system equal to 1/2 and 3/2
respectively, and $\sigma^{TL}$ denotes the interference term between the
transverse and longitudinal photon-nucleon amplitudes.
In the limit $m^2\ll Q^2$ Eqs.~(\ref{eq:asyrel}) reduce to $D=A_\parallel/A_1$,
$d=A_\perp/A_2$, thereby providing a physical interpretation of
$d$ and $D$ as depolarization factors.
 
Using Eqs.~(\ref{eq:asyrel}) in Eqs.~(\ref{g1toA}-\ref{g2toA}) we may
express the structure functions in terms of $A_1$ and $A_2$ instead:
\begin{align}
\label{g1toA1}
g_1(x,Q^2) &= \frac{F_1(x,Q^2)}{1+\gamma^2} \left[ A_1(x,Q^2) 
+ \gamma A_2 (x,Q^2) \right],\\\label{g2toA2}
g_2(x,Q^2)&=
\frac{F_1(x,Q^2)}{1+\gamma^2}
\left[\frac{A_2}{\gamma}- A_1\right] .
\end{align}

We are interested  in the structure function $g_1(x,Q^2)$,
whose moments are proportional to nucleon matrix elements of twist-two
longitudinally polarized quark and gluon operators, and therefore can
be expressed in terms of longitudinally polarized quark and gluon distributions.
Using Eqs.~(\ref{g1toA}-\ref{g2toA}) 
we may obtain an expression of it in terms of
the two asymmetries $A_{\parallel}$, $A_{\perp}$, or, using 
Eqs.~(\ref{g1toA1}-\ref{g2toA2}),  in terms of
the two asymmetries $A_1$, $A_2$. Clearly, up to corrections of
${\mathcal O}\left(\frac{m}{Q}\right)$, $g_1$ is fully determined by
$A_{\parallel}$, which coincides with $A_1$ up to
${\mathcal O}\left(\frac{m}{Q}\right)$ terms, while $g_2$ is
determined by $A_{\perp}$ or $A_2$.
It follows that, even though in principle a measurement of both
asymmetries is necessary for the determination of $g_1$, in practice
most of the information comes from $A_{\parallel}$ or $A_1$, with the
other asymmetry only providing a relatively small correction unless
$Q^2$ is very small. 

It may thus be convenient to express $g_1$ in terms of
$A_{\parallel}$
and $g_2$:
\be
\label{eq:g1tog2}
g_1(x,Q^2)
=
\frac{F_1(x,Q^2)}{1+\gamma\eta}\frac{A_{\parallel}}{D}
+\frac{\gamma(\gamma-\eta)}{\gamma\eta+1}g_2(x,Q^2),
\ee
or, equivalently, in terms of  $A_1$ and $g_2$:
\be
\label{eq:g1tog2p}
g_1(x,Q^2) = A_1(x,Q^2) F_1(x,Q^2) + \gamma^2 g_2(x,Q^2).
\ee
It is then possible to use Eq.~(\ref{eq:g1tog2}) or
Eq.~(\ref{eq:g1tog2p}) to determine $g_1(x,Q^2)$ from a dedicated
measurement of the longitudinal asymmetry, and an independent
determination of $g_2(x,Q^2)$.

In practice, experimental information on the transverse asymmetry and
structure function $g_2$ is
scarce~\cite{Abe:1998wq,Anthony:2002hy,Airapetian:2011wu}. 
However, the Wilson expansion for polarized DIS implies 
that the structure function $g_2$ can be
written as the sum of a twist-two
and a twist-three contribution~\cite{Wandzura:1977qf}:
\be
g_2(x,Q^2)=g_2^{\rm t2}(x,Q^2)+g_2^{\rm t3}(x,Q^2).
\ee
The twist-two contribution to $g_2$ is simply related
to $g_1$. One finds
\be
g_2^{\rm t2}(x,Q^2)= -g_1(x,Q^2)+\int_x^1\frac{dy}{y} g_1(y,Q^2)
\label{wweq}
\ee
which in Mellin space becomes
\be
g_2^{\rm t2}(N,Q^2)= -\frac{N-1}{N}g_1(N,Q^2).
\label{wweqN}
\ee
It is important to note that $g_2^{\rm t3}$ is
not suppressed by a power of $\frac{m}{Q}$  in comparison to
$g_2^{\rm t2}$, because in the polarized case the availability of the spin
vector allows the construction of an extra scalar
invariant. Nevertheless, experimental evidence 
suggests that $g_2^{\rm t3}$ is compatible with zero at low scale
$Q^2\sim m^2$. Fits to $g_2^{\rm t3}$~\cite{Accardi:2009au,Blumlein:2012se}, 
as well as theoretical
estimates of it~\cite{Accardi:2009au,Braun:2011aw} support the
conclusion that
\be
g_2(x,Q^2)\approx g_2^{\rm t2}(x,Q^2)\equiv g_2^{\rm WW}(x,Q^2),
\label{eq:wwrel}
\ee
which is known as the  Wandzura-Wilczek~\cite{Wandzura:1977qf}
relation. 

We will thus determine $g_1$, using Eq.~(\ref{eq:g1tog2}) or
Eq.~(\ref{eq:g1tog2p}), from an experimental determination of the
longitudinal asymmetry, and using the approximate  Wandzura-Wilczek
form Eq.~(\ref{eq:wwrel}) of $g_2$. 
In order to test the dependence of results on this approximation, we will
also consider the opposite assumption that $g_2=0$ identically. 

\subsection{The dataset: observables, kinematic cuts, uncertainties and
correlations}
\label{sec:datasetl}

We use deep-inelastic lepton-nucleon scattering (DIS) data 
coming from all relevant
experiments~\cite{Ashman:1989ig,Anthony:2002hy,Adeva:1998vv,Adeva:1999pa,Abe:1998wq,
Abe:1997cx,Anthony:2000fn,
Alexakhin:2006vx,Alekseev:2010hc,Ackerstaff:1997ws,Airapetian:2007mh}
performed at CERN, SLAC and DESY. 
The experiments use different nucleon targets (protons, neutrons or
deuterons).
The main
features of these data sets are summarized in
Tab.~\ref{tab:exps-sets}, where we show, for each experiment,
the number of available data points, the kinematic range covered by the
experiment, and the quantity which is published and which we
use for the extraction of $g_1$. This quantity 
is not the same for all
experiments: the primary observable can be one of the many asymmetries or structure functions
discussed in Sect.~\ref{sec:asysf}, as we now summarize (individual
experiments are labeled as in Tab.~\ref{tab:exps-sets}).

\begin{itemize}

\item {\bf EMC, SMC, SMClowx, COMPASS, HERMES97}

All these experiments have performed a measurement of
$A_\parallel$. They have then determined $A_1$ from it using
Eq.~(\ref{eq:asyrel}), under the assumption $\eta\approx0$. Therefore,
what these experiments actually publish is a measurement of
$\frac{A_\parallel}{D}$. We determine $g_1$ from $\frac{A_\parallel}{D}$
using Eq.~(\ref{eq:g1tog2}). This is possible because $D$
is completely fixed by Eq.~(\ref{eq:Ddef}) in terms of the unpolarized
structure function ratio Eq.~(\ref{eq:Rdef}) and of the kinematics. We
determine the unpolarized structure function ratio using as primary
inputs $F_2$, for which we use the parametrization of
Ref.~\cite{Forte:2002fg,DelDebbio:2004qj}, and $F_L$, which we
determine from its
expression in terms of parton distributions, using the \texttt{NNPDF2.1 NNLO}
parton set~\cite{Ball:2011uy}. 

\item{\bf HERMES} 

This experiment has performed a measurement of  
$A_\parallel$, and it
publishes both $A_\parallel$ and
  $A_1$ (which is determined using Eq.~(\ref{eq:asyrel}) and a
parametrization of $A_2$). We use the published values of
$A_\parallel$, which are closer to the experimentally measured
quantity, to determine $g_1$ through Eq.~(\ref{eq:g1tog2}).

\item{\bf E143} 

This experiment has taken data with three different
  beam energies,  $E_1=29.1$ GeV,
$E_2=16.2$ GeV, $E_3=9.7$ GeV. For the highest energy both
  $A_\parallel$ and $A_\perp$ are independently measured and $A_1$ is
  extracted from them using Eq.~(\ref{eq:asyrel}); for the two
  lowest energies only   $A_\parallel$ is measured and 
$A_1$ is extracted from it using Eqs.~(\ref{g1toA1}-\ref{g2toA2})
  while assuming the form Eq.~(\ref{eq:wwrel}) for $g_2$. The values of
  $A_1$ obtained with the three beam energies are combined into a
  single determination of $A_1$; radiative corrections are applied at
  this combination stage. Because of this, we must use this combined
  value of $A_1$, from which we then determine $g_1$
using Eq.~(\ref{eq:g1tog2p}). In order to determine  $y$
Eq.~(\ref{eq:ydef}), which depends on the beam energy, we use the
mean of the three energies.

\item{\bf E154} 

This experiment measures $A_\parallel$ and $A_\perp$
independently, and then extracts a determination of $A_1$. 
We use these values of $A_1$ to determine $g_1$ by means of Eq.~(\ref{eq:g1tog2p}).

\item{\bf E155} 

This experiment only measures $A_\parallel$, from
which $\frac{g_1}{F_1}$ is extracted using Eq.~(\ref{g1toA1}) with
the Wandzura-Wilczek form of $g_2$ Eq.~(\ref{eq:wwrel}). In this
case, we use these values of 
$\frac{g_1}{F_1}$, and we extract $g_1$ using
Eq.~(\ref{eq:fonedef}) for $F_1$, together with
the parametrization of
Ref.~\cite{Forte:2002fg,DelDebbio:2004qj} for $F_2$ and
the expression in terms 
of parton distributions and the \texttt{NNPDF2.1 NNLO}
parton set~\cite{Ball:2011uy} for $F_L$, as in the other cases. 
\end{itemize}

\begin{table}
\begin{center}
 \footnotesize
  \begin{tabular}{|ll|c|c|c|c|c|c|c|}
  \hline
  Experiment & Set                             & $N_{\rm dat}$               & $x_{\rm min}$ 
             &  $x_{\rm max}$                  & $Q^2_{\rm min}$ [GeV$^2$]   &  $Q^2_{\rm max}$  [GeV$^2$] 
                                               & $F$                         & Ref.\\ 
 \hline
 \hline
 \multicolumn{2}{|l|}{EMC} & & & & & & & \\
             &  EMC-A1P                        &  10                         & .0150 
             & .4660                           &  3.5                        & 29.5   
                                               & $A_{\parallel}^p/D$         & \cite{Ashman:1989ig}\\
 \hline
 \multicolumn{2}{|l|}{SMC} & & & & & & & \\
             &  SMC-A1P                        &  12                         & .0050 
             & .4800                           &  1.3                        & 58.0    
                                               & $A_{\parallel}^p/D$         & \cite{Adeva:1998vv}\\
             &  SMC-A1D                        &  12                         & .0050 
             & .4790                           &  1.3                        & 54.8   
                                               & $A_{\parallel}^d/D$         & \cite{Adeva:1998vv}\\
 \hline
 \multicolumn{2}{|l|}{SMClowx} & & & & & & & \\
             &  SMClx-A1P                      &  15 (8)                     & .0001 (.0043)
             & .1210                           &  0.02 (1.09)                & 23.1    
                                               & $A_{\parallel}^p/D$         & \cite{Adeva:1999pa}\\
             &  SMClx-A1D                      &  15 (8)                     & .0001 (.0043) 
             & .1210                           &  0.02 (1.09)                & 22.9   
                                               & $A_{\parallel}^d/D$         & \cite{Adeva:1999pa}\\
 \hline
 \multicolumn{2}{|l|}{E143} & & & & & & & \\
             &  E143-A1P                       &  28 (25)                    & .0310 
             & .5260                           &  1.27                       & 9.52 (7.72)    
                                               & $A_1^p$                     & \cite{Abe:1998wq}\\
             &  E143-A1D                       &  28 (25)                    & .0310 
             & .5260                           &  1.27                       & 9.52 (7.72)    
                                               & $A_1^d$                     & \cite{Abe:1998wq}\\
 \hline
 \multicolumn{2}{|l|}{E154} & & & & & & & \\
             &  E154-A1N                       &  11                         & .0170 
             & .5640                           &  1.2                        & 15.0    
                                               & $A_1^n$                     & \cite{Abe:1997cx}\\
 \hline
 \multicolumn{2}{|l|}{E155} & & & & & & & \\
             &  E155-G1P                       &  22 (20)                    & .0150 
             & .7500 (.5000)                   &  1.22                       & 34.72 (26.86)   
                                               & $g_1^p/F_1^p$               & \cite{Anthony:2000fn}\\
             &  E155-G1N                       &  22 (20)                    & .0150 
             & .7500 (.5000)                   &  1.22                       & 34.72 (26.86)  
                                               & $g_1^n/F_1^n$               & \cite{Anthony:2000fn}\\
 \hline
 \multicolumn{2}{|l|}{COMPASS-D} & & & & & & & \\
             &  CMP07-A1D                      &  15                         & .0046 
             & .5660                           &  1.10                       & 55.3   
                                               & $A_{\parallel}^d/D$         & \cite{Alexakhin:2006vx}\\
 \hline
 \multicolumn{2}{|l|}{COMPASS-P} & & & & & & & \\
             &  CMP10-A1P                      &  15                         & .0046 
             & .5680                           &  1.10                       & 62.1   
                                               & $A_{\parallel}^p/D$         & \cite{Alekseev:2010hc}\\
 \hline
 \multicolumn{2}{|l|}{HERMES97} & & & & & & & \\
             &  HER97-A1N                      &  9 (8)                      & .0330 
             & .4640 (.3420)                   &  1.22                       & 5.25 (3.86)    
                                               & $A_{\parallel}^n/D$         & \cite{Ackerstaff:1997ws}\\
 \hline
 \multicolumn{2}{|l|}{HERMES} & & & & & & & \\
             &  HER-A1P                        &  38 (28)                    & .0264 
             & .7311 (.5823)                   &  1.12                       & 14.29 (11.36)    
                                               & $A_{\parallel}^p$           & \cite{Airapetian:2007mh}\\
             &  HER-A1D                        &  38 (28)                    & .0264 
             & .7311 (.5823)                   &  1.12                       & 14.29 (11.36)    
                                               & $A_{\parallel}^d$           & \cite{Airapetian:2007mh}\\
 \hline
 \multicolumn{2}{|l|}{Total}     &  290 (245) & \multicolumn{6}{|c}{}\\
 \cline{1-3}
 \end{tabular}

\end{center}
\caption{\small 
Experimental data sets  included in the present analysis. 
For each experiment we 
show the number of points before and after (in parenthesis) 
applying kinematic cuts, the kinematic range  and the measured observable.
\label{tab:exps-sets}}
\end{table}

\begin{figure}[t]
\begin{center}
\epsfig{width=0.7\textwidth,figure=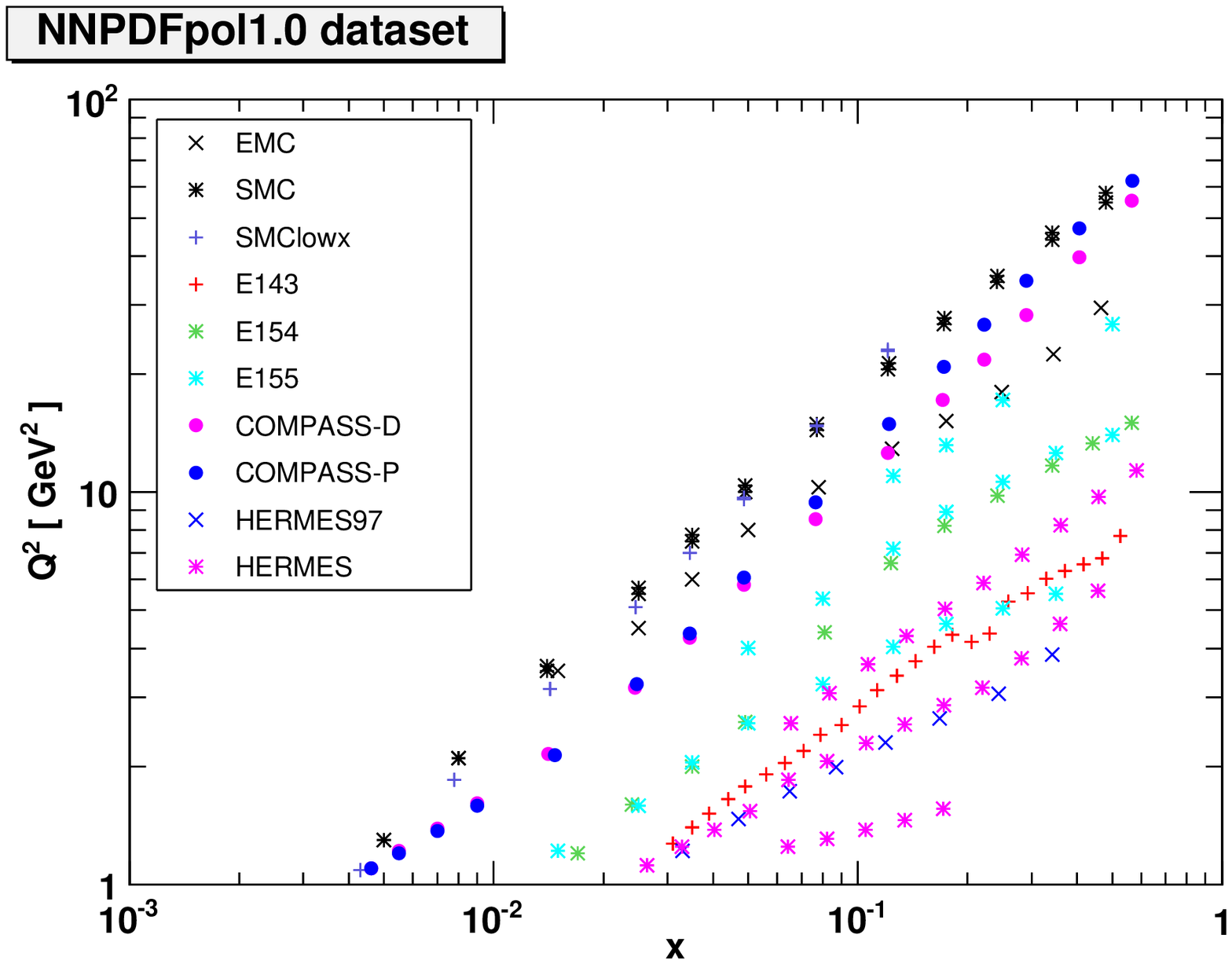}  
\caption{\small Experimental data in the $(x,Q^2)$ plane (after kinematic cuts).}
\label{fig:dataplot}
\end{center}
\end{figure}
We have excluded from our analysis all data points with
$Q^2 \le Q^2_{\rm cut}=1$ GeV$^2$, since below such energy scale
perturbative QCD cannot be considered reliable. A similar choice
of cut was made in Refs.~\cite{Ball:1995td,Altarelli:1996nm,Altarelli:1998nb,
deFlorian:2009vb,Blumlein:2010rn,Hirai:2008aj}.
We further impose a cut on the squared invariant mass of the 
hadronic final state $W^2=Q^2(1-x)/x$
in order to remove points which may be affected by sizable
higher-twist corrections. The cut is chosen based on
a  study presented in
Ref.~\cite{Simolo:2006iw}, where higher twist terms were added to the
observables, with a coefficient fitted to the data, and it was shown
that the higher twist contribution becomes compatible with zero if one
imposes the cut $W^2 \ge W^2_{\rm
  cut}=6.25$ GeV$^2$. We will follow  this choice, which
excludes data points with large Bjorken-$x$ at moderate values of
the squared momentum transfer $Q^2$, roughly corresponding to the
bottom-right corner of the $(x,Q^2)$-plane, see
Fig.~\ref{fig:dataplot}: in particular, it excludes
all available JLAB data~\cite{Zheng:2004ce,Fatemi:2003yh,Dharmawardane:2006zd}.
The 
number of data points surviving the kinematic cuts for each data set
is given in parenthesis in Tab.~\ref{tab:exps-sets}.

As can be seen from the scatter plot in Fig.~\ref{fig:dataplot}, the
region of the $(x,Q^2)$-plane where data are available  
after kinematic cuts is
roughly restricted to $4\cdot 10^{-3}\lesssim x\lesssim 0.6$ and 
$1$~GeV$^2\leq Q^2\lesssim 60$~GeV$^2$. In recent years,
the coverage of the low-$x$ region has been
improved by a complementary set of SMC data~\cite{Adeva:1999pa} and by
the more recent COMPASS
data~\cite{Alexakhin:2006vx,Alekseev:2010hc}. In the
large-$x$ region, information is provided at rather high $Q^2$ by the
same COMPASS data and at lower energy by the latest HERMES
measurements~\cite{Airapetian:2007mh}. In comparison  to the dataset used
in Refs.~\cite{Ball:1995td,Altarelli:1996nm,Altarelli:1998nb}
several new datasets are being used, in particular
the  SMC~\cite{Adeva:1999pa},
HERMES~\cite{Airapetian:2007mh} and
COMPASS~\cite{Alexakhin:2006vx,Alekseev:2010hc} data. 
The dataset used in this paper is the same as that of 
Ref.~\cite{Blumlein:2010rn}, and also the same as the DIS data of the
fit of Ref.~\cite{deFlorian:2009vb}, which however has a wider data
set which extends beyond inclusive DIS.

Each experimental collaboration provides uncertainties on the
measured quantities listed in the next-to-last column of
Tab.~\ref{tab:exps-sets}.
Correlated systematics are only provided by EMC and
E143, which give the values of the
systematics due to the uncertainty in the beam and target 
polarizations, while all other experiments do not provide any
information on the covariance matrix. For each experiment, we determine 
the uncorrelated uncertainty on $g_1$ by combining the uncertainty on
the experimental observable with that of the unpolarized structure
function using standard error propagation. We include all available correlated
systematics. These are provided by the experimental collaboration as
a percentage correction to $g_1$ (or, alternatively, to the asymmetry
$A_1$): we apply the percentage uncertainty on $g_1$ to the structure
function determined by us as discussed in Sect.~\ref{sec:datasetl}
(which, of course, is very close to the value determined by the
experimental collaboration).

We then  construct a covariance matrix
\be
\label{eq:covmat}
{\rm cov}_{pq}=\left(\sum_i
\sigma^{(c)}_{i,p}\sigma^{(c)}_{i,q} + \delta_{pq} \sigma^{(u)}_{p}\sigma^{(u)}_{q}
\right)
g_{1,p}g_{1,q},
\ee
where $p$ and $q$ run over the experimental data points, 
$g_{1,p}\equiv g_1(x_p,Q_p^2)$
($g_{1,q}\equiv g_1(x_q,Q_q^2)$),  $\sigma^{(c)}_{i,p}$ are the
various sources of  correlated uncertainty,  and $\sigma^{(u)}_{p}$
the uncorrelated  uncertainties, which are in turn found as a sum in
quadrature of all uncorrelated sources of statistical
$\sigma^{\rm (stat)}_{i,p}$ and systematic $\sigma^{\rm (syst)}_{i,p}$
uncertainty on each point:
\begin{equation}
\label{uncsum}
\left(\sigma^{(u)}_{p}\right)^2=\sum_i \left(\sigma^{\rm (stat)}_{i,p}\right)^2 +\sum_j
\left(\sigma^{\rm (syst)}_{j,p}\right)^2.
\end{equation}
The correlation matrix is defined as
\be
\label{eq:cormatr}
\rho_{pq} = \frac{{\rm cov}_{pq}}{\sigma^{\rm (tot)}_{p}\sigma^{\rm (tot)}_{q}g_{1,p}g_{1,q}}
\mbox{ ,}
\ee
where the total uncertainty $\sigma^{\rm (tot)}_{p}$ on the $p$-th data point is
\be
\label{eq:sigmatot}
\left(\sigma^{\rm (tot)}_{p}\right)^2 =
(\sigma^{(u)}_{p})^2+\sum_i\left(\sigma^{(c)}_{i,p}\right)^2 
\mbox{ .}
\ee

We show in Tab.~\ref{tab:exps-err} the
average experimental uncertainties for each dataset, with
uncertainties separated into statistical  and correlated systematics. 
All values are given as absolute uncertainties and refer to the
structure function $g_1$, which has been reconstructed for each experiment
as discussed above.
As in the case of Tab.~\ref{tab:exps-sets}, we provide
the values before and after kinematic cuts (if different).

In Tab.~\ref{tab:exps-sets}, we distinguish between
experiments, defined as groups of data which cannot be correlated to
each other, and datasets within a given experiment, which could in
principle be correlated with each other, as they
correspond to measurements of
different observables in the same experiment, or measurements of the
same observable in different years. Even though, in practice, only two
experiments provide such correlated systematics (see Tab.~\ref{tab:exps-err}), 
this distinction will be useful in the 
minimization strategy, see Sect.~\ref{sec:minim} below. 

\begin{table}
\begin{center}
 \footnotesize
  \begin{tabular}{|ll|c|c|c|}
 \hline
 Experiment & Set
 & $\langle\delta{g_1}_{\mbox{\tiny s}}\rangle$
 & $\langle\delta{g_1}_{\mbox{\tiny c}}\rangle$
 & $\langle\delta{g_1}_{\mbox{\tiny tot}}\rangle$\\
 \hline
 \hline
 \multicolumn{2}{|l|}{EMC       } & & & \\
 & EMC-A1P    &        0.144         &        0.037 &        0.150\\
 \hline
 \multicolumn{2}{|l|}{SMC       } & & & \\
 & SMC-A1P    &        0.098         &        --      &        0.098\\
 & SMC-A1D    &        0.116         &        --      &        0.116\\
 \hline
 \multicolumn{2}{|l|}{SMClowx   } & & & \\
 & SMClx-A1P  &       18.379 (0.291) &        -- (--) &       18.379 (0.291)\\
 & SMClx-A1D  &       22.536 (0.649) &        -- (--) &       22.536 (0.649)\\
 \hline
 \multicolumn{2}{|l|}{E143      } & & & \\
 & E143-A1P   &        0.042 (0.046) &  0.009 (0.009) &        0.043 (0.047)\\
 & E143-A1D   &        0.053 (0.058) &  0.004 (0.005) &        0.054 (0.059)\\
 \hline
 \multicolumn{2}{|l|}{E154      } & & & \\
 & E154-A1N   &        0.044         &        --      &        0.044\\
 \hline
 \multicolumn{2}{|l|}{E155      } & & & \\
 & E155-G1P   &        0.040 (0.043) &        -- (--) &        0.040 (0.043)\\
 & E155-G1N   &        0.124 (0.135) &        -- (--) &        0.124 (0.135)\\
 \hline
 \multicolumn{2}{|l|}{COMPASS-D } & & & \\
 & CMP07-A1D  &        0.061 &                --      &        0.061\\
 \hline
 \multicolumn{2}{|l|}{COMPASS-P } & & & \\
 & CMP10-A1P  &        0.101 &                --      &        0.101\\
 \hline
 \multicolumn{2}{|l|}{HERMES97  } & & & \\
 & HER97-A1N  &        0.087 (0.093) &        -- (--) &        0.087 (0.093)\\
 \hline
 \multicolumn{2}{|l|}{HERMES    } & & & \\
 & HER-A1P    &        0.067 (0.062) &        -- (--) &        0.067 (0.062)\\
 & HER-A1D    &        0.040 (0.034) &        -- (--) &        0.040 (0.034)\\
 \hline
 \end{tabular}

\end{center}
\caption{\small Averaged statistical, correlated systematic and total
uncertainties before and after (in parenthesis) kinematic cuts for each of 
the experimental sets included in the present analysis. Uncorrelated systematic
uncertainties are considered as part of the statistical uncertainty
and they are added in quadrature. 
All values are absolute uncertainties and refer to the structure function $g_1$,
which has been reconstructed for each experiment as discussed in the text.
Details on the number of points and
the kinematics of each dataset are provided in Tab.~\ref{tab:exps-sets}.
\label{tab:exps-err}}
\end{table}

\subsection{Monte-Carlo generation of the pseudo-data sample}

Error propagation from experimental data to the fit is handled by a
Monte Carlo sampling of the probability distribution defined by
data. The statistical sample is obtained by generating
$N_\mathrm{rep}$ pseudodata replicas, according to a multigaussian
distribution centered at the data points and with a covariance equal
to that of the original data.
Explicitly, given an
experimental data point $g_{1,p}^{(\mathrm{exp})}\equiv
g_1(x_p,Q_p^2)$, we generate $k=1,\dots,N_\mathrm{rep}$ artificial
points $g_{1,p}^{(\mathrm{art}),(k)}$ according to \be
\label{eq:MCgeneration}
g_{1,p}^{(\mathrm{art}),(k)} (x,Q^2)
= 
\left[
1+\sum_i r_{(c),p}^{(k)}\sigma_{i,p}^{(c)} + r_{(u),p}^{(k)}\sigma_p^{(u)}
\right]
g_{1,p}^{(\mathrm{exp})} (x,Q^2),
\ee
where $r_{(c),p}^{(k)}$, $r_{(u),p}^{(k)}$ are univariate gaussianly distributed 
random numbers, and $\sigma_{i,p}^{(c)}$ and $\sigma_{p}^{(u)}$ are respectively
the relative correlated systematic and
statistical uncertainty. Unlike in the unpolarized case, 
Eq.~(\ref{eq:MCgeneration}) receives no contribution from normalization uncertainties, given that
all polarized observables are obtained as cross section asymmetries.

\begin{figure}[t]
\begin{center}
\epsfig{width=0.4\textwidth,figure=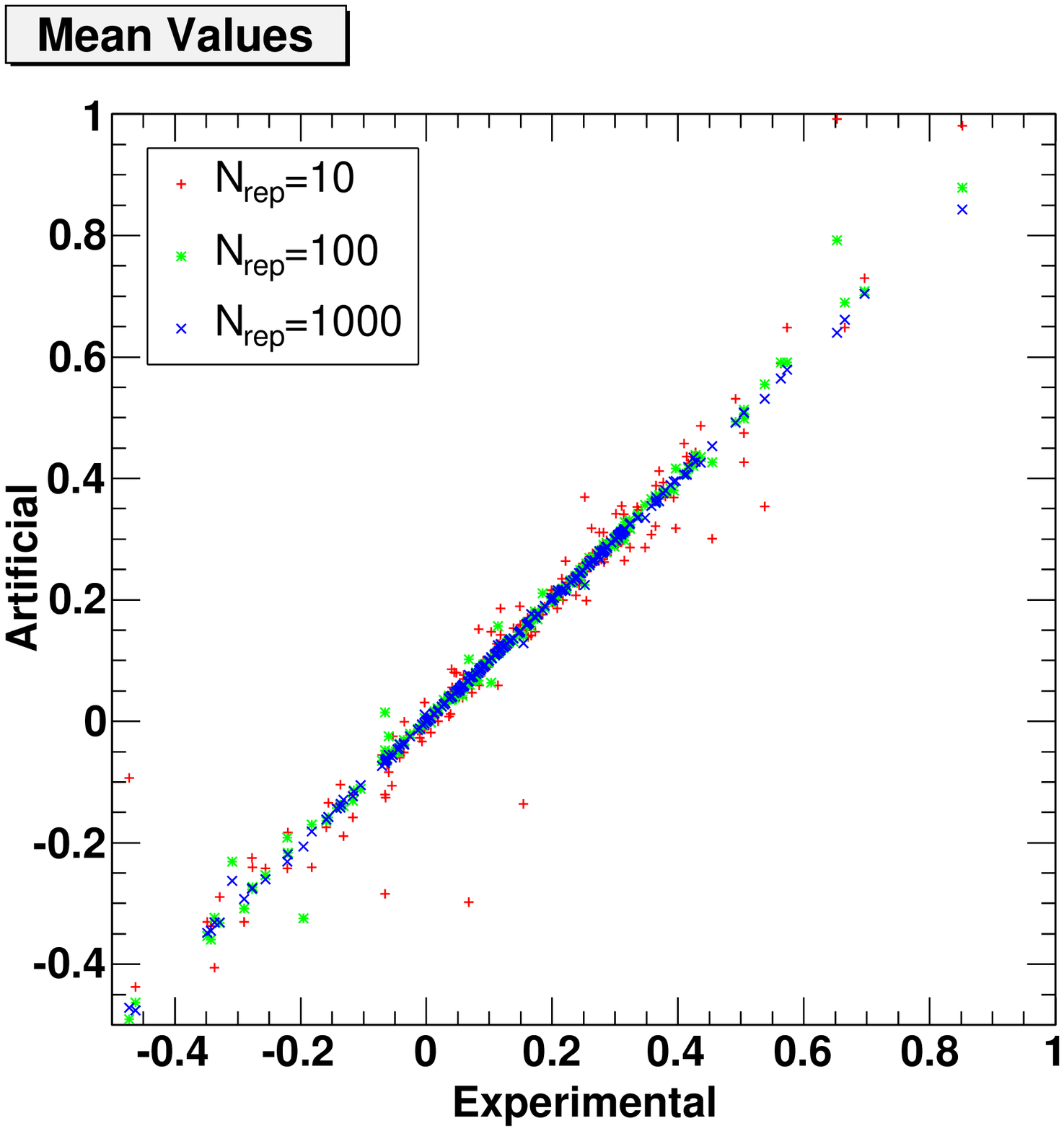}  
\epsfig{width=0.4\textwidth,figure=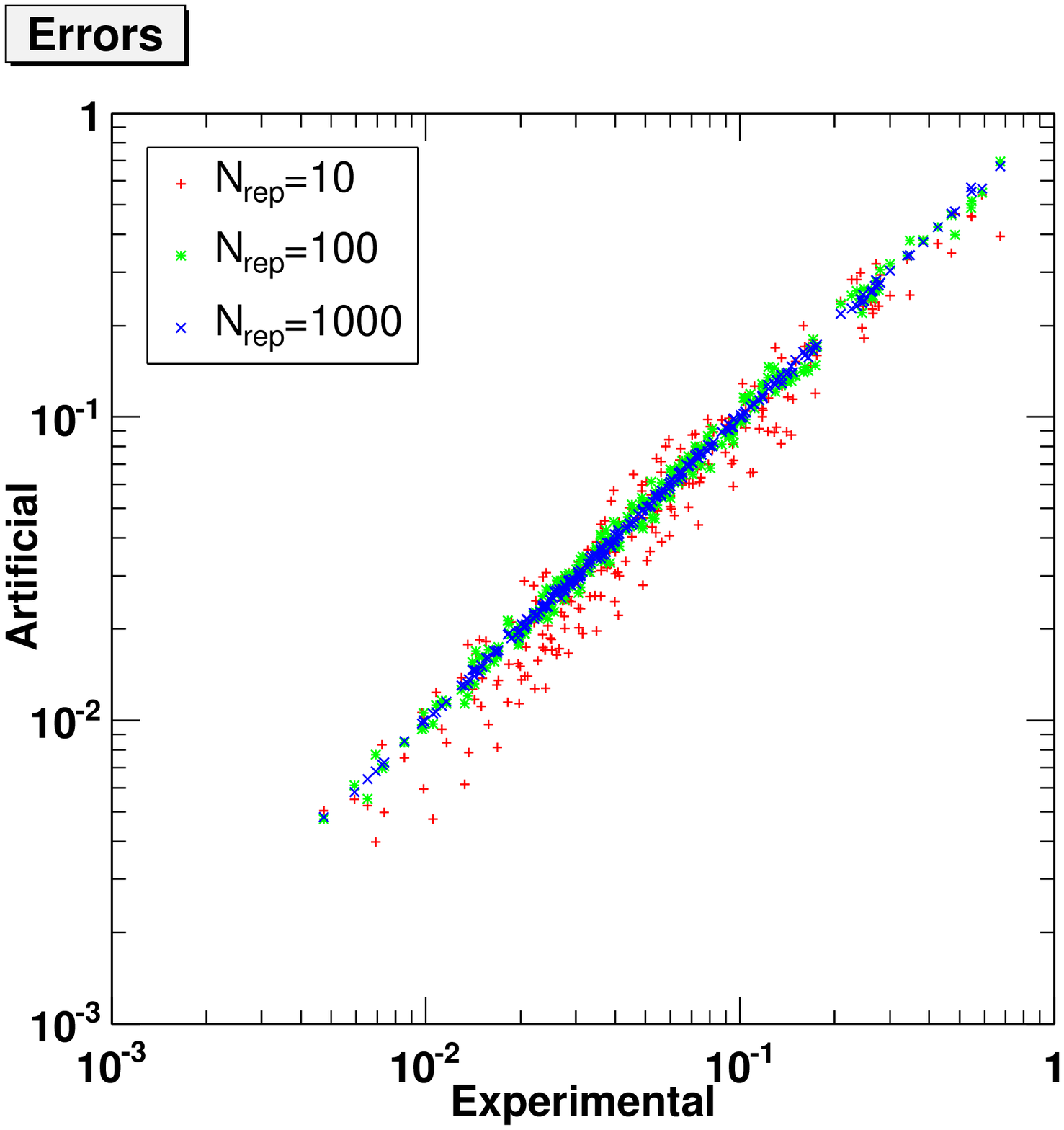} 
\caption{\small Scatter-plot of experimental versus artificial Monte Carlo mean central values and absolute uncertainties of polarized
structure functions computed from 
ensembles made of $N_{\mbox{\scriptsize{rep}}}=10,100,1000$ replicas.}
\label{fig:splots}
\end{center}
\end{figure}

The number of Monte Carlo replicas of the data is determined by
requiring that the central values, uncertainties and  correlations of the
original experimental data can be reproduced to a given accuracy by
taking  averages, variances and
covariances over the replica sample. 
A comparison between expectation values and variances of the Monte
Carlo set and the corresponding input experimental values as a
function of the number of replicas is shown in Fig.~\ref{fig:splots},
where we display scatter-plots of the central values and errors for
samples of $N_{\mbox{\scriptsize{rep}}}=10,100$ and $1000$ replicas.
A more quantitative comparison can be performed by defining suitable
statistical estimators (see, for example, Appendix B of
Ref.~\cite{DelDebbio:2004qj}).

We show in Tabs.~\ref{tab:est1gen}--\ref{tab:est2gen} the percentage
error and the scatter correlation $r$  (which is crudely speaking the
correlation between the input value and the value computed from the
replica sample) for central values and errors respectively . We do not compute values for correlations, as these,
as seen in Tab.~\ref{tab:exps-err},
are only available for a very small number of data points from two
experiments. 
 Note that
some large values of the percentage uncertainty are due to the fact
that $g_1$ for some experiments can take values which are very close
to zero. It is clear from both the tables and the plots that a Monte Carlo 
sample of pseudo-data with $N_\mathrm{rep}=100$ is
sufficient to reproduce the mean values and the errors of experimental
data to an accuracy which is better than 5\%, while the improvement in
going up to $N_\mathrm{rep}=1000$ is moderate. Therefore, we will
henceforth use a $N_\mathrm{rep}=100$ replica sample as a default 
in the remainder of this paper.

\begin{table}[t]
\centering
\footnotesize
\begin{tabular}{|c|l|ccc|ccc|}
\hline
& Estimator
& \multicolumn{3}{c|}{$\left\langle \mbox{PE} \left[\langle g_1^{\mbox{\scriptsize{(art)}}}\rangle \right] \right\rangle$ [\%]}
& \multicolumn{3}{c|}{$r\lc g_1^{\mbox{\scriptsize{(art)}}}\rc$} \\
\hline
\hline
& $N_{\mbox{\scriptsize{rep}}}$ & 10 & 100 & 1000 & 10 & 100 & 1000 \\
\hline
\hline
\multirow{10}{*}{\begin{sideways}Experiment\end{sideways}}
& EMC          
 &   23.7 &    3.5 & 2.9 
 & .76037   & .99547   & .99712 \\
& SMC         
 &   19.4 & 5.6    & 1.2   
 & .94789   & .99908   & .99993 \\
& SMClowx     
 &  183 & 25.8  & 15.4 
 & .80370   & .99239   & .99960 \\
& E143        
 &   18.5 & 5.7    & 2.1 
 & .99159   & .99860   & .99984 \\
& E154        
 &   239 & 44.0   & 21.9
 & .99635   & .99981   & .99994 \\
& E155        
 &   37.3 & 13.4   & 4.3   
 & .99798   & .99993   & .99998 \\
& COMPASS-D   
 &   26.4 & 8.6    & 3.2
 & .96016   & .98774   & .99917 \\
& COMPASS-P   
 & 16.4   & 1.9    & 1.5  
 & .91942   & .99829   & .99902 \\
& HERMES97    
 & 22.5   & 6.2    & 2.2 
 & .96168   & .99762   & .99979 \\
& HERMES      
 & 10.5   & 5.8    & 1.2 
 & .98564   & .99916   & .99973 \\
\hline
\end{tabular}

\caption{\small Table of statistical estimators for the mean value computed from
the Monte Carlo sample with $N_\mathrm{rep}=10,100,1000$ replicas.
Estimators refer to individual experiments and are defined in Appendix B of Ref.~\cite{DelDebbio:2004qj}.}
\label{tab:est1gen}
\end{table}
\begin{table}[t]
\centering
\footnotesize
\begin{tabular}{|c|l|ccc|ccc|}
\hline
& Estimator
& \multicolumn{3}{c|}{$\left\langle \mbox{PE} \left[\langle \delta g_1^{\mbox{\scriptsize{(art)}}}\rangle \right] \right\rangle$ [\%]}
& \multicolumn{3}{c|}{$r\lc \delta g_1^{\mbox{\scriptsize{(art)}}}\rc$} \\
\hline
\hline
& $N_{\mbox{\scriptsize{rep}}}$ & 10 & 100 & 1000 & 10 & 100 & 1000 \\
\hline
\hline
\multirow{10}{*}{\begin{sideways}Experiment\end{sideways}}
& EMC          
 &   12.8 &    4.9 & 2.0 
 & .97397   & .99521   & .99876 \\
& SMC         
 &   22.4 & 5.4    & 1.7    
 & .96585   & .99489   & .99980 \\
& SMClowx     
 &  16.9 & 6.2  & 2.1 
 & .97959   & .99490   & .99905 \\
& E143        
 &   16.0 & 7.4    & 2.0 
 & .95646   & .98684   & .99946 \\
& E154        
 &   19.1 & 3.7   & 1.3
 & .99410   & .99871   & .99992 \\
& E155        
 &   21.2 & 5.6   & 1.8   
 & .99428   & .99971   & .99997 \\
& COMPASS-D   
 &   15.5 & 5.2    & 1.6
 & .99375   & .99687   & .99993 \\
& COMPASS-P   
 & 18.4   & 7.4    & 1.5  
 & .99499   & .99005   & .99988 \\
& HERMES97    
 & 17.9   & 6.4    & 1.6 
 & .89065   & .97318   & .99894 \\
& HERMES      
 & 19.5   & 6.0    & 1.6 
 & .91523   & .99237   & .99942 \\
\hline
\end{tabular}

\caption{\small Table of statistical estimators for the errors computed from
the Monte Carlo sample with $N_\mathrm{rep}=10,100,1000$ replicas.
Estimators refer to individual experiments and are defined in Appendix B of Ref.~\cite{DelDebbio:2004qj}.}
\label{tab:est2gen}
\end{table}

\section{From polarized PDFs to observables}
\label{sec:polpdfs}

\subsection{Leading-twist factorization of the structure functions}
\label{sec:fact}

At leading twist, the 
polarized structure function $g_1$ for neutral-current virtual photon
DIS is given in terms of the  polarized quark
and gluon distributions by
\beq
g_1(x,Q^2)=\frac{\langle e^2 \rangle}{2} [C_{NS}\otimes \Delta q_{NS}
+C_S\otimes \Delta \Sigma + 2n_fC_g \otimes \Delta g]\ .
\label{1}
\eeq
Here $n_f$ is the number of active flavors, the average charge is given by
$\langle e^2 \rangle=n_f^{-1}\sum_{i=1}^{n_f} e^2_i$ in terms of the
electric charge $e_i$ of the $i$-th quark flavor, $\otimes$ denotes the
convolution with respect to $x$, and the nonsinglet and singlet quark
distributions are defined as
\beq
\Delta q_{NS}\equiv \sum_{i=1}^{n_f}
\left(\frac{e^2_i}{\langle e^2 \rangle} - 1\right)
(\Delta q_i+\Delta \bar q_i),\qquad
\Delta \Sigma\equiv \sum_{i=1}^{n_f}(\Delta q_i+\Delta \bar q_i),
\label{2}
\eeq
where $\Delta q_i$ and $\Delta \bar q_i$ are the polarized quark and antiquark
distributions of flavor $i$ and $\Delta g$ is the polarized gluon PDF. 

In the parton model, Eq.~(\ref{1}) reduces to 
\beq
\label{g1p-parton}
g_1(x,Q^2)=\frac{1}{2}\sum_{i=1}^{n_f}   e_i^2 \lp \Delta q_i(x,Q^2)
+\Delta \bar q_i (x,Q^2) \rp,
\eeq
but in perturbative QCD the parton model expression is not recovered
even when $\alpha_s\to0$ because at large $Q^2$ the first moment of the gluon
distribution $\int_0^1 \!dx\, \Delta g\sim ({\alpha_s(Q^2))^{-1}}$, 
so the gluon does not decouple from $g_1$ asymptotically. 
Be that as it may, below charm threshold, with $n_f=3$, Eq.~(\ref{1}) can be
rewritten as
\beq
g_1(x,Q^2)=\frac{1}{9}\Delta\Sigma(x,Q^2)
+\frac{1}{12}\Delta T_3(x,Q^2)+\frac{1}{36}\Delta T_8(x,Q^2),
\eeq
in terms of the singlet quark-antiquark distribution $\Delta\Sigma(x,Q^2)$, defined in 
Eq.~(\ref{2}), the isospin triplet  combination
\begin{equation}\label{tripdef}
\Delta T_3(x,Q_0^2)=\Delta u(x,Q_0^2) + \Delta \bar{u}(x,Q_0^2)
-\left[\Delta d(x,Q_0^2)+\Delta \bar{d}(x,Q_0^2)\right],
\end{equation}
and the SU(3) octet combination
\begin{equation}
\Delta T_8(x,Q_0^2) =\Delta u(x,Q_0^2)+\Delta \bar{u}(x,Q_0^2)
+\Delta d(x,Q_0^2)+\Delta \bar{d}(x,Q_0^2)
-2\left[\Delta s(x,Q_0^2)+\Delta \bar{s}(x,Q_0^2)\right].
\end{equation}

It is clear from Eqs.~(\ref{1}-\ref{g1p-parton}) that neutral current
$g_1$ data only allow for a direct determination of the four polarized
PDF combinations $\Delta g$, $\Delta\Sigma$, $\Delta T_3$ and $\Delta
T_8$. In principle, an intrinsic polarized component could also be present
for each heavy flavour. However, we will neglect it here and assume that heavy quark PDFs are dynamically generated above
threshold by (massless) Altarelli-Parisi evolution, in a zero-mass
variable-flavor number (ZM-VFNS) scheme. In such a scheme all heavy
quark mass effects are neglected. While they can be introduced for instance
through the \texttt{FONLL} method~\cite{Forte:2010ta}, these effects have been 
shown to be relatively small already on the scale of present-day unpolarized 
PDF uncertainties, and thus are most likely negligible in the polarized case 
where uncertainties are rather larger.

The proton and neutron PDFs are related to each other by isospin,
which we will assume to be exact, thus yielding
\begin{equation}
\Delta u^p=\Delta d^n,\quad \Delta d^p=\Delta u^n, \quad \Delta s^p=\Delta s^n,
\end{equation}
and likewise for the polarized anti-quarks. In the following we will
always assume that PDFs refer to the proton.  The first moment of all
non-singlet combinations of quark and antiquark distributions are
scale-independent because of axial current conservation, while the
first moment of the singlet quark distribution is not.  Because of the
axial anomaly, the first moment of the singlet quark distribution is
scale-dependent in the $\overline{\rm MS}$ scheme. However, it may be
convenient to choose a factorization scheme in which the first moment
of the singlet quark distribution is also scale independent so that all
the individual quark and antiquark spin fractions are scale
independent. Several such schemes, including the so-called
Adler-Bardeen (AB) scheme, were discussed in Ref.~\cite{Ball:1995td},
where the transformation connecting them to the $\overline{\rm MS}$
scheme was constructed explicitly.

By means of the SU(2) or SU(3) flavour symmetry it is possible to relate the first moments of the nonsinglet 
$C$-even combinations ($\Delta T_3$ and $\Delta T_8$)  to the baryon octet decay constants $a_3$ and 
$a_8$:
\begin{align}
\label{eq:t3sr}
a_3&= \int_0^1 dx~ \Delta T_3(x,Q^2),
\\
\label{eq:t8sr}
a_8&=\int_0^1 dx~ \Delta T_8(x,Q^2),
\end{align}
whose current experimental values are~\cite{Nakamura:2010zzi}
\begin{equation}\label{eq:a3}
a_3=g_A=1.2701\pm0.0025,
\end{equation}
\begin{equation}
\label{eq:a8}
a_8 = 0.585 \pm 0.025.
\end{equation}
A much larger uncertainty on the octet axial charge,  up to about 30\%, is found if SU(3) symmetry is 
violated~\cite{FloresMendieta:1998ii}. Even though a detailed phenomenological analysis does not 
seem to support this conclusion~\cite{Cabibbo:2003cu}, we will 
take as default this more conservative 
uncertainty estimation
\begin{equation}
\label{eq:a8p}
a_8 = 0.585 \pm 0.176 .
\end{equation}
The impact of replacing this with the more aggressive determination given in Eq.~(\ref{eq:a8}) will be 
studied in Sect.~\ref{sec:srres}.

Structure functions will be computed in terms of polarized parton
distributions using the so-called NNPDF {\tt FastKernel} method,
introduced in Ref.~\cite{Ball:2010de}. In short, in this method the
PDFs at scale $Q^2$ are obtained by convoluting the parton
distributions at the parametrization scale $Q_0^2$ with a set of Green's
functions, which are in turn obtained by solving the QCD evolution
equations in Mellin space.  These Green's functions are then convoluted
with coefficient functions, so that the structure function can be
directly expressed in terms of the PDFs at the parametrization scale
through suitable kernels $K$.  In terms of the polarized PDFs at the
input scale we have
\be
\label{Kg1p}
g_1^p=\lbrace 
        K_{{\rm g1},\Delta\Sigma}\otimes \Delta \Sigma_0 
+K_{{\rm g1},\Delta g} \otimes \Delta g_0 
+K_{{\rm g1},+} \otimes  \left(\Delta T_{3,0}
                + \smallfrac{1}{3}\Delta T_{8,0} \right)\rbrace\ ,
\ee
where the kernels $K_{{\rm g1},\Delta\Sigma},K_{{\rm g1},\Delta g},
K_{{\rm g1},+}$ take into account both the coefficient functions and
$Q^2$ evolution. This way of expressing structure
functions is amenable to numerical optimization, because all kernels
can then be precomputed and stored, and convolutions may be reduced to
matrix multiplications by projecting onto a set of suitable basis
functions. 

The neutron polarized structure function $g_1^n$ is given
in terms of the proton and deuteron ones as
\be
  \label{eq:g1n}
  g_1^n = 2\frac{g_1^d}{1-1.5\omega_D}-g_1^p\ ,
\ee
with $\omega_D=0.05$ the probability that the deuteron is found
in a D state.
Under the assumption of exact isospin symmetry, the expression
of $g_1^n$ in terms of parton densities is obtained from Eq.~\eqref{Kg1p}
by interchanging the up and down quark PDFs, which amounts
to changing the sign of $\Delta T_3$.

The implementation of the polarized PDF evolution up to NLO
has been benchmarked against the {\tt HOPPET} evolution
code~\cite{Salam:2008qg} using the settings of the Les Houches PDF
evolution benchmark tables~\cite{Jung:2009eq}. This benchmarking is
discussed in more detail in Appendix~\ref{sec:apppdfevol}.  We will
assume the values $\alpha_s\lp M_Z^2\rp=0.119$ for the strong coupling
constant and $m_c=1.4$~GeV and $m_b=4.75$~GeV for the charm and bottom
quark masses respectively.

\subsection{Target mass corrections to $g_1$}
\label{sec:tmc}

The leading twist expressions of structure functions given in Sect.~\ref{sec:fact} are corrected both by dynamical 
and kinematic higher-twist terms. The former are related to the contribution of higher twist operators to the Wilson 
expansion, and are generally expected to be small. The latter are related to target-mass corrections (TMCs),
and because of their kinematical origin they can be included
exactly: we do this following Ref.~\cite{Piccione:1997zh}. 
As discussed in Sect.~\ref{sec:asysf}, we thus consistently include all nucleon mass effects, both in the relation between 
measured asymmetries and structure functions, and in the relation between the latter and parton distributions.

The target mass corrections  are especially simple in Mellin space, where they take the form~\cite{Piccione:1997zh}
\bea
&&\tilde g_1(N,Q^2) = g_1(N,Q^2)+\frac{m^{2}}{Q^2}\frac{N(N+1)}{(N+2)^2}
\left[(N+4)~g_1(N+2,Q^2)+4\frac{N+2}{N+1}~g_2(N+2,Q^2)\right]+{\mathcal O}\left(\frac{m^2}{Q^2}\right)^2\ ,
\nonumber\\
\label{g1n1bis}
\\
&&\tilde g_2(N,Q^2)=g_2(N,Q^2)+\frac{m^2}{Q^2}\frac{N(N-1)}{(N+2)^2}
\left[N\frac{N+2}{N+1}g_2(N+2,Q^2)-g_1(N+2,Q^2)\right]+{\mathcal O}\left(\frac{m^2}{Q^2}\right)^2\ .
\label{g2n1bis}
\eea
We denote by $\tilde g_{1,2}(N,Q^2)$ the Mellin space
structure functions with TMCs included, while $g_{1,2}(N,Q^2)$
are the structure functions determined in the  $m=0$ limit.

As discussed in Sect.~\ref{sec:asysf}, in the absence of precise data
on the structure function $g_2$, we will either determine it using the
Wandzura-Wilczek approximation Eq.~(\ref{eq:wwrel}) (which is
uncorrected by target-mass effects~\cite{Piccione:1997zh}), or, as a
cross-check, simply setting it to zero. In either case, we may then
determine $\tilde g_1$ Eq.(\ref{g1n1bis}) in terms of $g_1$. 

In the former (Wandzura-Wilczek) case, substituting  Eq.~(\ref{wweqN}) 
in Eq.~(\ref{g1n1bis}) and taking the inverse 
Mellin transform, we get
\beq
\tilde g_1(x,Q^2)=
\frac{1}{2\pi i}\int dN\,x^{-N}\left[1
+\frac{m^2x^2}{Q^2}
\frac{(N-2)^2(N-1)}{N^2}\right]g_1(N,Q^2)\ ,
\label{g1tmc1ww}
\eeq
where we have shifted $N\to N-2$ in the term proportional to $m^2$.
Inverting the Mellin transform we then obtain
\beq
\tilde g_1(x,Q^2)=g_1(x,Q^2)
+\frac{m^2x^2}{Q^2}
\left[-5g_1(x,Q^2)-x\frac{dg_1(x,Q^2)}{dx}
+\int_x^1\frac{dy}{y}\left(8g_1(y,Q^2)
+4g_1(y,Q^2)\log\frac{x}{y}\right)\right]\ .
\label{g1xWW}
\eeq

If instead $g_2=0$,
\beq
\tilde g_1(x,Q^2)=
\frac{1}{2\pi i}\int dN\,x^{-N}\left[1
+\frac{m^2x^2}{Q^2}\frac{(N^2-4)(N-1)}{N^2}\right]
g_1(N,Q^2)\ ,
\label{g1tmc10}
\eeq
whence
\beq
\tilde g_1(x,Q^2)=g_1(x,Q^2)
+\frac{m^2x^2}{Q^2}
\left[-g_1(x,Q^2)-x\frac{dg_1(x,Q^2)}{dx}
-\int_x^1\frac{dy}{y}\left(4g_1(y,Q^2)
+4g_1(y,Q^2)\log\frac{x}{y}\right)\right].
\label{g1x0}
\eeq 

The numerical implementation of  Eqs.~(\ref{g1xWW}) or
Eq.~(\ref{g1x0}) is difficult, because of the presence
of the first derivative of $g_1$ in the correction term.
Therefore, we will  include target mass
effects in an iterative way: 
we start by performing a fit in which we set $m=0$ and
at each iteration the target mass corrected $g_1$ structure function
is computed by means of Eqs.~(\ref{g1xWW}--\ref{g1x0}) using the   
$g_1$ obtained in the previous minimization step.

\section{Neural networks and fitting strategy}
\label{sec:minim}

We will now briefly review the NNPDF methodology for parton
parametrization in terms of neural networks, and their optimization
(fitting) through a genetic algorithm. The details of the procedure 
have been discussed in previous NNPDF papers, in particular
Refs.~\cite{Ball:2008by,Ball:2010de,Rojo:2004iq}. Here we summarize
the main steps of the whole strategy, and discuss in greater detail 
some points which are specific to the polarized case.

\subsection{Neural network parametrization}
\label{sec:net-param}

Each of the independent polarized PDFs in the evolution basis
introduced in Sect.~\ref{sec:fact}, $\Delta\Sigma,\Delta g,\Delta T_3$
and $\Delta T_8$, is parametrized using a multi-layer feed-forward
neural network~\cite{Ball:2011eq}. All neural networks have the same
architecture, namely 2-5-3-1, which corresponds to 37 free parameters
for each PDF, and thus a total of 148 free parameters. This is to be
compared to about 10-15 free parameters for all other available
determinations of polarized PDFs. This parametrization has been
explicitly shown to be redundant in the unpolarized case, in that
results are unchanged when a smaller neural network architecture is
adopted: this ensures that results do not depend on the
architecture~\cite{Ball:2011eq}. Given that polarized data are much
less abundant and affected by much larger uncertainties than
unpolarized ones, this architecture is adequate also in the
polarized case.

The neural network parametrization is supplemented with a
preprocessing function. In principle, large enough neural networks
can reproduce any functional form given sufficient training
time. However, the training can be made more efficient by adding a
preprocessing step, i.e. by multiplying the output of the neural
networks by a fixed function. The neural network then only fits the
deviation from this function, which improves the speed of the
minimization procedure if the preprocessing function is suitably
chosen.  We thus write the input PDF basis in terms of preprocessing
functions and neural networks ${\rm NN}_{\rm \Delta pdf}$ as follows
\begin{eqnarray}
\label{eq:PDFbasisnets}
\Delta \Sigma(x,Q_0^2)
&=&{(1-x)^{m_1}}{x^{-n_1}}{\rm NN}_{\Delta
\Sigma}(x)\ ,
\nonumber\\
\Delta T_3(x,Q_0^2)&=&A_3{(1-x)^{m_3}}{
x^{-n_3}}
{\rm NN}_{ \Delta T_3}(x)  \ , \nonumber\\
\Delta T_8(x,Q_0^2)&=&A_8{(1-x)^{m_8}}{
x^{-n_{ \Delta T_8}}}
{\rm NN}_{ \Delta T_3}(x)  \ , \\
\Delta g(x,Q_0^2)&=&{(1-x)^{m_g}}{x^{-n_g}}{\rm NN}_{\Delta g}(x).
\nonumber
\end{eqnarray}

Of course, one should check that no bias is introduced in the choice
of preprocessing functions. To this purpose, we first select a
reasonable range of values for the large and small--$x$
preprocessing exponents $m$ and $n$, and produce a PDF determination
by choosing for each replica a value of the exponents at random with
uniform distribution within this range. We then determine effective 
exponents for each replica, defined as 
\be
\label{eq:effexp2}
m_{\rm eff}(Q^2)\equiv
\lim_{x\to1}\frac{ \ln \Delta f(x,Q^2) }{\ln(1-x)}
\mbox{ ,}
\ee
\be
\label{eq:effexp1}
n_{\rm eff}(Q^2)\equiv
\lim_{x\to0} \frac{\ln \Delta f(x,Q^2)}{\ln\frac{1}{x}}
\mbox{ ,}
\ee
where $\Delta f = \Delta\Sigma\mbox{, }\Delta T_3\mbox{, }\Delta T_8\mbox{, }\Delta g$.
Finally, we check that the range of variation of the preprocessing
exponents is wider than the range of effective exponents for each PDF.
If it is not, we enlarge the range of variation of preprocessing, then
repeat the PDF determination, and iterate
until the condition is satisfied. 
This ensures that the range of effective large- and
small-$x$ exponents found in the fit is not biased, and in particular not
restricted, by the range of preprocessing exponents. Our final values
for the preprocessing exponents are summarized in
Tab.~\ref{tab:prepexps}, while the effective exponents obtained in
our fit will be discussed in Sect.~\ref{sec:prepexp}.
It is apparent from Tab.~\ref{tab:prepexps} that
the allowed range  of preprocessing exponents is rather 
wider than in the unpolarized case, as a
consequence of the limited amount of experimental information. 
It is enough to perform this check at the input evolution
scale, $Q_0^2=1$~GeV$^2$.

\begin{table}[t]
  \begin{center}
    \begin{tabular}{c|c|c}
      \hline PDF & $m$ & $n$ \\
      \hline
\hline
      $\Delta\Sigma(x,Q_0^2)$  & $\lc 1.5,3.5\rc$ & 
      $\lc 0.2,0.7\rc$ \\
      \hline
      $\Delta g(x,Q_0^2)$  & $\lc  2.5,5.0\rc$ & 
      $\lc 0.4,0.9\rc$ \\
      \hline
      $\Delta T_3(x,Q_0^2)$  & ·$\lc  1.5,3.5\rc$ & 
      $\lc 0.4,0.7\rc$ \\
      \hline
      $\Delta T_8(x,Q_0^2)$  & ·$\lc  1.5,3.0\rc$ & 
      $\lc 0.1,0.6\rc$ \\
      \hline
    \end{tabular}
    \caption{\small \label{tab:prepexps} Ranges for the small and
      large $x$
 preprocessing exponents Eq.~(\ref{eq:PDFbasisnets}).}
  \end{center}
\end{table}

Two of the PDFs in the parametrization basis
Eq.~(\ref{eq:PDFbasisnets}), namely the nonsinglet
triplet and octet $\Delta T_3$ and $\Delta T_8$, are supplemented by
a prefactor. This is because these PDFs 
must satisfy the sum rules Eqs.~(\ref{eq:t3sr}, \ref{eq:t8sr}),
which are enforced by letting
\begin{eqnarray}
A_3&=&\frac{a_3}
{\int_0^1 dx\,(1-x)^{m_3}x^{-n_3} {\rm NN}_{\Delta T_3}(x)} ,
\nonumber\\ 
A_8&=&\frac{a_8}
{\int_0^1 dx\, (1-x)^{m_8}x^{-n_8} {\rm NN}_{\Delta T_8}(x)  } . 
\label{eq:sumrules1}
\end{eqnarray}
The integrals are computed numerically each time the parameters of 
the PDF set are modified.  The values of $a_3$ and $a_8$ are chosen
for each replica as gaussianly distributed numbers, with central value
and width given by the corresponding experimental values,
Eqs.~(\ref{eq:a3},\ref{eq:a8p}).

\subsection{Genetic algorithm minimization}

As discussed at length in Ref.~\cite{Ball:2008by}, minimization
with a neural network parametrization of PDFs must be performed through an
algorithm which explores the very wide functional space efficiently.
This is done by means of a genetic algorithm, which is
used to minimize  a suitably defined figure
of merit, namely the error function~\cite{Ball:2008by},
\begin{equation}
  \label{eq:errfun}
  E^{(k)}=\frac{1}{N_{\mathrm{dat}}}\sum_{I,J=1}^{N_{\rm dat}}
                 \left(g_I^{(\mathrm{art})(k)}-g_I^{(\mathrm{net})(k)}\right)
                 \left(\left({\mathrm{cov}}\right)^{-1}\right)_{IJ}
                 \left(g_J^{(\mathrm{art})(k)}-g_J^{(\mathrm{net})(k)}\right) \ .
\end{equation}
Here $g_I^{\rm (art)(k)}$ is the value of the observable $g_I$ at the
kinematical point $I$ corresponding to the Monte Carlo replica $k$,
and
$g_I^{(\rm net)(k)}$ is the same observable computed from the neural
network PDFs; the covariance matrix $\left({\rm cov}\right)_{IJ}$ is defined in
Eq.~(\ref{eq:covmat}).  

The minimization procedure we adopt follows
closely that of Ref.~\cite{DelDebbio:2007ee}, to which we refer for a more
general discussion. Minimization is perfomed by means of a genetic
algorithm, which  minimizes the
figure of merit, Eq.~(\ref{eq:errfun}) by creating, at each
minimization step,  a pool of new
neural nets, obtained by randomly mutating the parameters of the
starting set, and retaining the configuration which corresponds to the
lowest value of the figure of merit.

The parameters which characterize the behaviour of the genetic
algorithm are tuned  in order to optimize the efficiency of
the minimization procedure: here, we rely on  previous experience of
the development of unpolarized NNPDF sets. In particular, the
algorithm is characterized by a mutation rate, which we take to
decrease  as a function of the number of iterations $N_{\rm ite}$ of
the algorithm according to~\cite{Ball:2008by}
\begin{equation}
 \eta_{i,j}=\eta_{i,j}^{(0)}/N_{\rm ite}^{r_\eta} \ ,
\label{eq:etarate}
\end{equation}
so that  in the early stages of the training large mutations are
allowed, while they become less likely as one approaches the
minimum. The starting mutation rates are chosen to be larger for PDFs
which contain more information. We perform two mutations per PDF at
each step, with the starting rates given in Tab.~\ref{tab:etapars}.
The exponent $r_\eta$ has been
introduced in order to optimally span the whole range of possible
beneficial mutations and it is randomized between $0$ and $1$ at each
iteration of the genetic algorithm, as in Ref.~\cite{Ball:2010de}.

Furthermore, following Ref.~\cite{Ball:2010de}, we let the number of new
candidate solutions depend on the stage of the minimization.
 At earlier stages of the minimization, when 
the number of generations is smaller than $N^{\rm mut}$, we
use a large population of mutants, $N_{\rm mut}^{a}\gg 1$, so a larger
space of mutations is being explored. At later stages of the
minimization, as the minimum is  approached, a smaller
number of mutations $N_{\rm mut}^{b}\ll N_{\rm
  mut}^{a}$ is used. The values of the parameters $N_{\rm gen}^{\rm
  mut}$, $N_{\rm mut}^{a}$ and $N_{\rm mut}^{b}$ are collected in
Tab.~\ref{tab:mutpars}.

\begin{table}[t]
\begin{center}
\begin{tabular}{|c|c|c|c|}
\hline
$\eta^{(0)}_{i,\Delta\Sigma}$ & $\eta^{(0)}_{i,\Delta g}$ & $\eta^{(0)}_{i,\Delta T_3}$ & $\eta^{(0)}_{i,\Delta T_8}$ \\
\hline 
$5, 0.5$ & $5, 0.5$ & $2, 0.2$ & $2, 0.2$\\
\hline 
\end{tabular}
\caption{\small The initial values of the mutation rates for the two
  mutations of each PDF.}
\label{tab:etapars}
\end{center}
\end{table}

\begin{table}[t]
  \centering
  \begin{tabular}{|c|c|c|c|c|}
    \hline 
    $N_{\rm gen}^{\rm mut}$ & $N^a_{\rm mut}$ & $N^b_{\rm mut}$ & $N_{\rm gen}^{\rm wt}$ & $E^{\mathrm{sw}}$ \\
    \hline
    200 & 50 & 10 &  5000 & 2.5\\
    \hline
  \end{tabular}
  \caption{\small Values of the parameters of the genetic algorithm.}
  \label{tab:mutpars}
\end{table}

Because the minimization procedure stops the fit to all experiments at
once, we must make sure that the quality of the fit to different
experiments is approximately the same. This is nontrivial, because of
the variety of experiments and datasets included in the
fit. Therefore, the figure of merit per datapoint for a given set is 
not necessarily a reliable indicator of the quality of the fit to that set,
because some experiments may have systematically underestimated or
overestimated uncertainties. Furthermore, unlike for unpolarized PDF
fits, information on the experimental covariance matrix is only
available for a small subset of experiments, so for most experiments
statistical and systematic errors must be added in quadrature, thereby
leading to an overestimate of uncertainties: this leads to a wide
spread of values of the figure of merit, whose value depends on the
size of the correlated uncertainties which are being treated as
uncorrelated.

A methodology to deal with this situation was developed in
Ref.~\cite{Ball:2010de}. The idea is to first determine the optimal value of
the figure of merit for each experiment, i.e. a  set of target values
$E_{i}^{\rm targ}$ for each of the $i$ experiments, then
during the fit  give more weight to experiments for which the figure
of merit is further away from its target value, and stop training
experiments which have already reached the target value. This is done by
minimizing, instead of the figure of merit Eq.~(\ref{eq:errfun}), the
weighted figure of merit
\begin{equation}
  \label{eq:weight_errfun}
  E_{\rm wt}^{(k)}=\frac{1}{N_{\mathrm{dat}}}
  \sum_{j=1}^{N_{\mathrm{sets}}}p_j^{(k)} N_{\mathrm{dat},j}E_j^{(k)}\, ,
\end{equation}
where $E_j^{(k)}$ is the error function for the $j$-th dataset with
$N_{{\rm dat},j}$  points, and the weights  $p_j^{(k)}$ are given by 
\begin{enumerate}
\item If $E_{i}^{(k)} \ge E_{i}^{\rm targ}$, then $p_i^{(k)}=\lp E_{i}^{(k)}/E_{i}^{\rm targ}\rp^n$, 
\item If $E_{i}^{(k)} < E_{i}^{\rm targ}$, then $p_i^{(k)}=0$ \ ,
\end{enumerate}
with $n$ a free parameter which essentially determines the amount of
weighting. In the unpolarized fits of
Refs.~\cite{Ball:2010de,Ball:2011mu,Ball:2011uy,Ball:2012cx} the value
$n=2$ was used. Here instead we will choose $n=3$. This larger value,
determined by trial and error, is justified by the wider spread of
figures of merit in the polarized case, which in turn is related 
to the absence of correlated systematics for most
experiments. 

The target values $E_{i}^{\rm targ}$ are determined through an
iterative procedure: they are set to one at first, then a very long
fixed-length fit is run, and the values of  $E_{i}$ are taken as
targets for a new fit, which  is performed until stopping
(according to the criterion to be discussed in Sect.~\ref{sec-minim}
below). The values of  $E_{i}$ at the end of this fit are then taken
as new targets until convergence is reached, usually after a couple
iterations.

Weighted training stops after the first $N_{\rm
  gen}^{\rm wt}$ generations, unless the total error function
Eq.~(\ref{eq:errfun}) is above some threshold $E^{(k)}\geq E^{\rm
  sw}$. If it is, weighted training continues until $E^{(k)}$ falls
below the threshold value. Afterwards, the error function is just the
unweighted error function Eq.~(\ref{eq:errfun}) computed on
experiments. This ensures that the figure of merit behaves smoothly in
the last stages of training. The values for the parameters $N_{\rm
  gen}^{\rm wt}$ and $E^{\rm sw}$ are also given in
Tab.~\ref{tab:mutpars}.

\subsection{Determination of the optimal fit}
\label{sec-minim}

Because the neural network parametrization is very redundant, it may
be able to fit not only the underlying behaviour of the PDFs, but also
the statistical noise in the data. Therefore, the best fit does not necessarily coincide with the
absolute minimum of the figure of merit Eq.~(\ref{eq:errfun}). We
thus determine the best fit, as in Refs.~\cite{DelDebbio:2007ee,Ball:2008by},  using a 
cross-validation method~\cite{Bishop:1995}:
for each replica, the data are randomly divided in two sets, training
and validation, which include a fraction $f_{\rm tr}^{(j)}$ and $f_{\rm
  val}^{(j)}=1-f_{\rm tr}^{(j)}$ of the data points respectively.  The
figure of merit Eq.~(\ref{eq:errfun}) is then computed for both sets.
The training figure of merit function is minimized through the genetic
algorithm, while the validation figure of merit is monitored: when the
latter starts increasing while the former still decreases the fit is
stopped.  This means that the fit is stopped as soon as the neural
network is starting to learn the statistical fluctuations of the
points, which are different in the training and validation sets,
rather than the underlying law which they share.

In the unpolarized fits of
Refs.~\cite{DelDebbio:2007ee,Ball:2008by,Ball:2010de,Ball:2011mu,Ball:2011uy,Ball:2012cx} 
equal training and validation fractions were uniforlmly chosen,
$f_{\rm tr}^{(j)}=f_{\rm val}^{(j)}=1/2$.
However, in this case we have to face the problem that the number of
datapoints is quite small: most experiments include about ten
datapoints (see Tab.~\ref{tab:exps-sets}). Hence, it is difficult to
achieve a stable minimization if only half of them
are actually used for minimization, as we have explicitly verified. Therefore,  
we have chosen to include 80\% of the data in the training set,
i.e.  $f_{\rm tr}^{(j)}=0.8$ and $f_{\rm val}^{(j)}=0.2$. We have explicitly
verified that the fit quality which is obtained in this case is
comparable to the one achieved when including all data in the training set
(i.e. with $f_{\rm tr}^{(j)}=1.0$ and $f_{\rm val}^{(j)}=0.0$), but the presence of a
nonzero validation set allows for a satisfactory stopping, as we have
checked by explicit inspection of the profiles of the figure of merit
as a function of training time.

In practice, in order to implement cross-validation we must determine
a stopping criterion, namely, give conditions which must be satisfied
in order for the minimization to stop.
First, we require that the weighted training stage
has been completed, i.e., that the genetic algorithm has been run for
at least  $N_{\rm gen}^{\rm wt}$ minimization steps. Furthermore,
we check that all experiments have reached a value of the figure of merit
below a minimal threshold $E_{\rm thr}$. Note that because stopping
can occur only after weighted training has been switched off, and this
in turn only happens when the figure of merit falls below the value
$E^{\rm sw}$, the total figure of merit must be below this value in
order for stopping to be possible.

We then compute moving averages
\begin{equation}
  \label{eq:smearing}
  \langle E_{\mathrm{tr,val}}(i)\rangle\equiv
  \frac{1}{N_{\mathrm{smear}}}
\sum_{l=i-N_{\mathrm{smear}}+1}^iE_{\mathrm{wt;\,tr,val}}(l)\,,
\end{equation}
of the figure of merit Eq.~(\ref{eq:weight_errfun}) for either the
training or the validation set at the  $l$-th genetic minimzation step.
The fit is then stopped  if 
\begin{equation}
\label{eq:trratcond}
r_{\tr} < 1-\delta_{\rm tr}\quad{\rm and}\quad r_{\val} > 1+\delta_{\rm val}\, ,
\end{equation}
where
\begin{equation}
  \label{eq:dec-train}
r_{\rm tr}\equiv  \frac{\langle E_{\mathrm{tr}}(i)\rangle}
  {\langle E_{\mathrm{tr}}(i-\Delta_{\mathrm{smear}})\rangle} 
\, ,
\end{equation}
\begin{equation}
  \label{eq:dec-valid}
 r_{\rm val}\equiv \frac{\langle E_{\mathrm{val}}(i)\rangle}
       {\langle E_{\mathrm{val}}(i-\Delta_{\mathrm{smear}})\rangle} \,.
\end{equation}

The parameter $N_{\rm smear}$ determines the width of the moving
average; the parameter $\Delta_{\rm smear}$ determines the distance
between the two points along the minimization path which are compared
in order to determine whether the figure of merit is increasing or
decreasing; and the parameters $\delta_{\rm tr}$, $\delta_{\rm val}$ are
the threshold values for the decrease of the training and increase of
the validation figure of merit to be deemed significant.
The optimal value of these parameters should be chosen in such a way
that the fit does not stop on a statistical fluctuation, yet it does
stop before the fit starts overlearning (i.e. learning statistical
fluctuation). As explained in Ref.~\cite{Ball:2010de}, this is done 
studying the profiles of the error functions for individual dataset
and for individual replicas. 
In order to avoid unacceptably long fits, training is stopped anyway
when a maximum number of iterations $N_{\rm gen}^{\rm max}$ is reached, even though the stopping
conditions Eq.~(\ref{eq:trratcond}) are not satisfied.  This leads to
a small loss of accuracy of the corresponding fits:  this is
acceptable provided it only happens for a small enough fraction of
replicas. If a fit stops at $N_{\rm gen}^{\rm max}$ without the
stopping criterion having been satisfied, we also check that the
total figure of merit is below the value $E^{\rm sw}$ at which
weighted training is switched off. If it hasn't, we conclude that the
specific fit has not converged, and we retrain the same replica, i.e.,
we perform a new fit to the same data starting with a different random
seed. This only occurs in about one or two percent of cases.

The full set of parameters which
determine the stopping criterion is given in
Tab.~\ref{tab:stopping_pars}.

\begin{table}[t]
  \centering
  \begin{tabular}{|c|c|c|c|c|c|}
    \hline 
    $N_{\rm gen}^{\rm max}$ & $E_{\rm thr}$ & $N_{\rm smear}$ & $\Delta_{\rm smear}$ & $\delta_{\rm tr}$ & $\delta_{\rm val}$ \\
    \hline
    $20000$ & $8$ & $100$ & $100$ & $5 \cdot 10^{-4}$ & $5 \cdot 10^{-4}$ \\
    \hline
  \end{tabular}
    \caption{\small Parameters for the stopping criterium.}
    \label{tab:stopping_pars}
\end{table}

\begin{figure}[t]
 \centering
 \epsfig{width=0.43\textwidth,figure=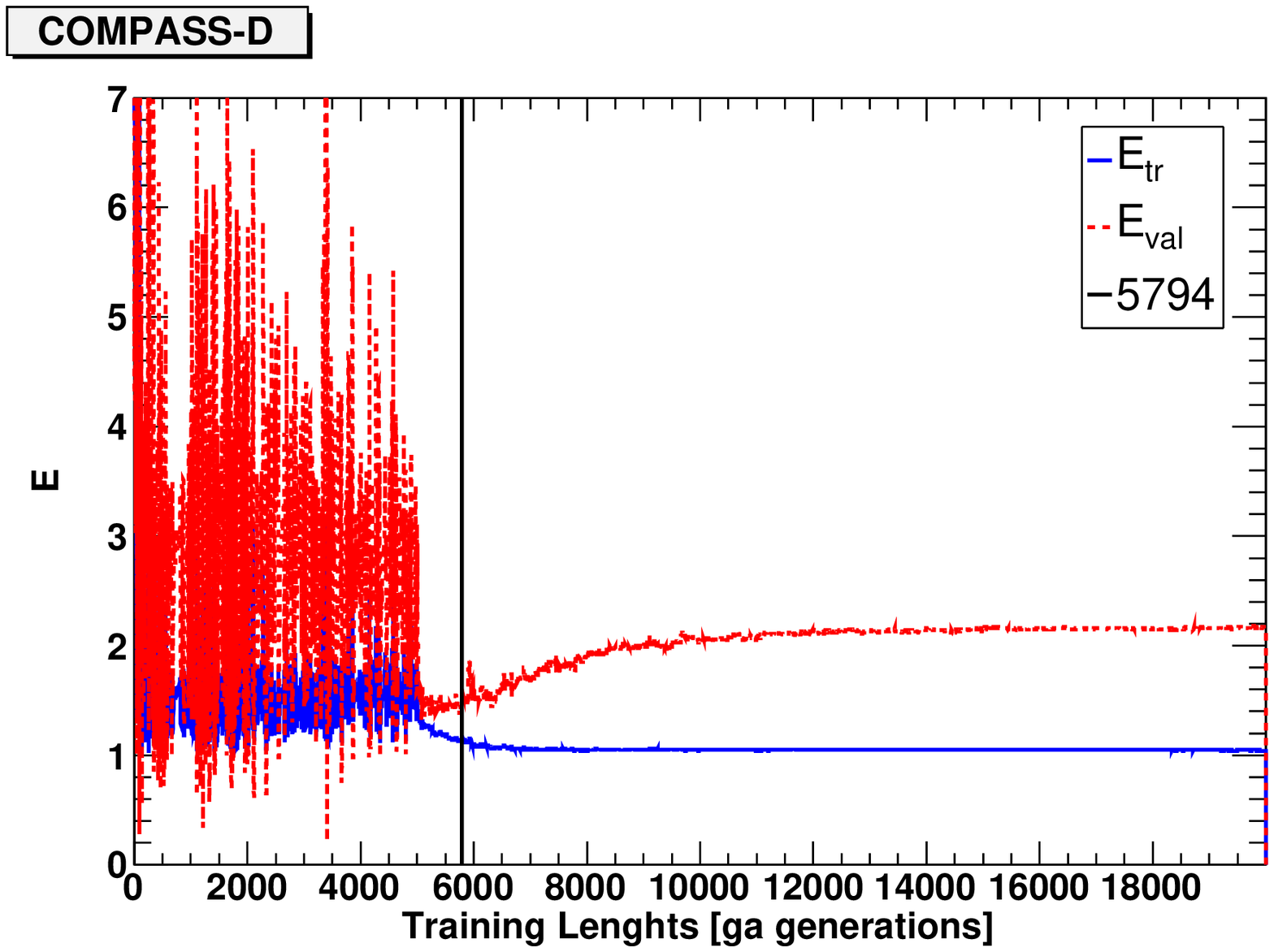}
 \epsfig{width=0.43\textwidth,figure=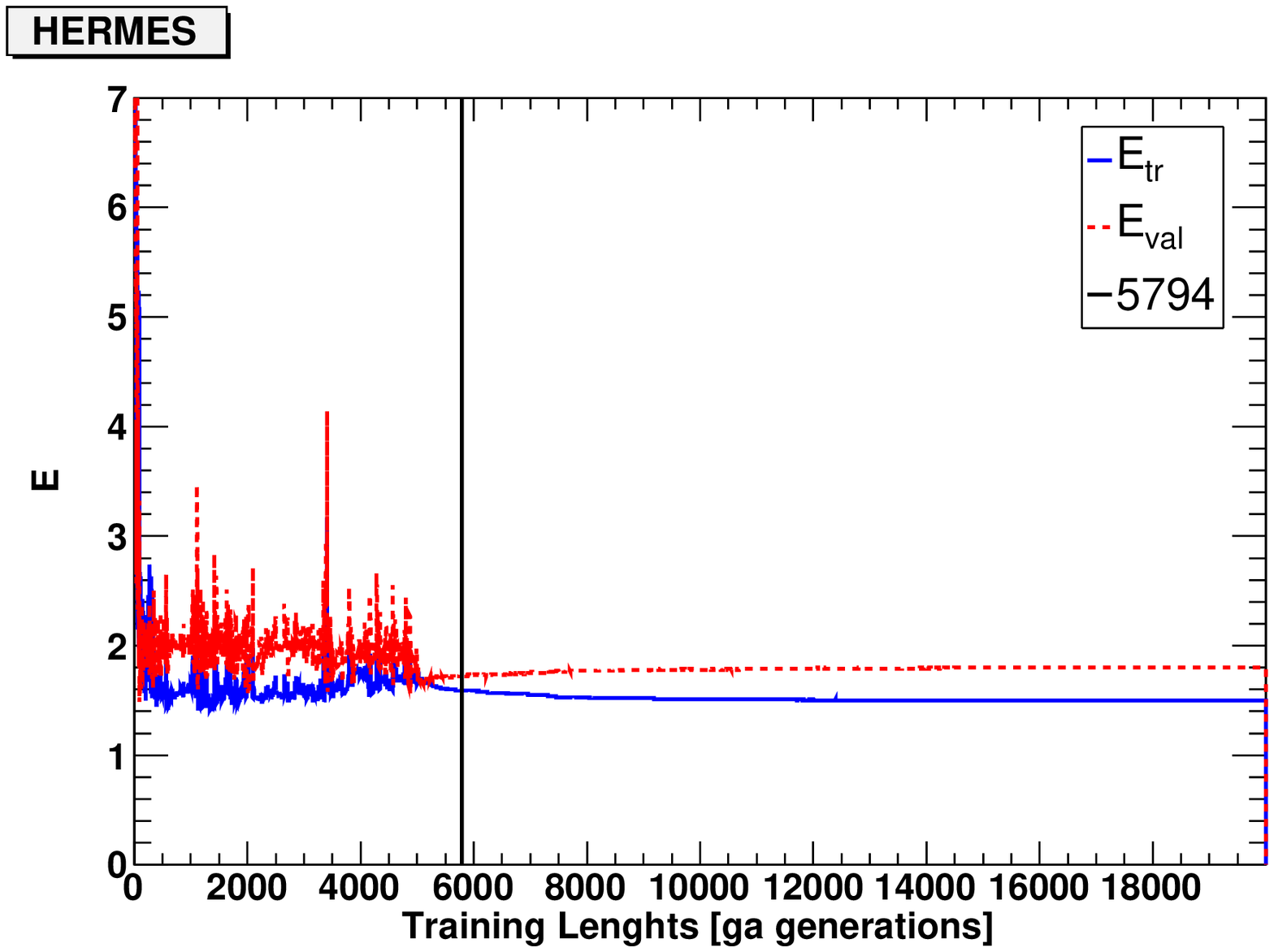}
 \caption{\small Behaviour of the moving average Eq.~(\ref{eq:smearing})
of the training and validation figure of merit for two different
datasets included in a global fit (\texttt{COMPASS-P} and
\texttt{HERMES}) 
as a function of training length. The  
 The straight vertical line
   indicates the point at which the  fit stops with the stopping parameters
   of Tab.~\ref{tab:stopping_pars}.  The weighted
   training is switched off at $N_{\rm gen}^{\rm wt}=5000$.}
 \label{fig:stop}
\end{figure}

An example of how the stopping criterium works in practice is
shown in Fig.~\ref{fig:stop}. We display the moving averages
Eq.~(\ref{eq:smearing}) of the
training and validation error functions $\langle E_{\rm tr,
  val}^{(k)}\rangle$,
computed with the parameter settings of Tab.~\ref{tab:stopping_pars}, and
plotted as a function of the number of iterations of the
genetic algorithm, for a particular replica and for two
of the experiments included in the fit. The wide fluctuations which
are observed in the first part of training, up to the $N_{\rm
  gen}^{\rm wt}$-th generation,  are due to the fact that the weights
which enter the definition of the figure of merit 
Eq.~(\ref{eq:weight_errfun}) are frequently adjusted. Nevertheless,
the downwards trend of the figure of merit is clearly visible.
Once the weighted training is switched off,
minimization proceeds smoothly. The vertical line denotes the point at
which the stopping criterion is satisfied. Here, we have let the
minimization go on beyond this point, and we see clearly that the
minimization has entered  an overlearning regime, in which
the validation
error function $E_{\rm val}^{(k)}$ is rising while the training
$E_{\rm tr}^{(k)}$ is still decreasing. Note that the stopping point,
which in this particular case occurs at
$N_{\rm gen}^{\rm stop}=5794$, is determined by verifying that the
stopping criteria are satisfied by the {\it total} 
figure of merit, not that of individual experiments shown here. 
The fact that the two different experiments
considered here both start overlearning at the same point shows that
the weighted training has been effective in synchronizing the fit
quality for different experiments.

\subsection{Theoretical constraints}
\label{sec:thconst}

Polarized PDFs are only loosely constrained by data, which are 
scarce and  not very accurate. Theoretical constraints are
thus especially important in reducing the uncertainty on the PDFs. We
consider in particular positivity and integrability. 

Positivity of the individual cross sections
which enter the polarized asymmetries Eq.~(\ref{eq:xsecasy}) implies
that, up to power-suppressed corrections, longitudinal polarized
structure functions are bounded by their unpolarized counterparts,
i.e.  \be
\label{eq:pos}
|g_1(x,Q^2)| \le F_1(x,Q^2) .
\ee
At leading order, structure functions are proportional to parton
distributions, so imposing Eq.~(\ref{eq:pos}) for any process (and a
similar condition on an asymmetry which is sensitive to polarized
gluons~\cite{Altarelli:1998gn}), would imply 
\be
\label{eq:pospdf}
|\Delta  f_i(x,Q^2)|\le f_i(x,Q^2)
\ee
for any pair of unpolarized and polarized PDFs $f$ and $\Delta f$, for
all quark flavors and gluon $i$, for all $x$, and for all $Q^2$.
Beyond leading order, the condition
Eq.~(\ref{eq:pos}) must still hold, but it does not necessarily imply 
Eq.~(\ref{eq:pospdf}). Rather, one should then impose at least a number of
conditions of the form of Eq.~(\ref{eq:pos}) on physically measurable
cross-sections which is equal to the number of independent polarized
PDFs. For example, in principle one may require that the condition
Eq.~(\ref{eq:pos}) is separately satisfied for each flavor, i.e.
when only contributions from the
$i$-th flavor are included in the polarized and unpolarized structure
function: this corresponds to requiring positivity of semi-inclusive
structure functions which could in principle be measured 
(and that fragmentation effects cancel in the ratio).
A condition on the gluon can be obtained by imposing
positivity of the polarized and unpolarized cross-sections for 
inclusive Higgs production in gluon-proton
scattering~\cite{Altarelli:1998gn}, again measurable in principle if
not in practice.

Because $g_1/F_1\sim x$ as $x\to0$~\cite{Ball:1995ye}, the positivity
bound Eq.~(\ref{eq:pos}) is only significant at large enough
$x\gtrsim10^{-2}$. On the other hand, at very large $x$ the NLO
corrections to the LO positivity bound become
negligible~\cite{Altarelli:1998gn,Forte:1998kd}. Therefore, the NLO
positivity bound in practice only differs from its LO counterpart
Eq.~(\ref{eq:pospdf}) in a small region $10^{-2}\sim x\lesssim0.3$, and
even there by an  amount of rather less that
10\%~\cite{Altarelli:1998gn}, which is negligible in comparison to the size
of PDF uncertainties, as we shall see explicitly  in
Sec.~\ref{sec:results}.

Therefore, we will impose the leading-order positivity bound
Eq.~(\ref{eq:pospdf}) on each flavor combination $\Delta q_i+\Delta
\bar q_i$ and on the gluon $\Delta g$ (denoted as $\Delta f_i$ below).
We do this by requiring 
\be
\label{eq:possigma}
|\Delta  f_i(x,Q^2)| \le  f_i(x,Q^2) + \sigma_i(x,Q^2)  \ ,
\ee
where $\sigma_i(x,Q^2)$ is the uncertainty on the corresponding unpolarized PDF
combination $f_i(x,Q^2)$ at the kinematic point $(x,Q^2)$. This choice is
motivated by two considerations. First, it is clearly meaningless to
impose positivity of the polarized PDF 
to an accuracy which is greater than that with which
the unpolarized PDF has been determined. Second, because the
unpolarized PDFs satisfy NLO positivity, they can become negative and
thus they may have nodes. As a consequence, the LO bound
Eq.~(\ref{eq:pospdf}) would imply that the polarized PDF must vanish
at the same point, which would be clearly meaningless.

As in Ref.~\cite{Ball:2010de} positivity is imposed during the
minimization procedure, thereby guaranteeing that the genetic
algorithm only explores the subspace of acceptable physical
solutions. This is done through a Lagrange multiplier $\lambda_{\rm
  pos}$, i.e.  by computing the polarized PDF at $N_{\rm dat,pos}$
fixed kinematic points $(x_p,Q_0^2)$ and then adding to the error function
 Eq.~(\ref{eq:errfun}) a contribution 
\begin{eqnarray}
 \label{eq:lagrmult}
&&E_{\rm pos}^{(k)}={\lambda_{\rm pos}}\sum_{p=1}^{N_{\rm dat,pos}}
  \Bigg\{
  \sum_{j=u+\bar{u},d+\bar{d},s+\bar{s},g} \Theta\left[\left|\Delta f_j^{(\rm net)(k)}(x_p,Q_0^2)\right| -\left(f_j + \sigma_j\right) (x_p,Q_0^2) \right] \nonumber\\
  &&\qquad\qquad \times
  \left[\left|\Delta f_j^{(\rm net)(k)}(x_p,Q_0^2)\right| - 
\left(f_j + \sigma_j\right)(x_p,Q_0^2) \right]  
  \Bigg\} \ .
\end{eqnarray}
This provides a penalty, proportional to the violation of positivity,
which enforces  Eq.~(\ref{eq:possigma}) separately for all
the non-zero quark-antiquark combinations. The values of the
unpolarized PDF combination $f_j(x,Q^2)$ and its uncertainty $\sigma_j(x,Q^2)$ are
computed using the
\texttt{NNPDF2.1 NNLO} PDF set~\cite{Ball:2011mu}, while $\Delta
f_j^{(\rm net)(k)}$ is the corresponding polarized PDF computed from
the neural network parametrization for the $k$-th replica.  The
polarized and unpolarized PDFs are evaluated at $N_{\rm dat,pos}=20$
points with $x$ equally spaced in the interval
\be 
x \in \lc
10^{-2},0.9\rc \ .  
\ee
Positivity
is imposed at the initial scale $Q_0^2=1$~GeV$^2$ since once 
positivity is enforced at low scales, it is automatically
satisfied at larger scales~\cite{Altarelli:1998gn,Forte:1998kd}.
After stopping, we finally test the positivity condition
Eq.~(\ref{eq:possigma}) is satisfied on a grid of $N_{\rm dat,pos}=40$ points in
the same intervals. Replicas for which positivity is violated in one
or more points are discarded and retrained.

In the unpolarized case, in which positivity only played a minor role
in constraining PDFs, a fixed value of the Lagrange multiplier
$\lambda_{\rm pos}$ was chosen. In the polarized case it turns out to
be necessary to vary the Lagrange multiplier along the minimization.
Specifically, we let
\begin{equation}
 \label{eq:lagrmult2}
 \left\{
 \begin{array}{rcll}
  \lambda_{\rm pos} & = & \lambda_{\rm max}^{(N_{\rm gen}-1)/(N_{\lambda_{\rm max}}-1)} & N_{\rm gen} < N_{\lambda_{\rm max}}\\
  \lambda_{\rm pos} & = & \lambda_{\rm max} & N_{\rm gen} \geq N_{\lambda_{\rm max}}.
 \end{array}
 \right.
\end{equation}
This means that  the Lagrange multiplier increases as
the minimization proceeds, starting from $\lambda_{\rm pos}=1$, at the
first minimization step, $N_{\rm gen}=1$, up to $\lambda_{\rm pos} =
\lambda_{\rm max}\gg 1$ when $N_{\rm gen} = N_{\lambda_{\rm max}}$. After
$N_{\lambda_{\rm max}}$ generations $\lambda_{\rm pos}$ is then kept
constant to $\lambda_{\rm max}$. The rationale behind this choice is
that the genetic algorithm can thus learn experimental data and
positivity at different stages of minimization.  During the early
stages, the contribution coming from the modified error function
Eq.~(\ref{eq:lagrmult}) is negligible, due to the moderate value of
the Lagrange multiplier; hence, the genetic algorithm will mostly
learn the basic shape of the PDF driven by experimental data. As soon
as the minimization proceeds, the contribution coming from the
Lagrange multiplier increases, thus ensuring the proper learning of
positivity: at this stage, most of the replicas which will not fulfill
the positivity bound will be discarded.

The final values of $N_{\lambda_{\rm max}}=2000$ and $\lambda_{\rm max}=10$
have been  determined as follows. 
First of all, we have performed a fit without
any positivity constraint and we have observed that data were mostly
learnt in about $2000$ generations: hence we have taken this value for
$N_{\lambda_{\rm max}}$. Then we have tried different values for
$\lambda_{\rm max}$ until we managed to reproduce the same $\chi^2$
obtained in the previous, positivity unconstrained, fit.  This ensures
that positivity is not learnt to the detriment of the global fit
quality.

Notice that the value of $\lambda_{\rm max}$ is rather small if
compared to the analogous Lagrange multiplier used in the unpolarized
case~\cite{Ball:2011mu}. This depends on the fact that, in this latter
case, positivity is learnt at the early stages of minimization, when
the error function can be much larger than its asymptotic value:
a large Lagrange multiplier is then needed to select the best
replicas. Also, unpolarized PDFs are quite well constrained by data
and positivity is almost automatically fulfilled, except in some
restricted kinematic regions; only a few replicas 
violate positivity and need to be penalized. This means that the behaviour
of the error function Eq.~(\ref{eq:errfun}), which governs the fitting
procedure, is essentially dominated by data instead of positivity.  

In the polarized case, instead, positivity starts to be effectively
implemented only after some minimizaton steps, when the error function
has already decreased to a value of a few units. Furthermore, we have
checked that, at this stage, most of replicas slightly violate the
positivity condition Eq.~(\ref{eq:possigma}): thus, a too large value
of the Lagrange multiplier on the one hand would penalize replicas which
are good in reproducing experimental data and only slightly worse in
reproducing positivity; on the other, it would promote replicas which
fulfill positivity but whose fit to data is quite bad. As a
consequence of this behaviour, the convergence of the minimization
algorithm would be harder to reach.  We also verified that, using a value 
for the Lagrange multiplier up to $\lambda_{\rm pos}=100$ leads to no 
significant improvement neither in the fulfillment of positivity requirement 
nor in the fit quality.  
We will show in detail the effects of the positivity bound Eq.~(\ref{eq:possigma}) 
on the fitted replicas and on polarized PDFs in Sect.~\ref{sec:results}.

Finally, as already mentioned, we impose an integrability constraint.
The requirement that polarized PDFs be integrable, i.e. that they have
finite first moments, corresponds to the assumption that the nucleon
matrix element of the axial current for the $i$-th flavor is finite.
The integrability condition is imposed by computing at each
minimization step the integral of each of the polarized PDFs
in a given interval, 
\be I(x_1,x_2)=\int_{x_1}^{x_2} dx~\Delta q_i(x,Q_0^2) \,
\qquad \Delta q_i=\Delta\Sigma, \Delta g, \Delta T_3, \Delta T_8 \ee 
with $x_1$ and $x_2$ chosen in the small $x$ region, well below the
data points, and verifying that in this region the growth of the
integral as $x_1$ decreases for fixed $x_2$ is less than logarithmic. 
In practice, we test for the condition
\be
\frac{I(x_1,x_2)}{I(x_1^\prime,x_2)} < 
\frac{\ln\frac{x_2}{x_1}}{\ln\frac{x_2}{x_1\prime}}, 
\ee
with $x_1<x_1^\prime$. 
Mutations which do not satisfy the 
condition are rejected during the minimization procedure. In our default fit,
we chose $x_1=10^{-5}$,  $x_1^\prime=2\cdot 10^{-5}$ and $x_2=10^{-4}$.

\clearpage

\section{Results}
\label{sec:results}

We now present the main result of this paper, namely the
first determination of a polarized PDF set based on the NNPDF
methodology, \texttt{NNPDFpol1.0}. We will first illustrate the
statistical features of our PDF fit, then compare the
\texttt{NNPDFpol1.0} PDFs to other recent polarized parton
sets~\cite{deFlorian:2009vb,Leader:2010rb,Hirai:2008aj,Blumlein:2010rn}.
We will finally discuss the stability of our results upon the 
variation of several 
theoretical and methodological assumptions: the treatment of target-mass 
corrections, the use of sum rules to fix the triplet and octet axial charges, 
the implementation  of positivity of PDFs, and preprocessing of neural 
networks and its impact on small and large $x$ behaviour.

We will not discuss here the  way predictions for PDFs and uncertainties
are obtained from NNPDF replica sets, for which we refer to general
reviews, such as Ref.~\cite{Alekhin:2011sk}.

\subsection{Statistical features}
\label{sec:stat_features}

The statistical features of the {\tt NNPDFpol1.0} analysis are summarized in 
Tabs.~\ref{tab:chi2tab1}-\ref{tab:chi2tab2}, for the full
dataset and for individual experiments and sets respectively. 
\begin{table}[h!]
\begin{center}
\small
\begin{tabular}{|c|c|}
\hline
\multicolumn{2}{c}{\texttt{NNPDFpol1.0}} \\
\hline
\hline 
$\chi^{2}_{\tot}$ &      0.77 \\
$\la E \ra \pm \sigma_{E} $   &       1.82 $\pm$       0.18    \\
$\la E_{\rm tr} \ra \pm \sigma_{E_{\rm tr}}$&       1.66 $\pm$       0.49    \\
$\la E_{\rm val} \ra \pm \sigma_{E_{\rm val}}$&       1.88 $\pm$       0.67    \\
$\la{\rm TL} \ra \pm \sigma_{TL}$ &    6927 $\pm$ 3839     \\
\hline
$\la \chi^{2(k)} \ra \pm \sigma_{\chi^{2}} $  &       0.91 $\pm$       0.12    \\
\hline
\end{tabular}

\end{center}
\caption{\small Statistical estimators for {\tt NNPDFpol1.0} with 
$N_\mathrm{rep}=100$ replicas.}
\label{tab:chi2tab1}
\end{table}
\begin{table}[t]
\centering

{
\footnotesize
\begin{tabular}{|ll|c|c|}
\hline 
Experiment  & Set  & $\chi^{2}_{\tot}$  & $\la E\ra \pm \sigma_{E}$ \\
\hline 
\hline
EMC       & &  0.44&  1.54 $ \pm $  0.64 \\
&EMC-A1P   &  0.44&  1.54 $ \pm$  0.64 \\
\hline
SMC       & &  0.93&  1.93 $ \pm $  0.51 \\
&SMC-A1P   &  0.40&  1.44 $ \pm$  0.54 \\
&SMC-A1D   &  1.46&  2.42 $ \pm$  0.82 \\
\hline
SMClowx   & &  0.97&  1.90 $ \pm $  0.67 \\
&SMClx-A1P &  1.40&  2.32 $ \pm$  1.13 \\
&SMClx-A1D &  0.53&  1.48 $ \pm$  0.69 \\
\hline
E143      & &  0.64&  1.68 $ \pm $  0.29 \\
&E143-A1P  &  0.43&  1.49 $ \pm$  0.34 \\
&E143-A1D  &  0.88&  1.90 $ \pm$  0.45 \\
\hline
E154      & &  0.40&  1.69 $ \pm $  0.61 \\
&E154-A1N  &  0.40&  1.69 $ \pm$  0.61 \\
\hline
E155      & &  0.89&  1.96 $ \pm $  0.36 \\
&E155-G1P  &  0.89&  2.00 $ \pm$  0.51 \\
&E155-G1N  &  0.88&  1.93 $ \pm$  0.47 \\
\hline
COMPASS-D & &  0.65&  1.72 $ \pm $  0.53 \\
&CMP07-A1D &  0.65&  1.72 $ \pm$  0.53 \\
\hline
COMPASS-P & &  1.31&  2.38 $ \pm $  0.72 \\
&CMP10-A1P &  1.31&  2.38 $ \pm$  0.72 \\
\hline
HERMES97  & &  0.34&  1.37 $ \pm $  0.69 \\
&HER97-A1N &  0.34&  1.37 $ \pm$  0.69 \\
\hline
HERMES    & &  0.79&  1.79 $ \pm $  0.30 \\
&HER-A1P   &  0.44&  1.49 $ \pm$  0.39 \\
&HER-A1D   &  1.13&  2.09 $ \pm$  0.50 \\
\hline
\end{tabular}
}

\caption{\small Same as Tab.~\ref{tab:chi2tab1} but for individual experiments.}
\label{tab:chi2tab2}
\end{table}
The error function $\langle E\rangle$ Eq.~(\ref{eq:errfun}) shown in
the tables both for the total, training and validation datasets
is the figure of merit for the quality of the fit of each PDF replica
to the corresponding data replica. The quantity which is actually
minimized during the neural network training is this figure of merit
for the training set, supplemented by weighting in the early stages of
training according to Eq.~(\ref{eq:weight_errfun}) and by a Lagrange
multiplier to enforce positivity according to Eq.~(\ref{eq:lagrmult}).
In the table we also show  the average over all replicas $\la
\chi_{\mathrm{tot}}^{2(k)}\ra$  of $\chi_{\mathrm{tot}}^{2(k)}$
computed for the $k$-th replica, which coincides with the figure of
merit  Eq.~(\ref{eq:weight_errfun}), but with the data replica $g_I^{\rm (art)(k)}$
replaced by the experimental data $g_I^{\rm (dat)}$. We finally show 
$\chi^2_{\mathrm {tot}}$,  which  coincides with the figure of
merit  Eq.~(\ref{eq:weight_errfun}), but again with  $g_I^{\rm
  (art)(k)}$ replaced by $g_I^{\rm (dat)}$, and also with
$g_I^{(\mathrm{net})(k)}$ replaced by $\la
g_I^{(\mathrm{net})(k)}\ra$, i.e. the average of the observable over
replicas, which provides our best prediction.
The average
number of iterations of the genetic algorithm at stopping, $\la \rm TL
\ra$, is also given in this table. 

\begin{figure}[t]
\begin{center}
\epsfig{width=0.49\textwidth,figure=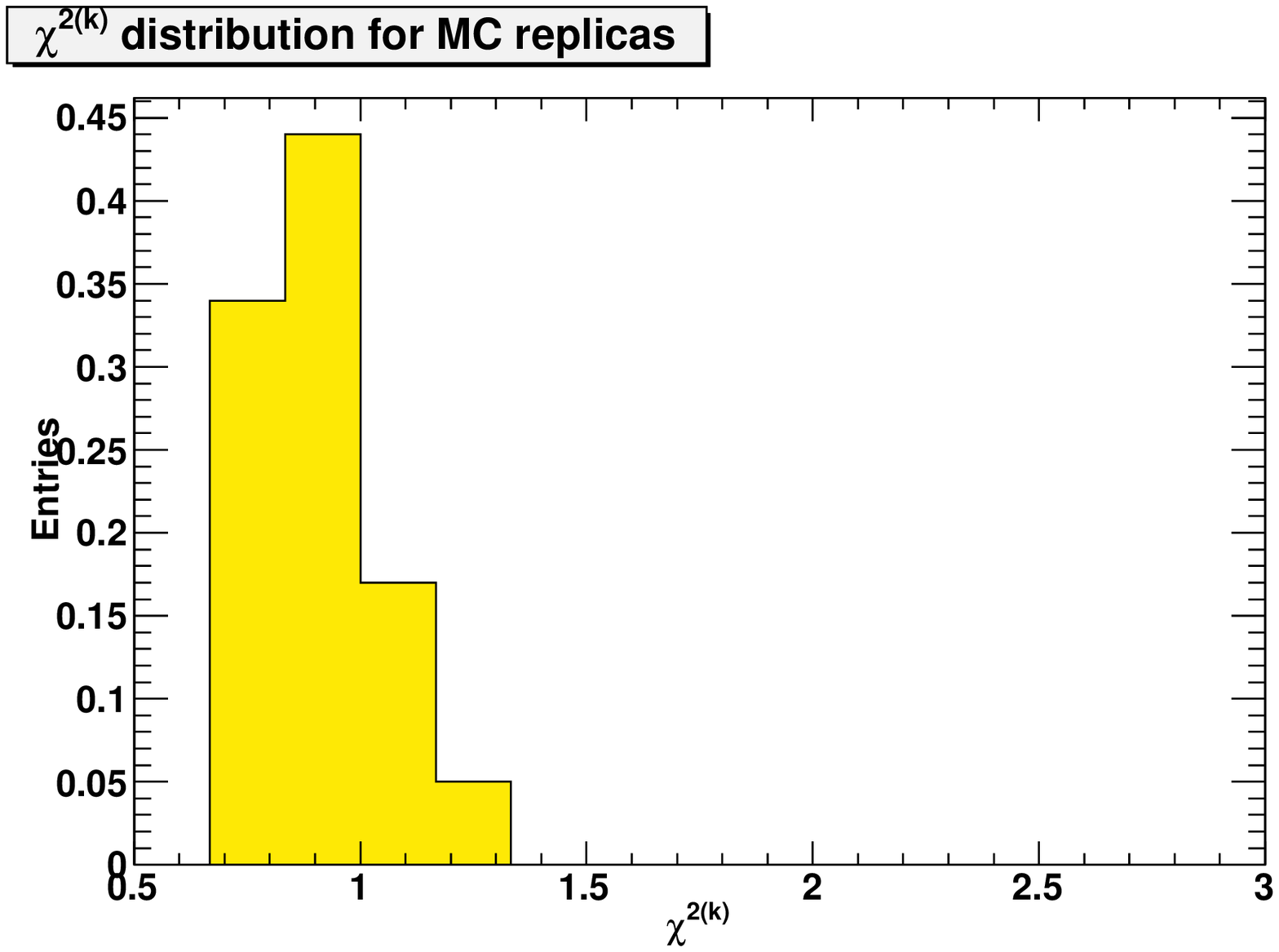}
\epsfig{width=0.49\textwidth,figure=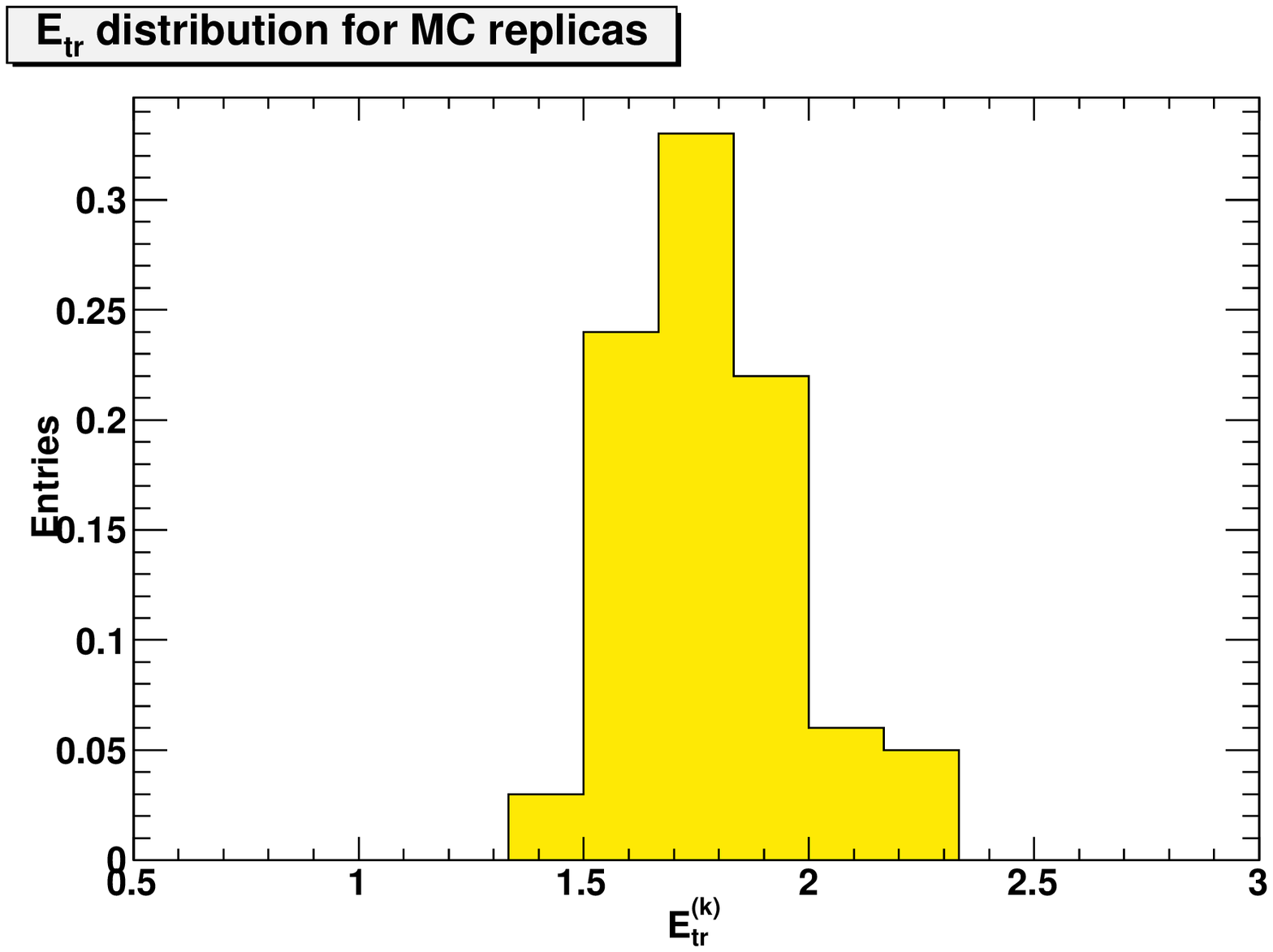}
\caption{\small Distribution of $\chi^{2(k)}$ and $E_{\mathrm{tr}}^{(k)}$ over 
the sample of $N_{\mathrm{rep}}=100$ replicas.} 
\label{fig:chi2-Etr-distr}
\end{center}
\end{figure}
\begin{figure}[t]
\begin{center}
\epsfig{width=0.49\textwidth,figure=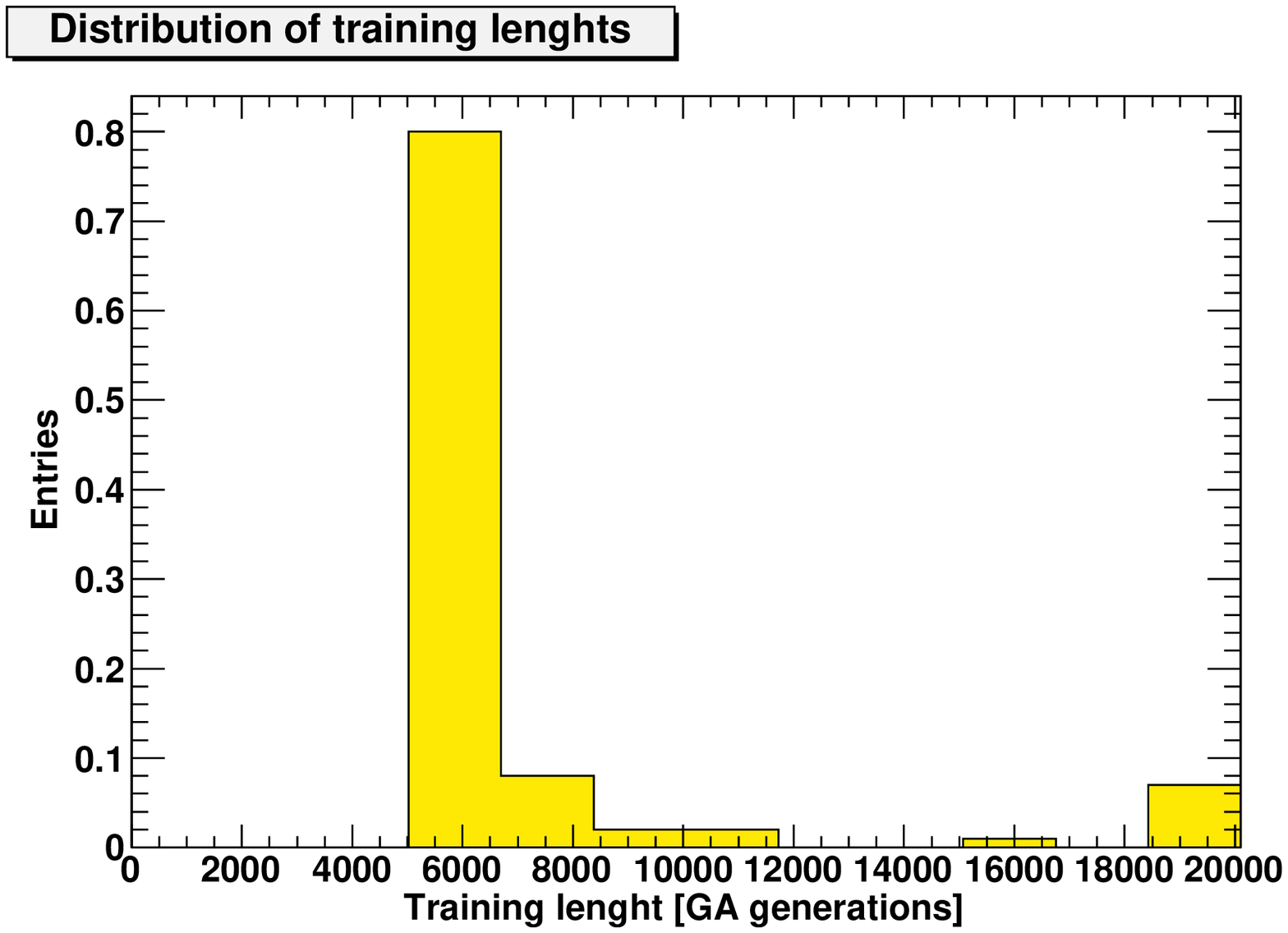}
\caption{\small Distribution of training lengths over the sample of 
$N_{\mathrm{rep}}=100$ replicas.} 
\label{fig:TL-distr}
\end{center}
\end{figure}
The distribution of $\chi^{2(k)}$, $E_{\mathrm{tr}}^{(k)}$, and
training lengths among the $N_{\mathrm{rep}}=100$ replicas are shown
in Fig.~\ref{fig:chi2-Etr-distr} and Fig.~\ref{fig:TL-distr}
respectively. Note that the latter has 
a long
tail which causes an accumulation of points at the maximum
training length, $N_{\mathrm{gen}}^{\mathrm{max}}$. This means that
there is a fraction of replicas that do not 
fulfill the stopping criterion. 
This may cause a loss in accuracy in outlier fits,
which however make up fewer than $10\%$ of the total sample. 

The features of the fit can be summarized as follows:
\begin{itemize}
\item The quality of the central fit, as measured by its
  $\chi_{\mathrm{tot}}^{2}=0.77$, is good. However, this value should
  be taken with care in view of the fact that uncertainties for all
  experiments but two are overestimated because  the
  covariance matrix is not available and thus correlations between
  systematics cannot be properly accounted for. This explains the
  value lower than one for this quantity, which would be very unlikely
  if it had included correlations.
\item The values of $\chi_{\mathrm{tot}}^{2}$ and $\la E \ra$ differ
  by approximately one unit. This is due to the fact that replicas
  fluctuate within their uncertainty about the experimental data, which in
  turn are gaussianly distributed about a true
  value~\cite{Forte:2002fg}: it shows that the neural net is correctly
  reproducing the underlying law thus being closer to the true
  value. This is confirmed by the fact that $\la \chi^{2(k)}\ra$ is of
  order one.
\begin{figure}[t]
\begin{center}
\epsfig{width=0.8\textwidth,figure=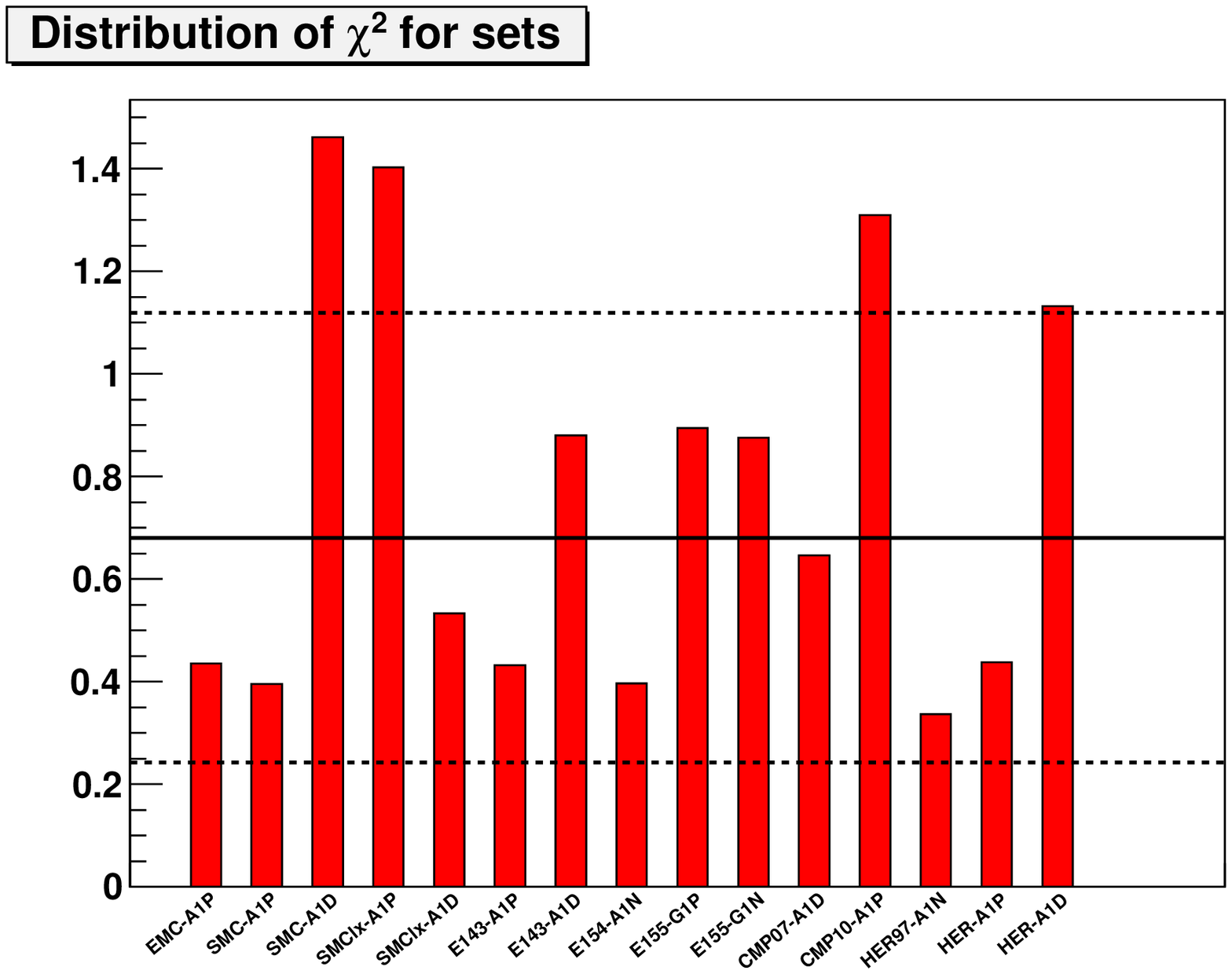}
\caption{\small Value of the $\chi^2$ per data
point for the datasets
included in the {\tt NNPDFpol1.0} reference fit, listed in Tab.~\ref{tab:chi2tab2}. 
The horizontal line is 
the unweighted average of these $\chi^2$ over the datasets and 
the black dashed lines give  the one-sigma interval about it.} 
\label{fig:chi2-dist}
\end{center}
\end{figure}
\item The distribution of $\chi^2$ for different experiments (also
  shown as a histogram in Fig.~\ref{fig:chi2-dist}) shows sizable
  differences, and indeed the standard deviation (shown as a dashed
  line in the plot) about the mean (shown as a solid line) is very
  large. This can be understood as a consequence of the lack of
  information on the covariance matrix: experiments where
  large correlated uncertainties are treated as uncorrelated will
  necessarily have a smaller value of the  $\chi^2$.
\end{itemize}

\subsection{Parton distributions}
\label{sec:pdfs}

\begin{figure}[p]
\begin{center}
\epsfig{width=0.43\textwidth,figure=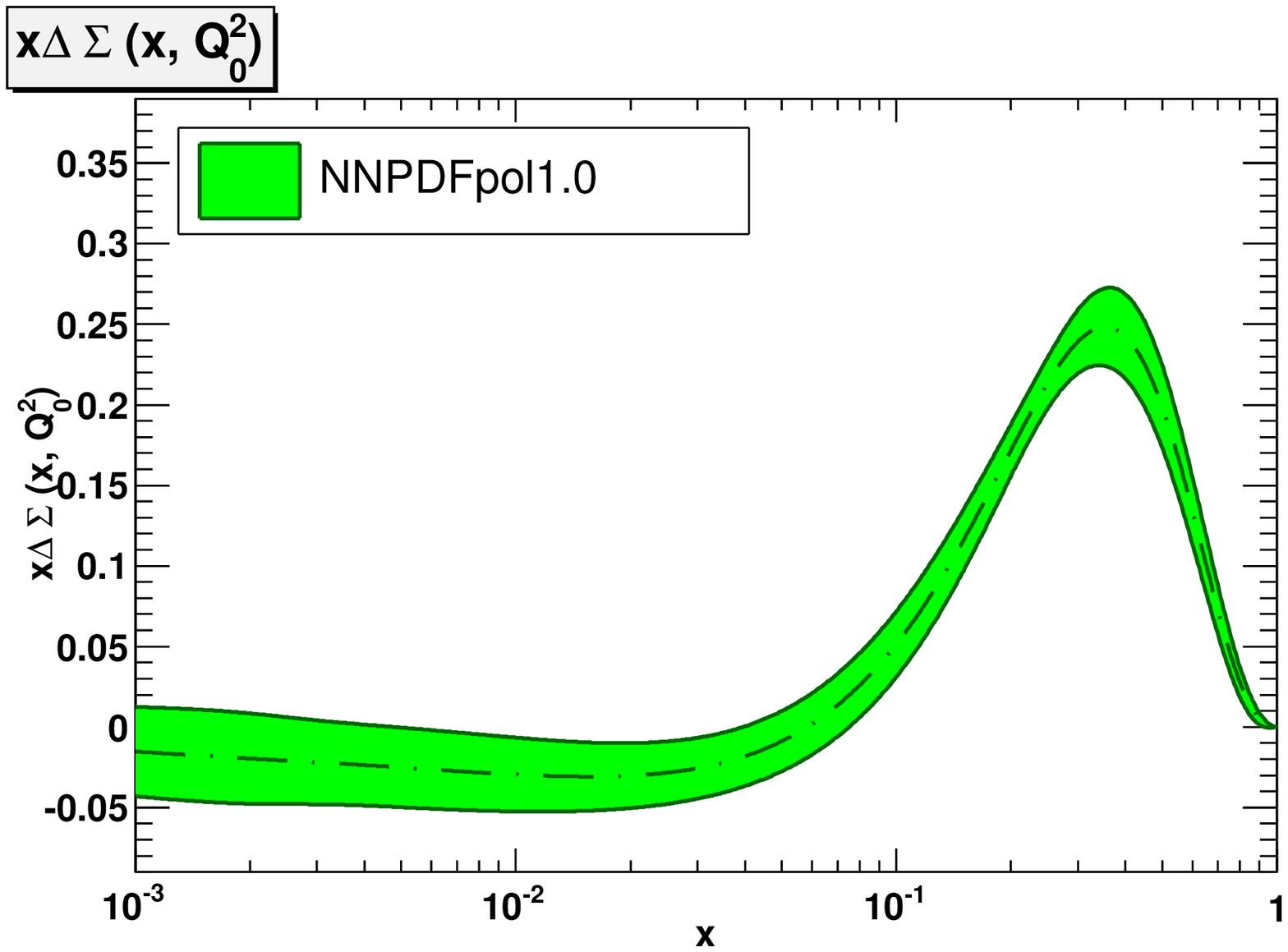}
\epsfig{width=0.43\textwidth,figure=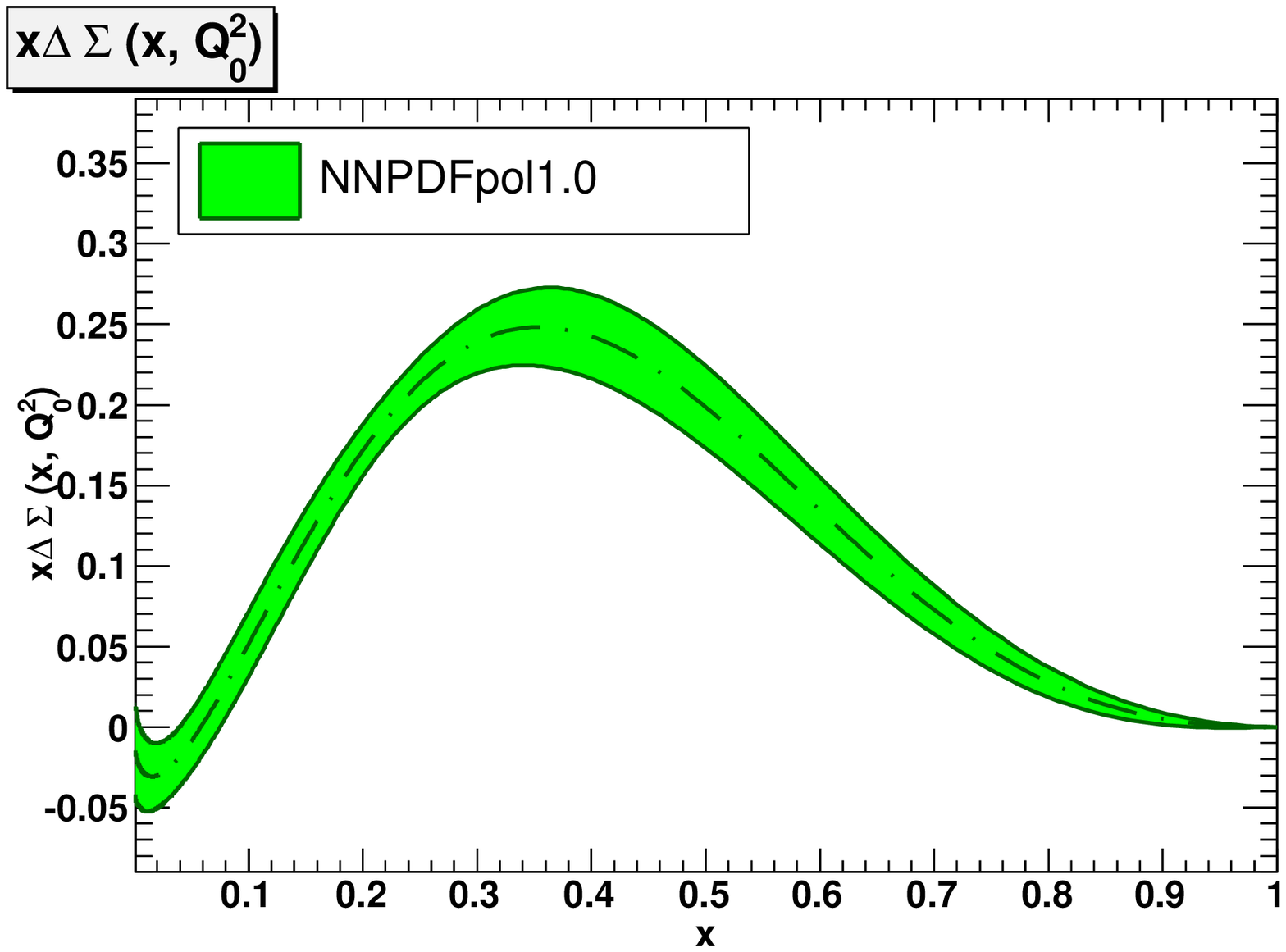}
\epsfig{width=0.43\textwidth,figure=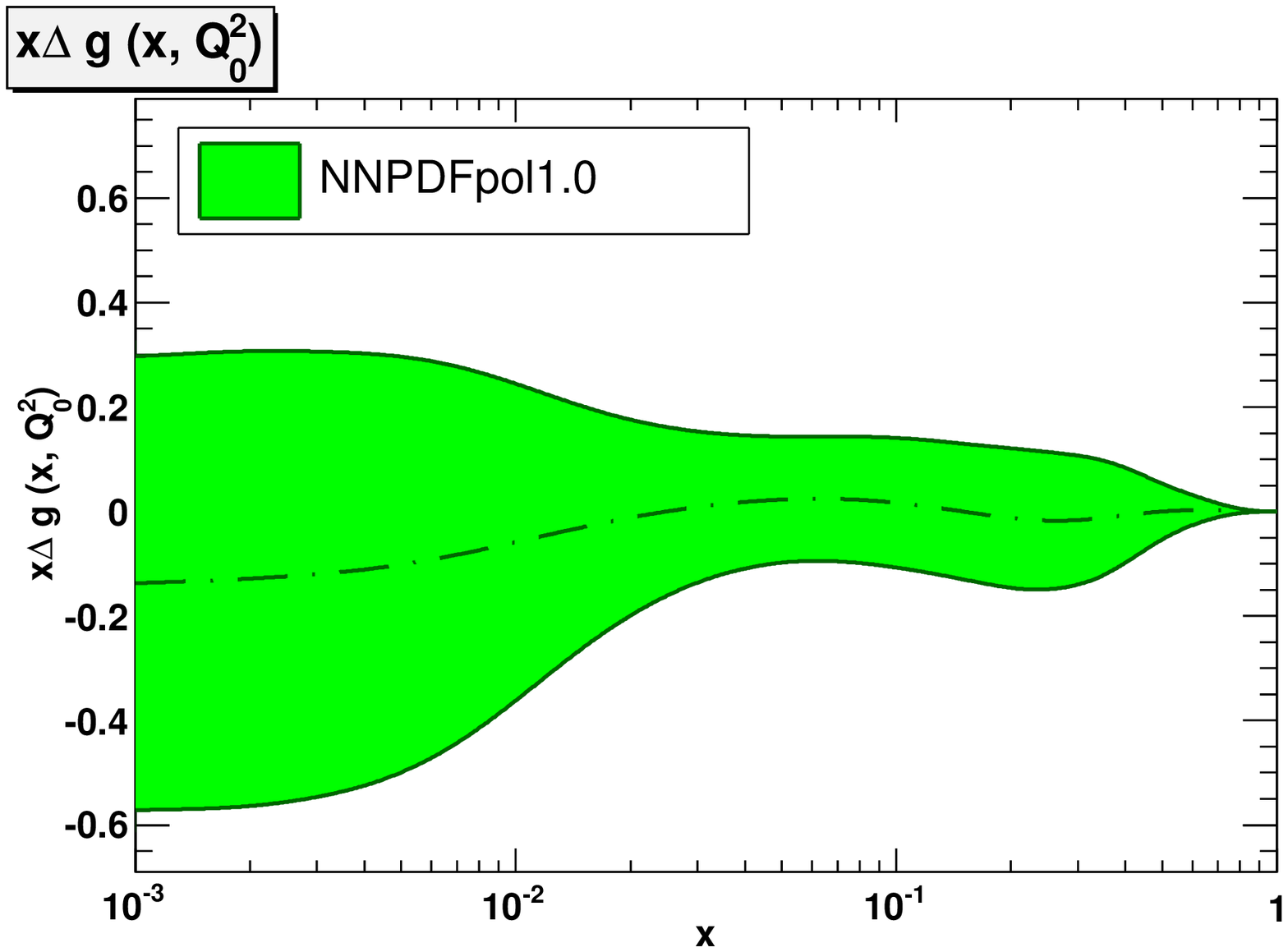}
\epsfig{width=0.43\textwidth,figure=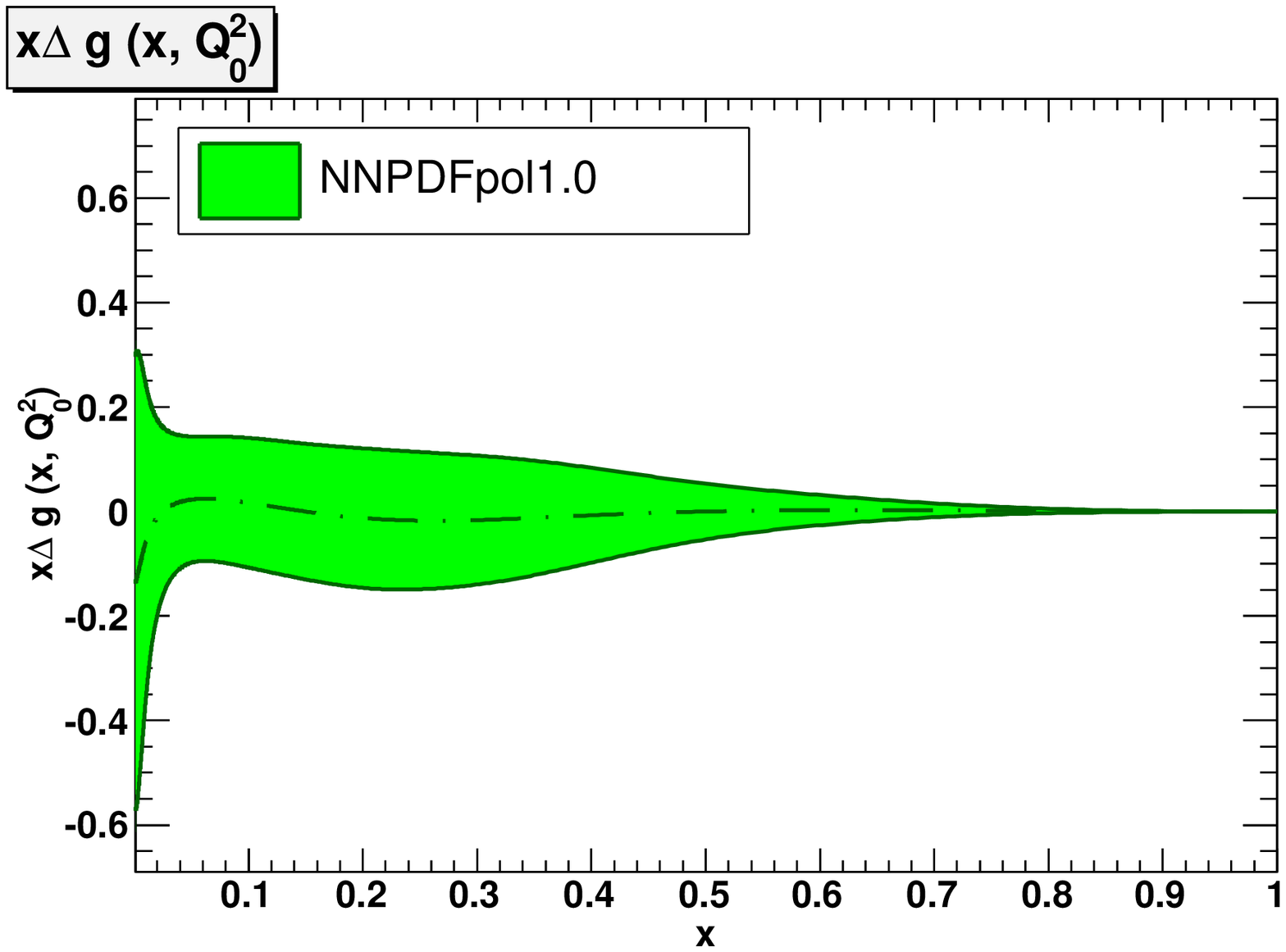}
\epsfig{width=0.43\textwidth,figure=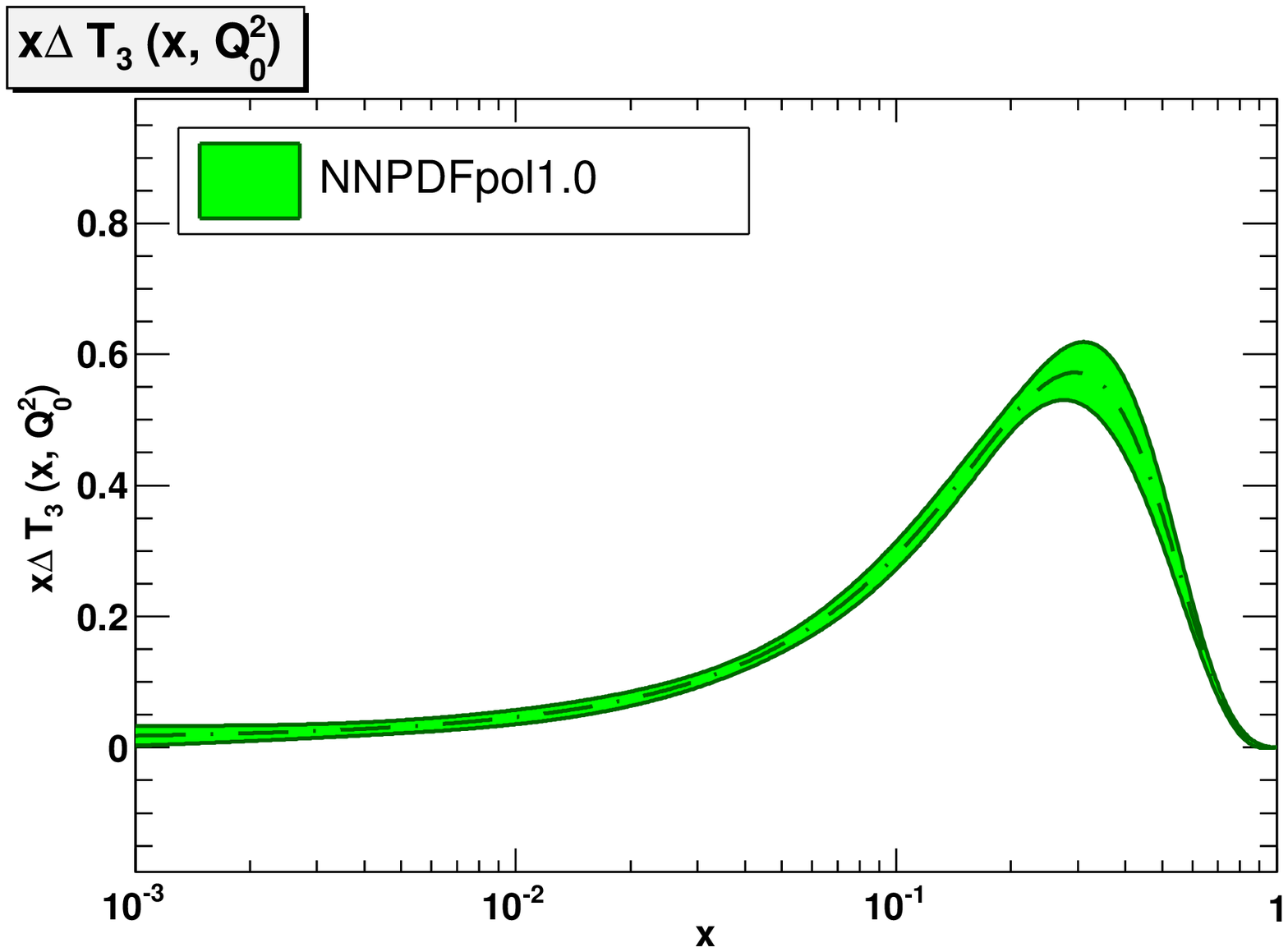}
\epsfig{width=0.43\textwidth,figure=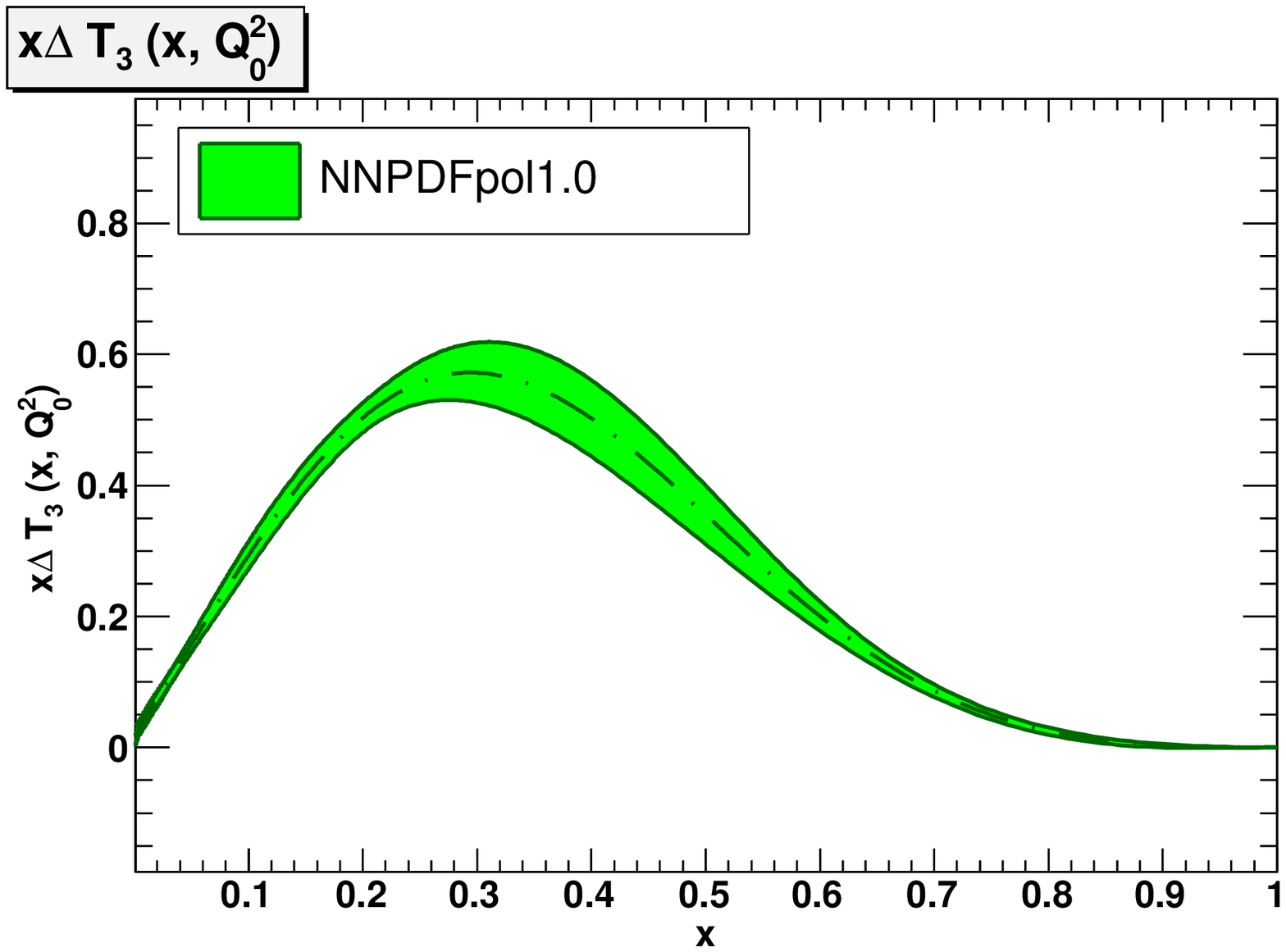}
\epsfig{width=0.43\textwidth,figure=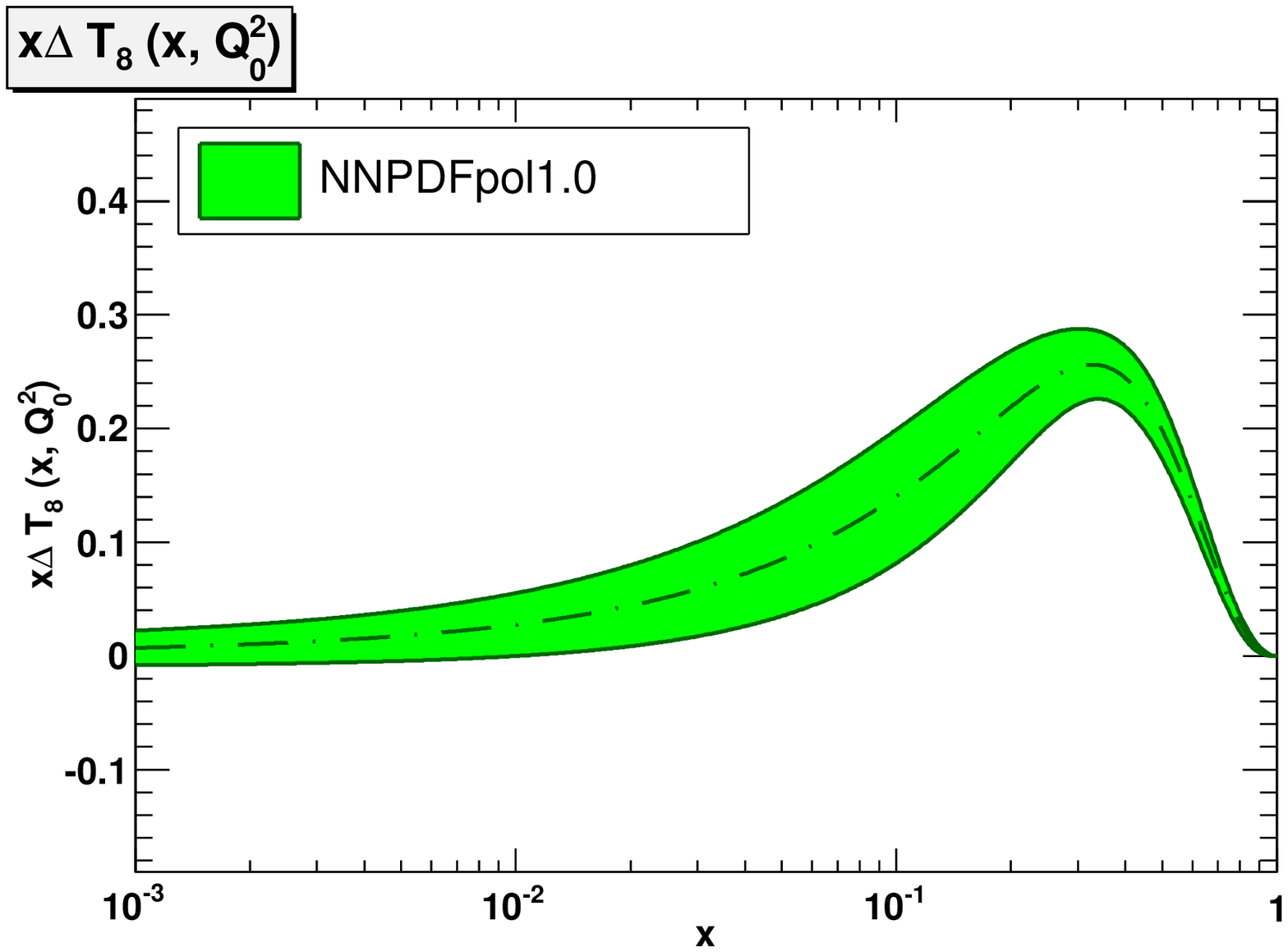}
\epsfig{width=0.43\textwidth,figure=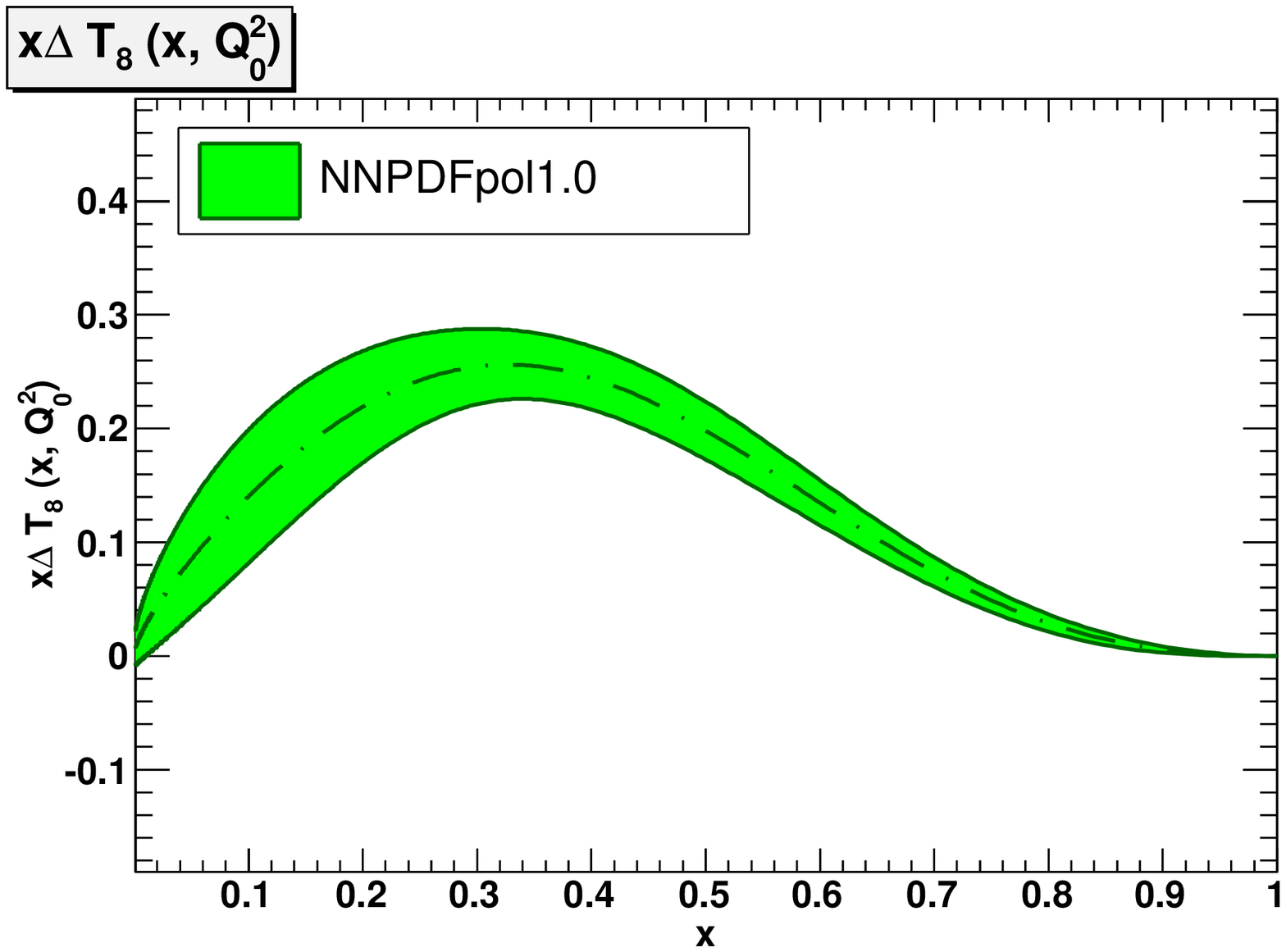}
\caption{\small The {\tt NNPDFpol1.0} polarized parton distributions at 
 $Q_0^2=1$ GeV$^2$ in the parametrization basis plotted as a function of $x$,
  on a logarithmic (left) and linear (right) scale.}
\label{fig:ppdfs-100}
\end{center}
\end{figure}

\begin{figure}[p]
\begin{center}
\epsfig{width=0.43\textwidth,figure=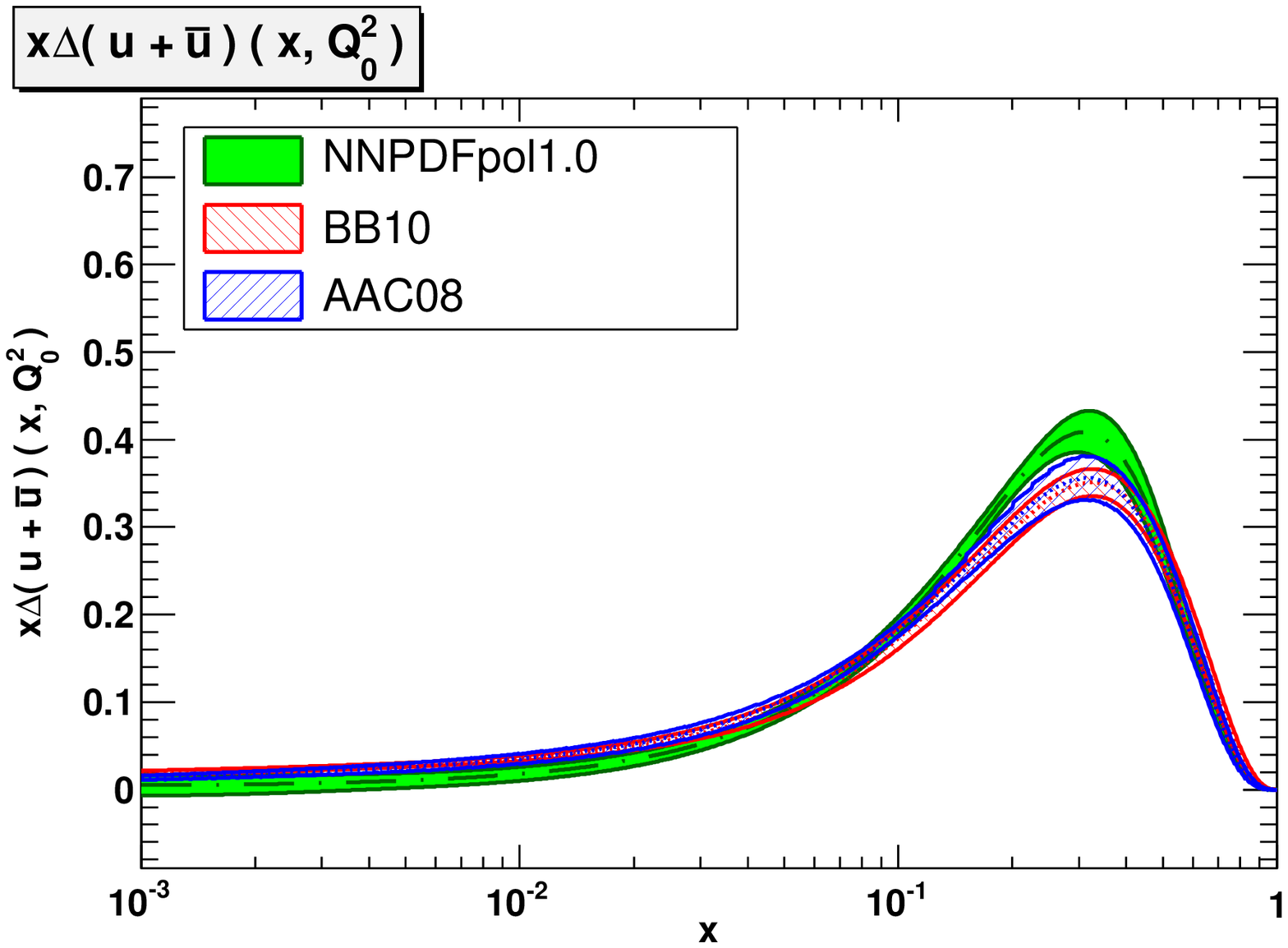}
\epsfig{width=0.43\textwidth,figure=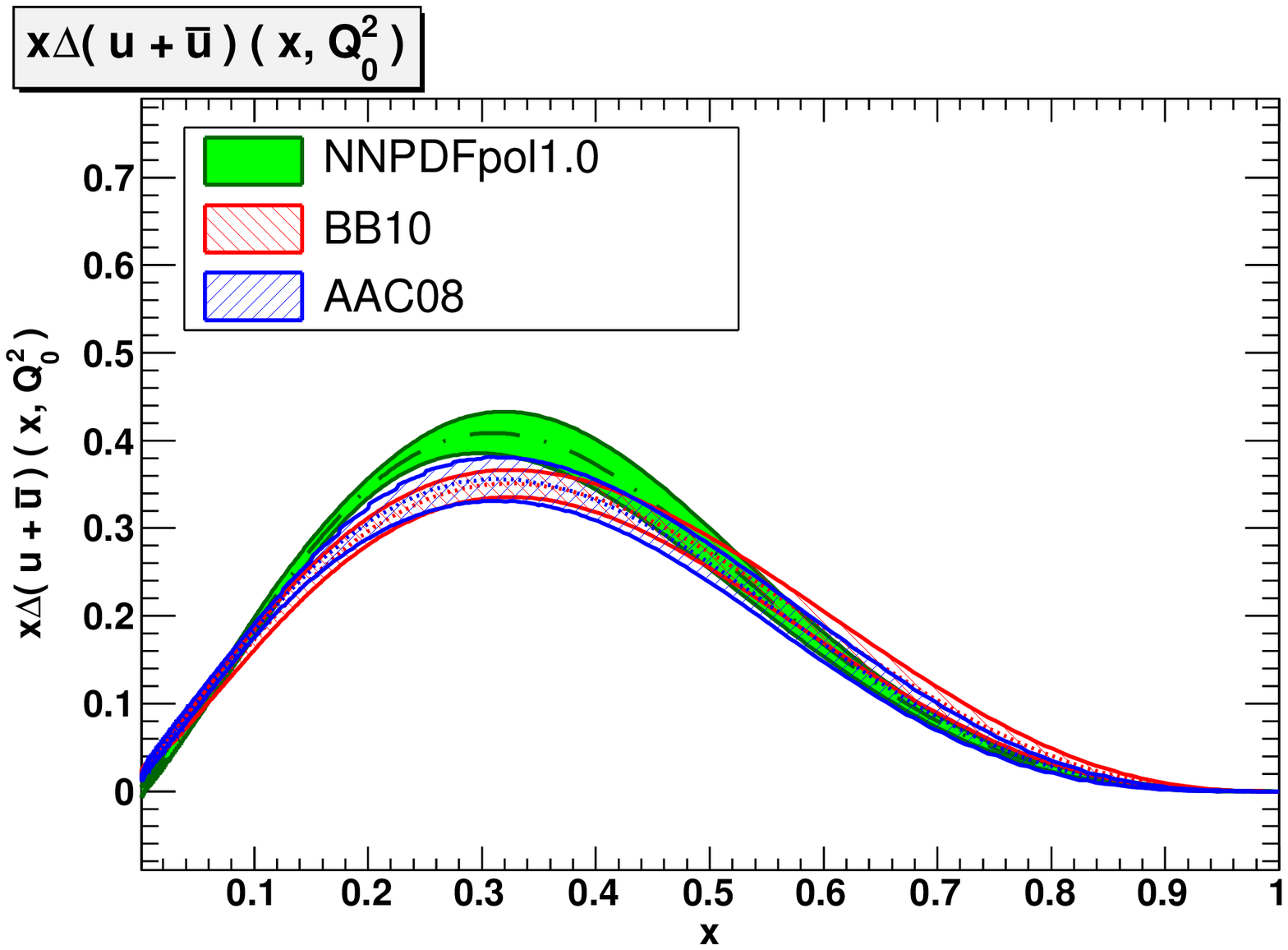}
\epsfig{width=0.43\textwidth,figure=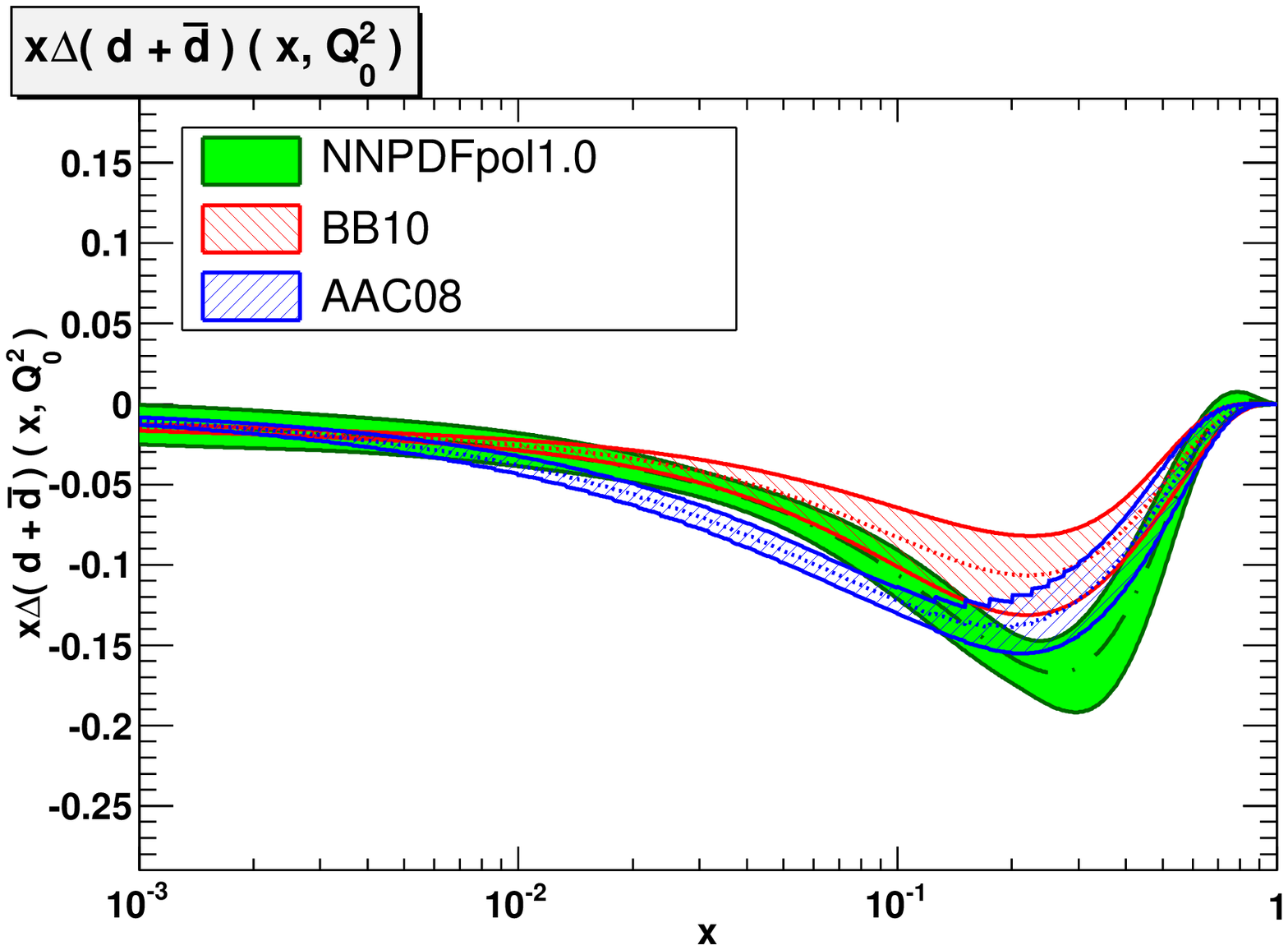}
\epsfig{width=0.43\textwidth,figure=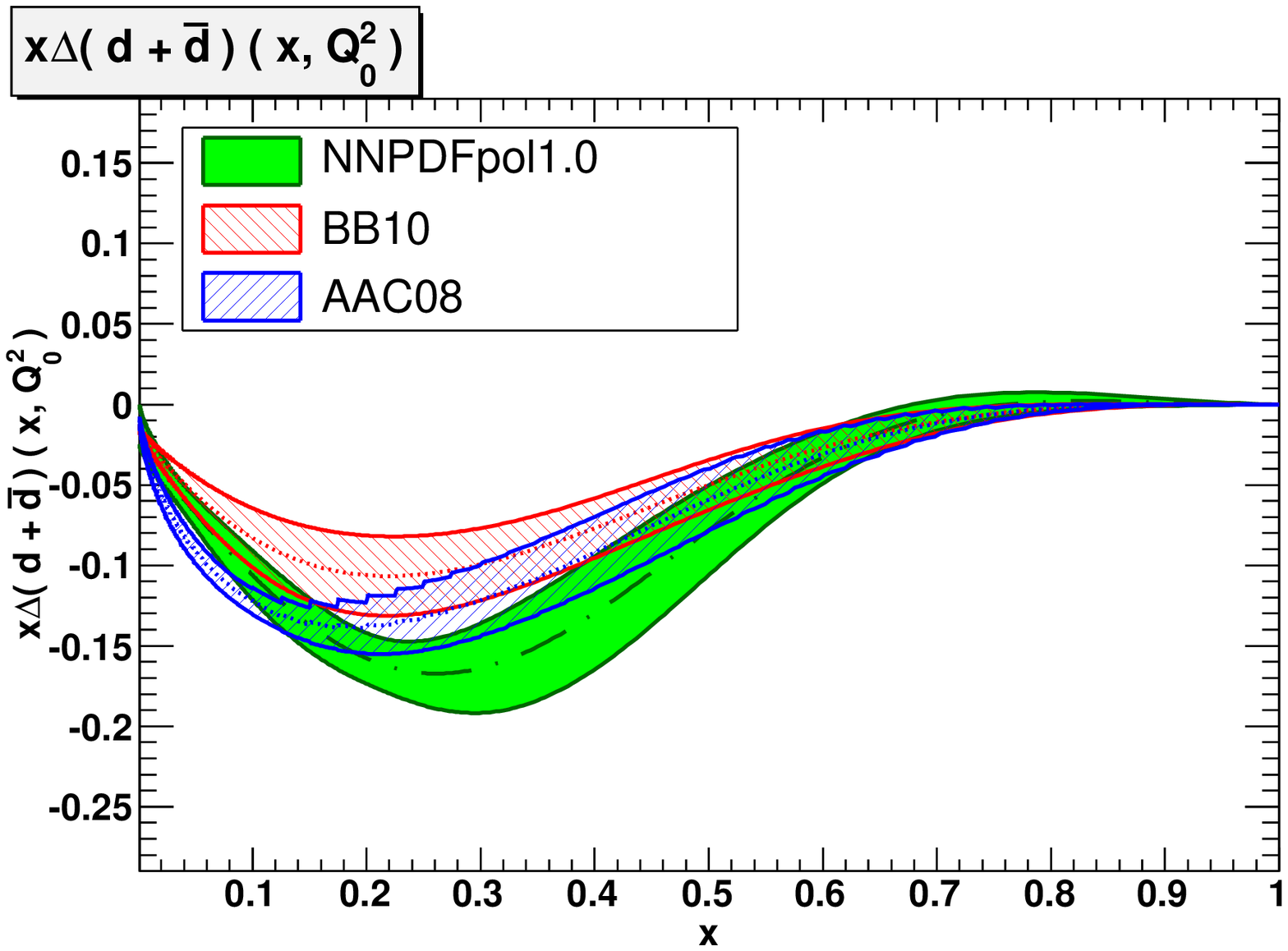}
\epsfig{width=0.43\textwidth,figure=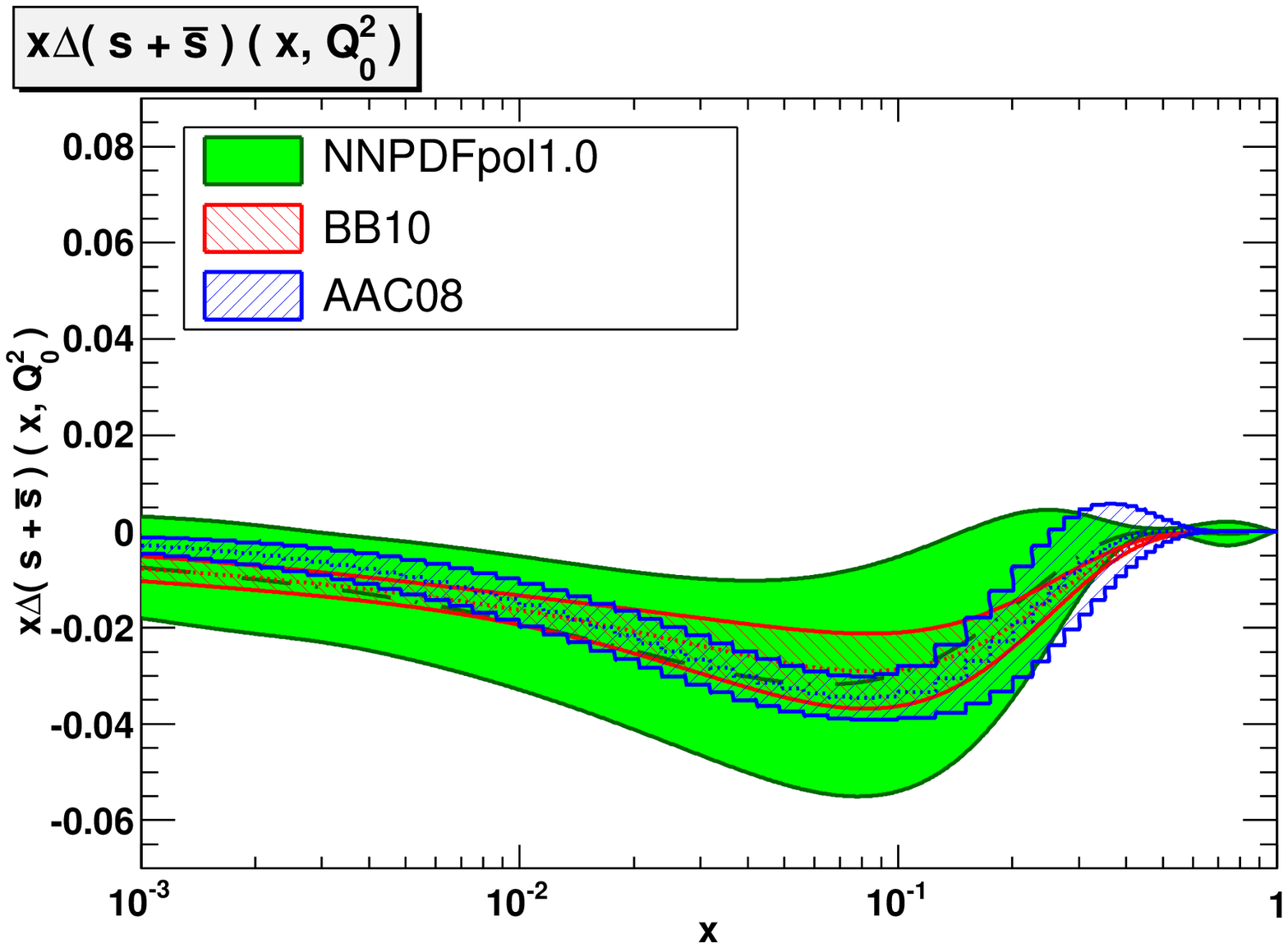}
\epsfig{width=0.43\textwidth,figure=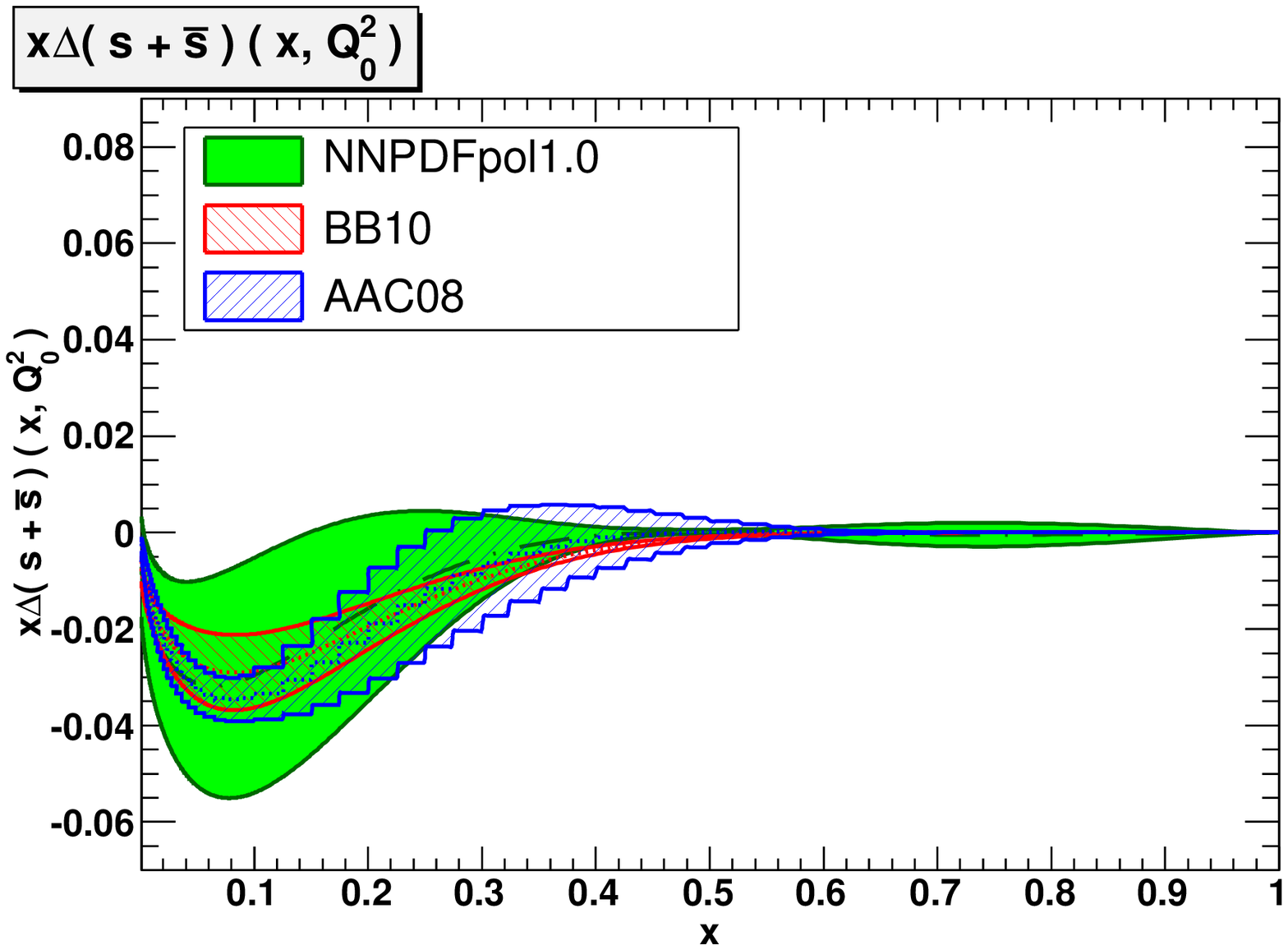}
\epsfig{width=0.43\textwidth,figure=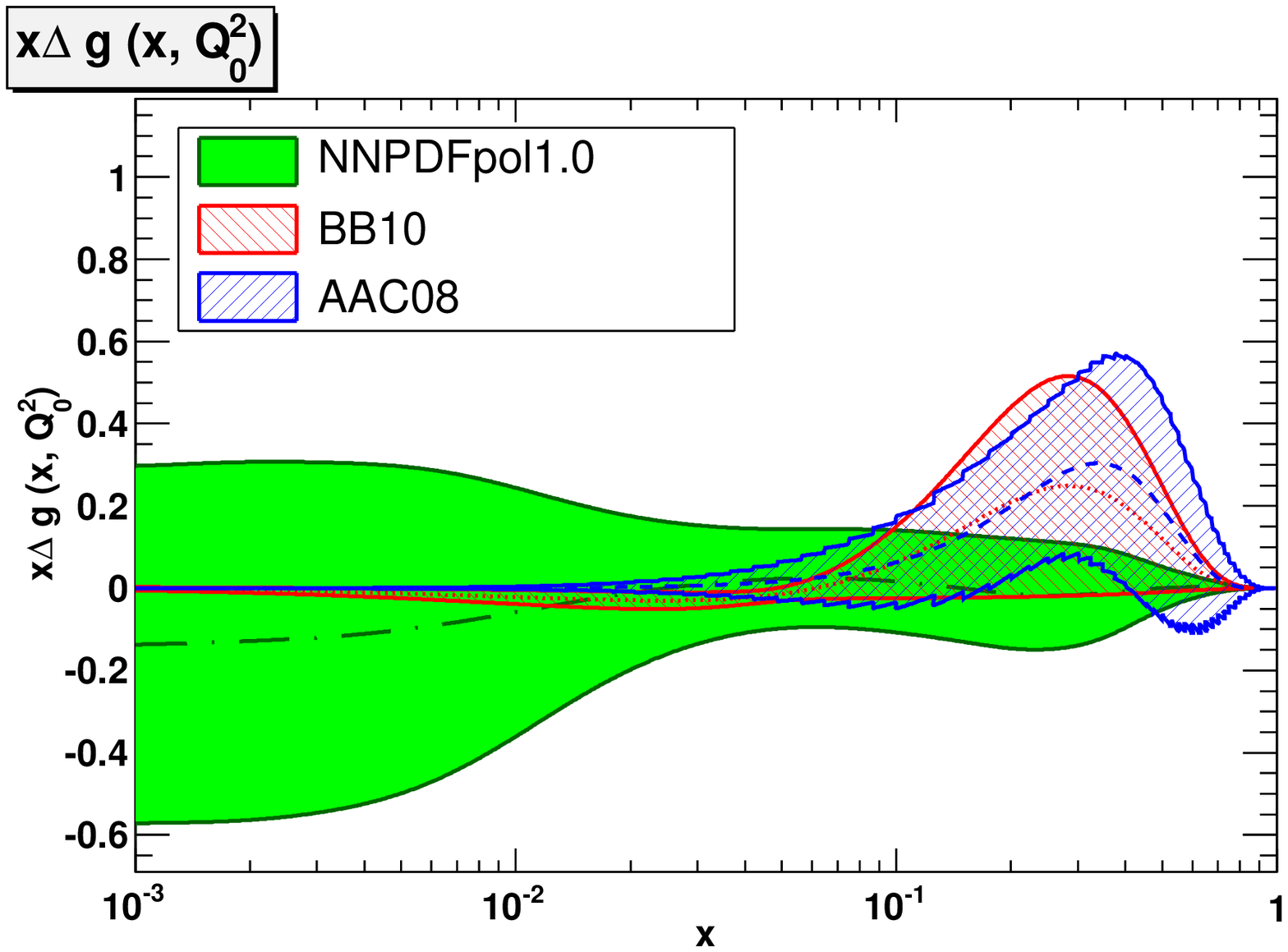}
\epsfig{width=0.43\textwidth,figure=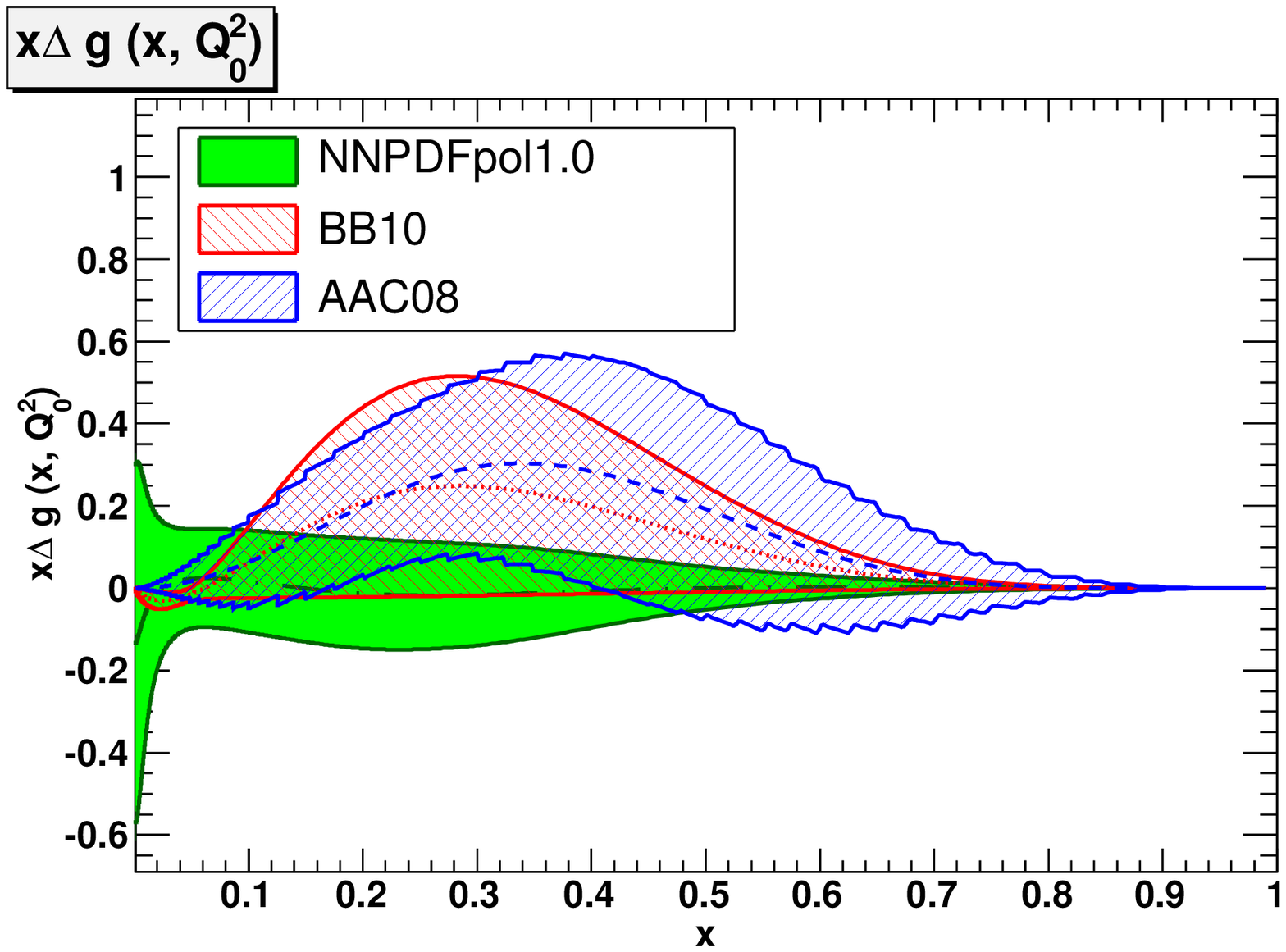}
\caption{\small Comparison of the {\tt NNPDFpol1.0} PDFs (in the
  flavour basis) and the
  BB10~\cite{Blumlein:2010rn} and AAC08~\cite{Hirai:2008aj}
  PDFs. \label{fig:ppdfs3}}
\end{center}
\end{figure}

\begin{figure}[p]
\begin{center}
\epsfig{width=0.43\textwidth,figure=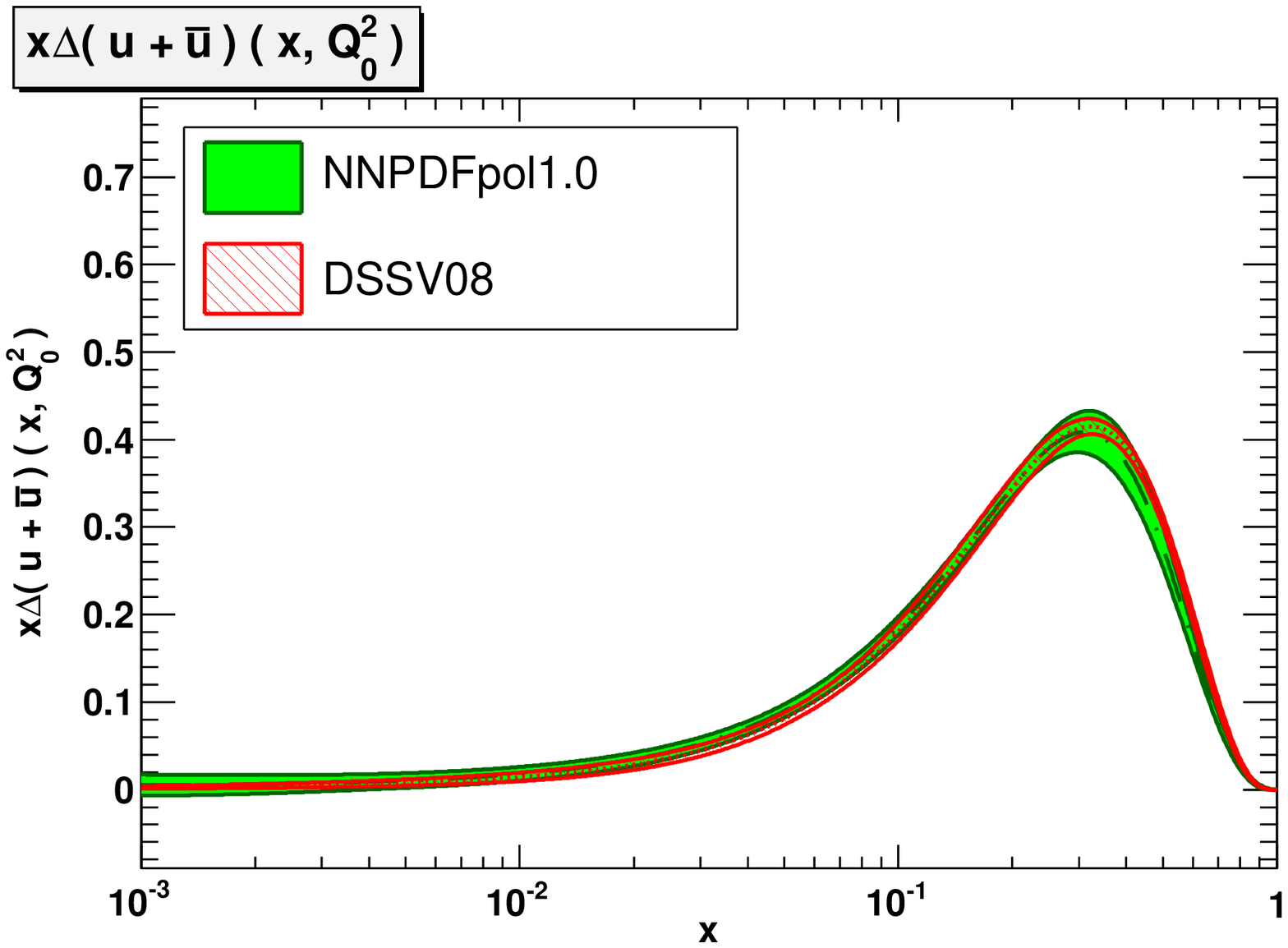}
\epsfig{width=0.43\textwidth,figure=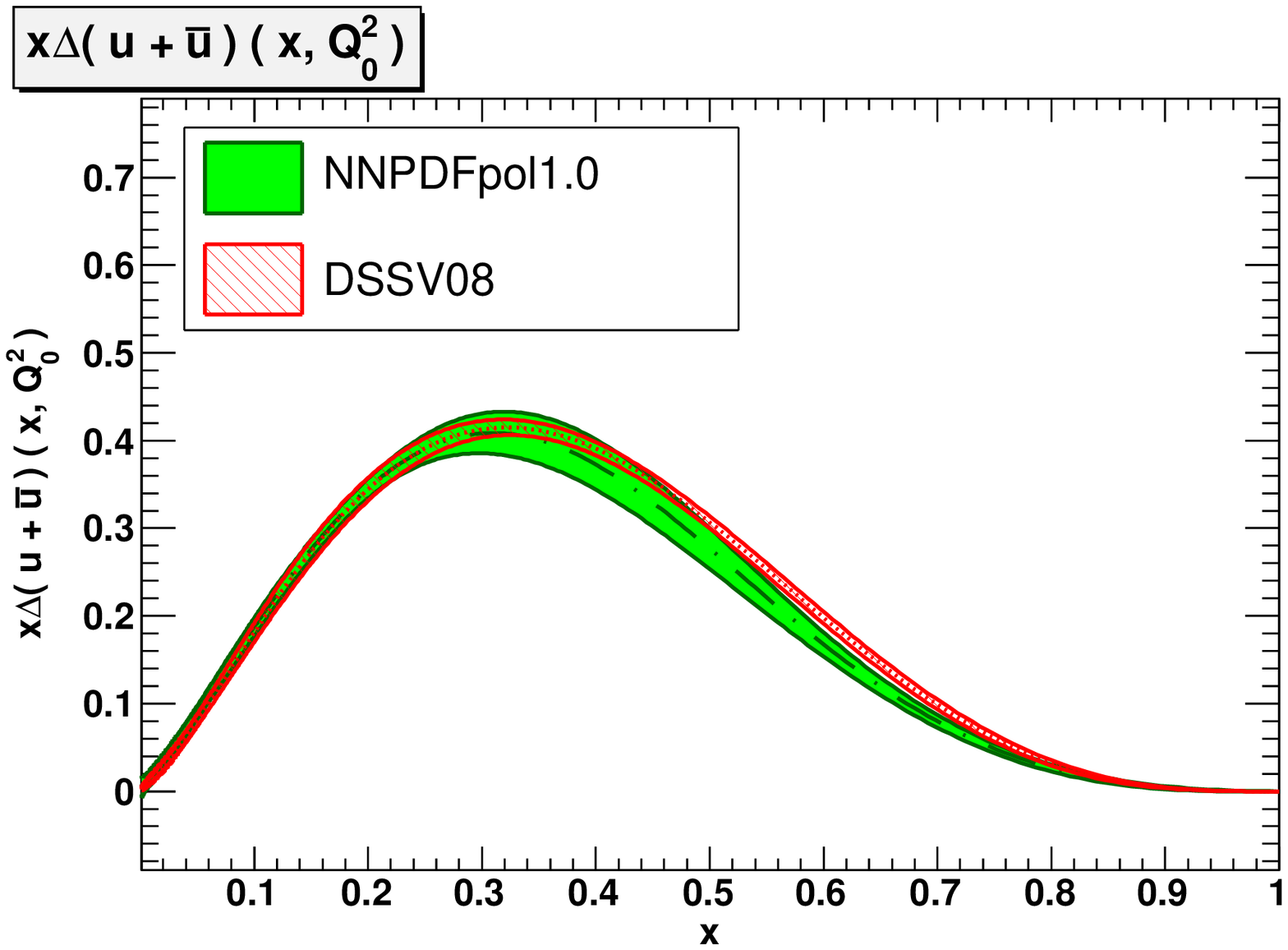}
\epsfig{width=0.43\textwidth,figure=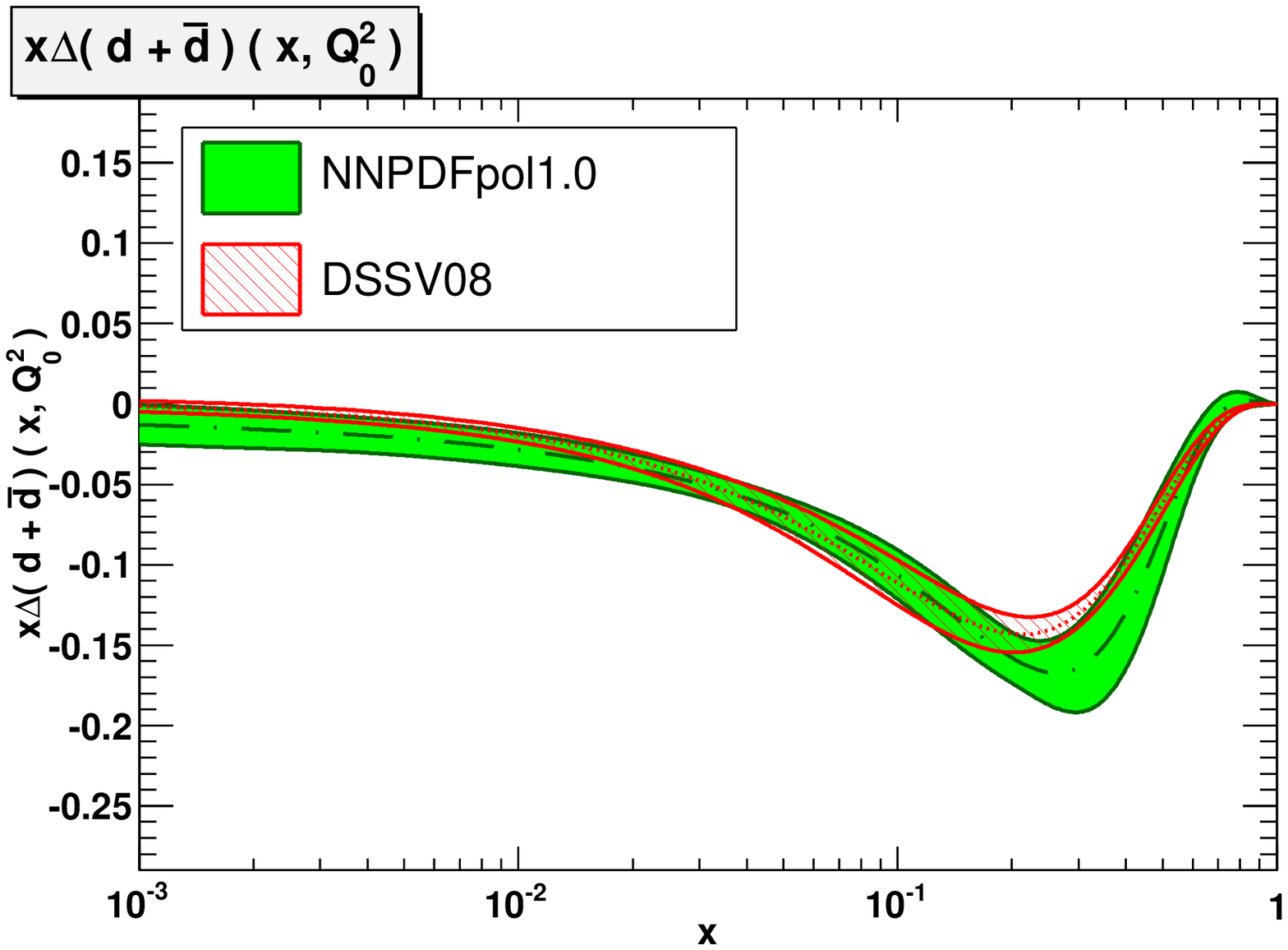}
\epsfig{width=0.43\textwidth,figure=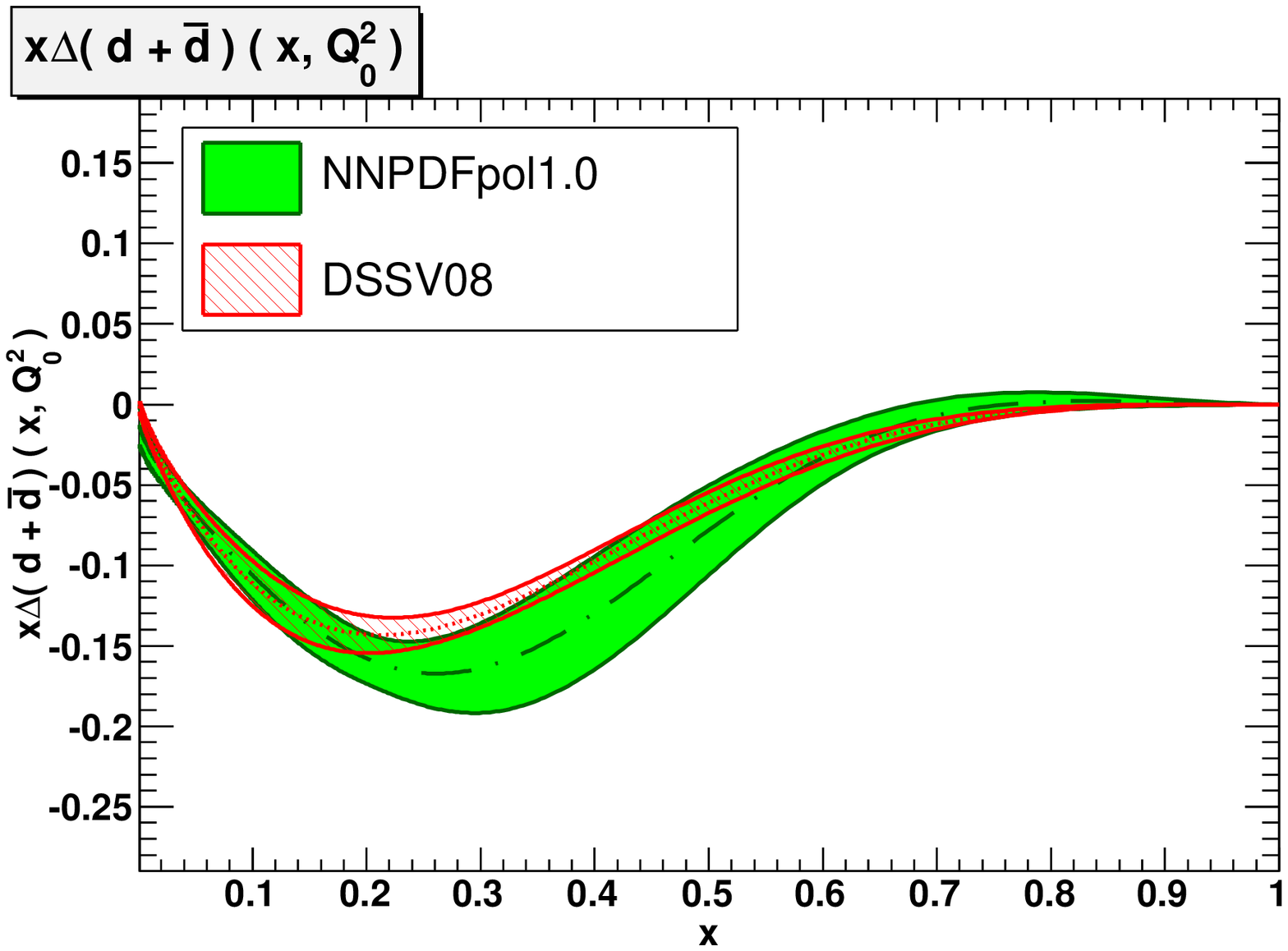}
\epsfig{width=0.43\textwidth,figure=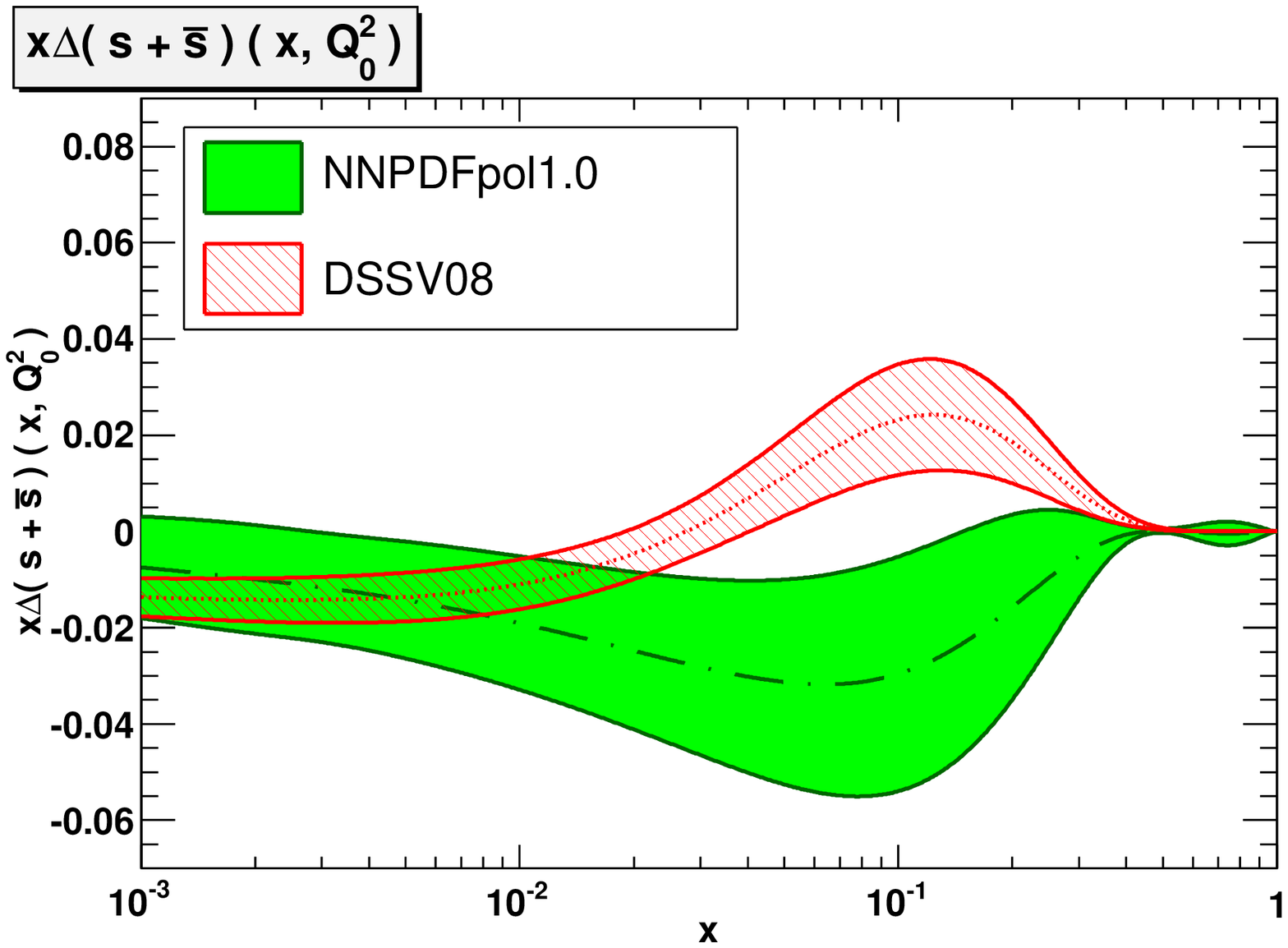}
\epsfig{width=0.43\textwidth,figure=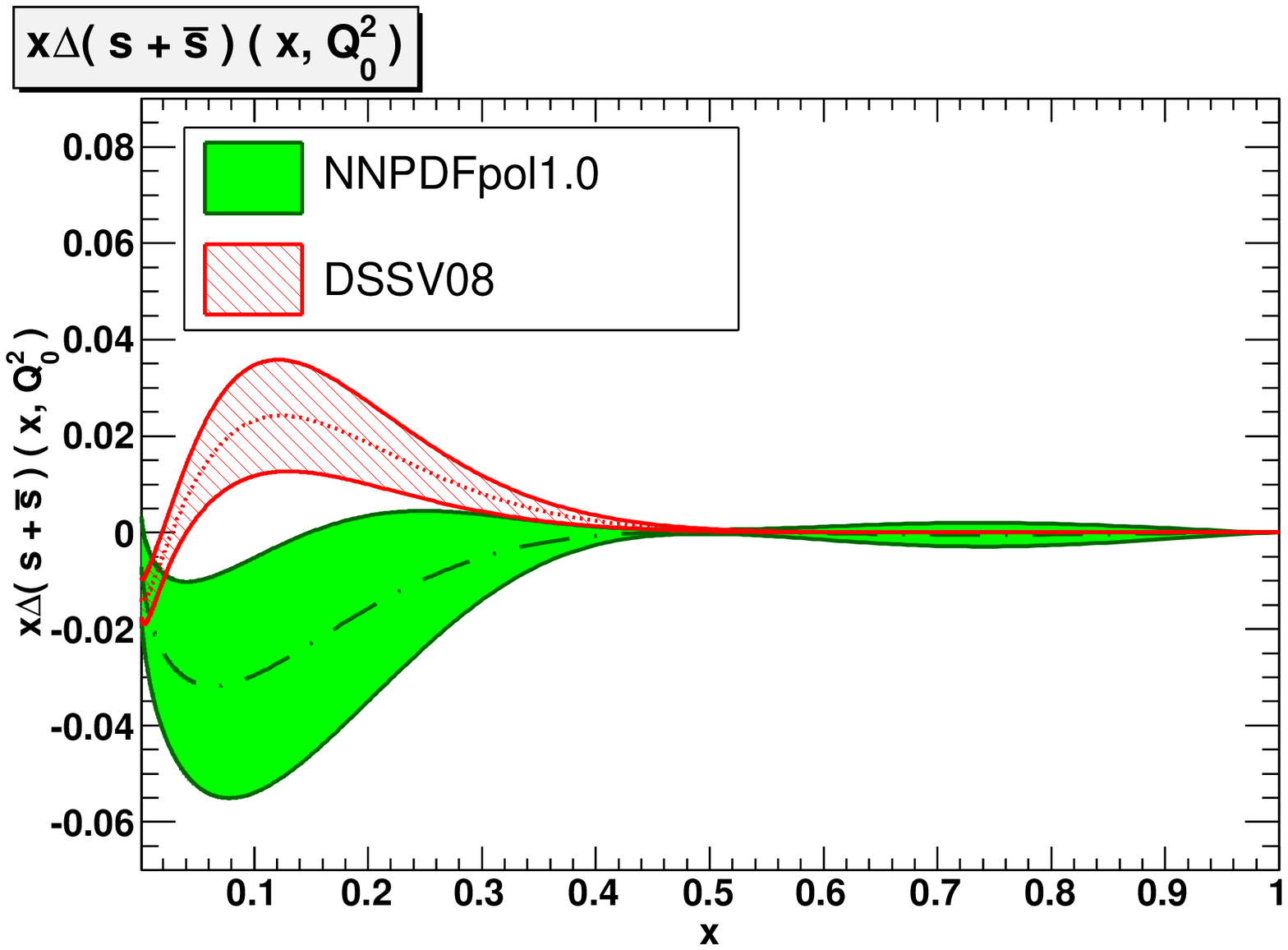}
\epsfig{width=0.43\textwidth,figure=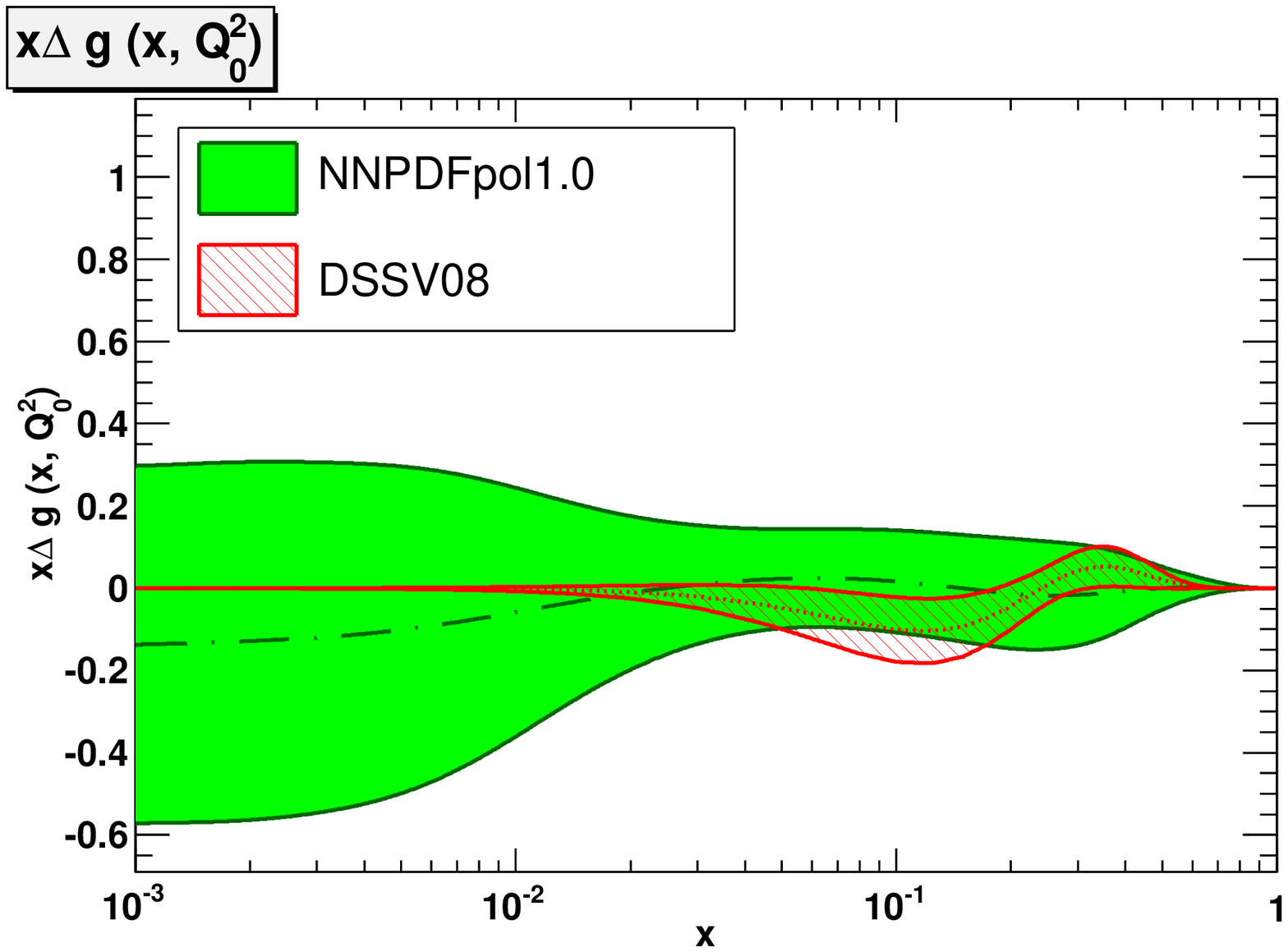}
\epsfig{width=0.43\textwidth,figure=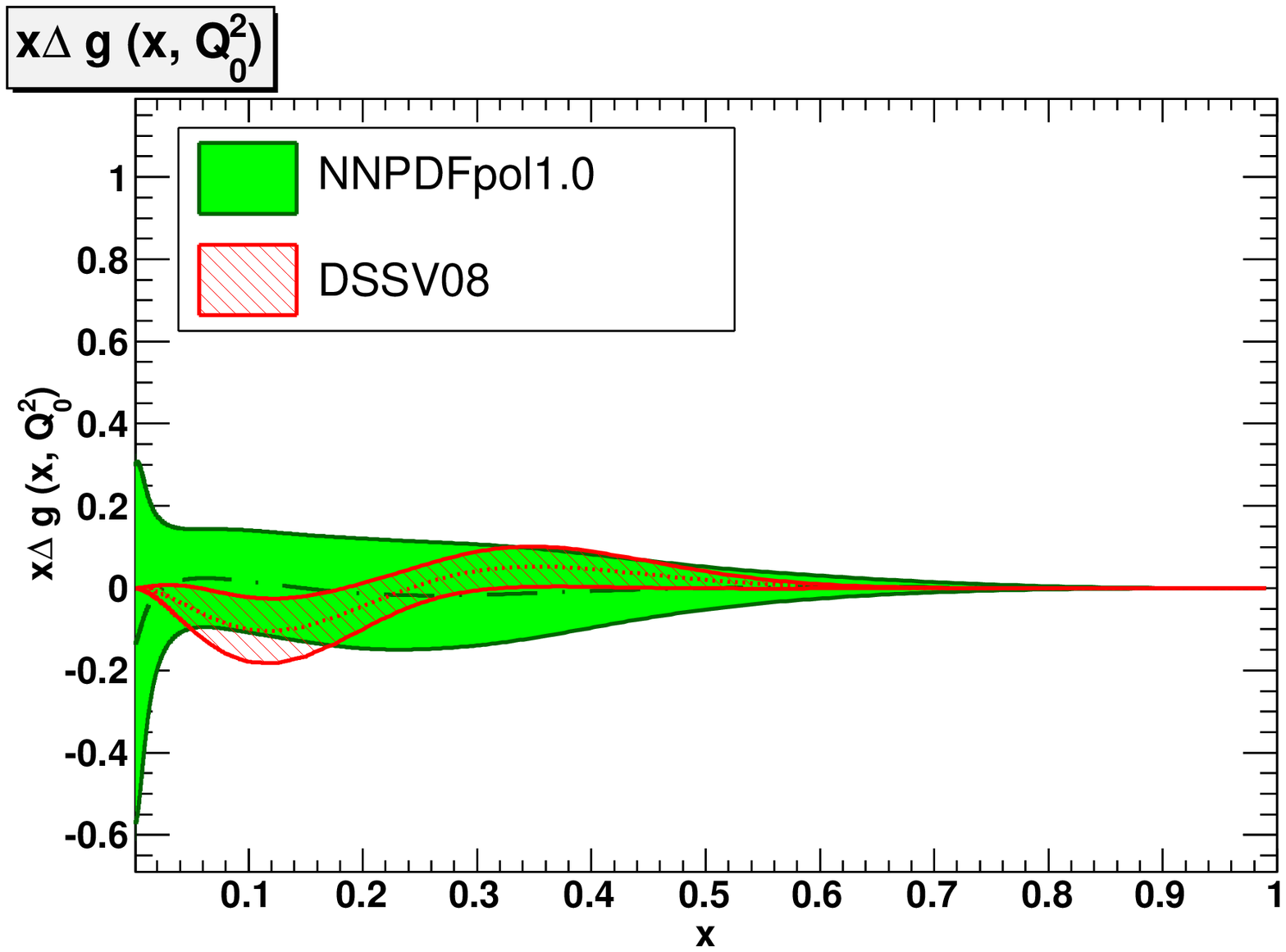}
\caption{\small Comparison of the {\tt NNPDFpol1.0} PDFs (in the
  flavour basis) and the DSSV08 PDFs~\cite{deFlorian:2009vb}. 
\label{fig:ppdfs2}}
\end{center}
\end{figure}

The {\tt NNPDFpol1.0} parton distributions, computed from a set of
$N_{\mbox{\scriptsize{rep}}}=100$ replicas, are displayed in
Fig.~\ref{fig:ppdfs-100} at the input scale $Q_0^2=1$ GeV$^2$, in the
PDF parametrization basis Eq.~(\ref{eq:PDFbasisnets}) as a function of
$x$ both on a logarithmic and linear scale.  In
Figs.~\ref{fig:ppdfs3}-\ref{fig:ppdfs2} the same PDFs are plotted in
the flavour basis, and compared to other available NLO PDF sets:
BB10~\cite{Blumlein:2010rn} and AAC08~\cite{Hirai:2008aj} in
Fig.~\ref{fig:ppdfs3}, and DSSV08~\cite{deFlorian:2009vb} in
Fig.~\ref{fig:ppdfs2}. We do not show a direct comparison to the LSS
polarized PDFs~\cite{Leader:2010rb} because there are no publicly
available routines for the computation of PDF uncertainties for this
set.  Note that the dataset used for the BB10 determination contains
purely DIS data, and that for
AAC contains DIS supplemented by some high-$p_T$ RHIC pion production data:
hence they are directly comparable to our PDF determination. The
DSSV08 determination instead includes, on top of DIS data, polarized
jet production data, and, more importantly, a large amount of
semi-inclusive DIS data which in particular allow for
flavour-antiflavour separation and a more direct handle on
strangeness.  All uncertainties in these plots correspond to the
nominal 1--$\sigma$ error bands.

The main conclusions of this comparison are the following:
\begin{itemize}
\item The central values of the $\Delta u + \Delta\bar{u}$ and the
  $\Delta d + \Delta\bar{d}$  are in
  reasonable agreement with those of other parton sets. The {\tt NNPDFpol1.0}
  results are in best agreement with DSSV08, in slightly worse
  agreement with AAC08, and in worst agreement with
  BB10. Uncertainties on these PDFs are generally slightly larger for
  NNPDF than for other sets, especially DSSV, which however is based
  on a much wider dataset. 
\item The {\tt NNPDFpol1.0} determination of  $\Delta s +
  \Delta\bar{s}$ is affected by a much larger uncertainty than BB10
  and AAC08,  for almost all
  values of $x$. The AAC08 and BB10 strange PDFs fall well within the
{\tt NNPDFpol1.0} uncertainty band.
\item  The {\tt NNPDFpol1.0} determination of $\Delta s +
  \Delta\bar{s}$ is inconsistent at the two sigma level 
  in the medium-small $x\sim0.1$ region with DSSV08, which is also
  rather more accurate, as one would expect as it includes
  semi-inclusive data (in particular for production of hadrons with
  strangeness). This suggests a tension between the inclusive analysis
  data and the semi-inclusive analysis.
\item The gluon PDF is affected by a large uncertainty,
rather larger than any other set, especially at small $x$. In
particular, the {\tt NNPDFpol1.0} polarized gluon distribution 
is compatible with zero for all values of $x$.
\item Uncertainties on the PDFs in the regions where no data are available
  tend to be larger than those of other sets. At very large
  values of $x$ the PDF uncertainty band is largely determined by the
  positivity constraint.
\end{itemize}

\begin{figure}[p]
\begin{center}
\epsfig{width=0.4\textwidth,figure=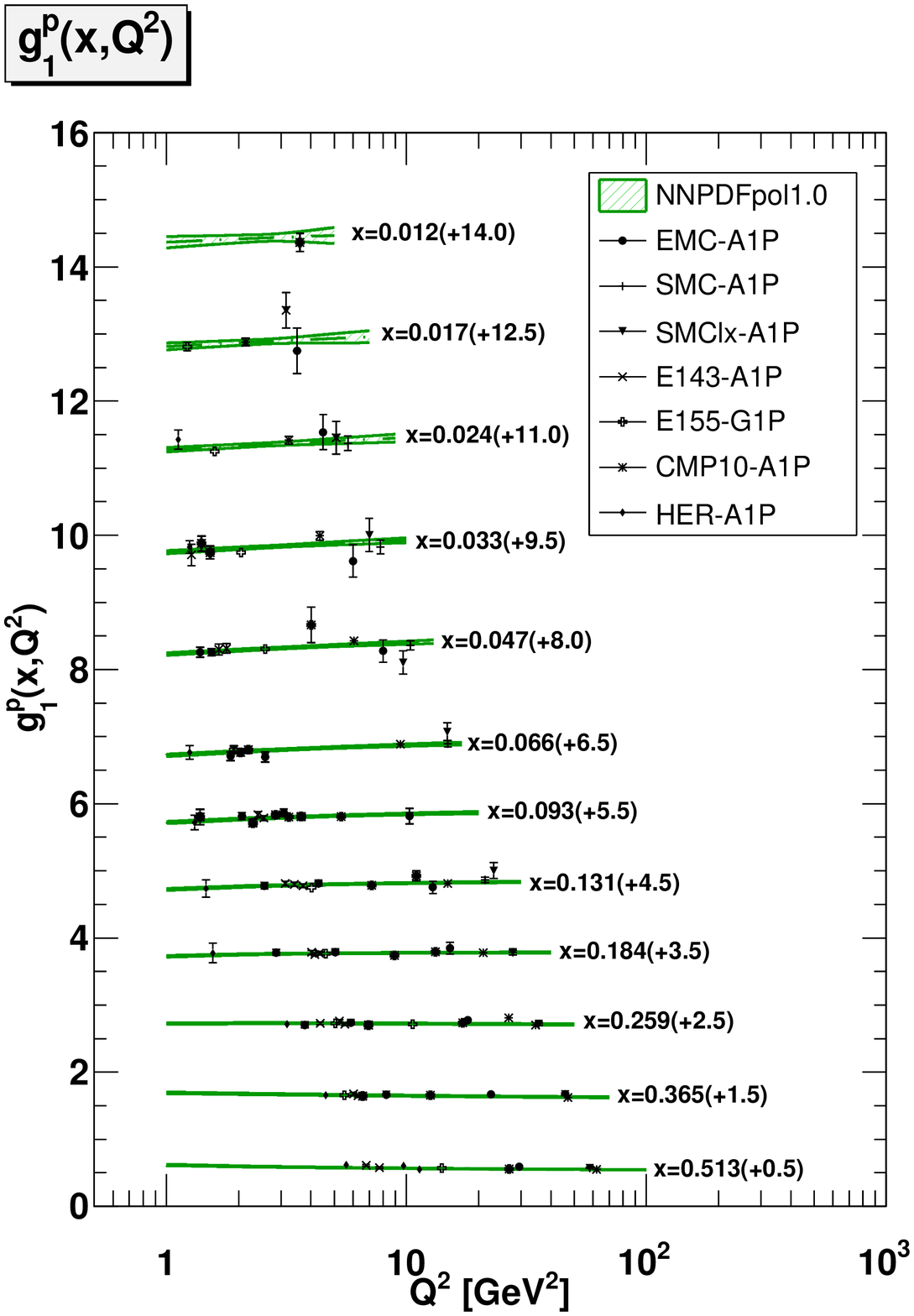}
\epsfig{width=0.4\textwidth,figure=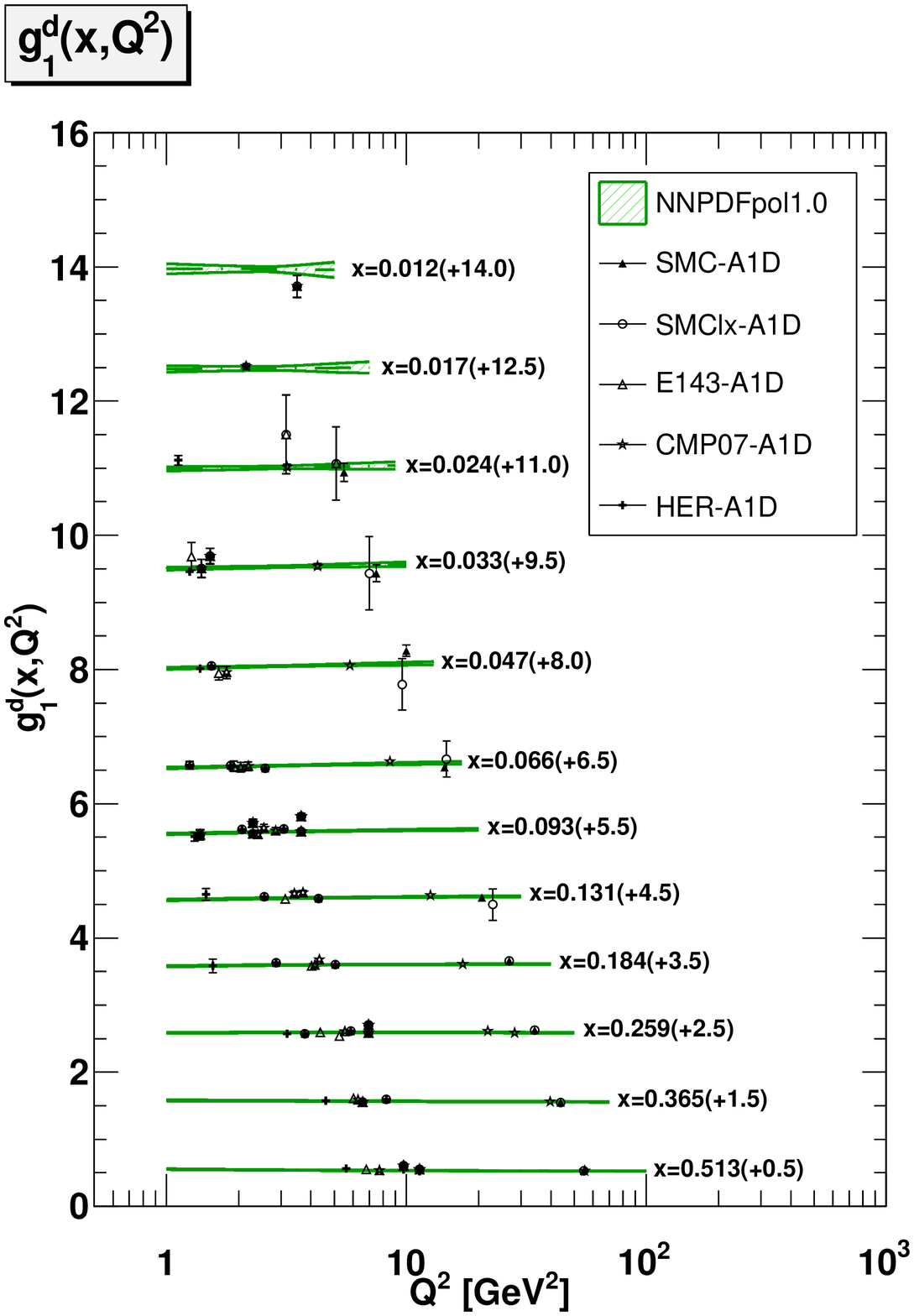}
\epsfig{width=0.4\textwidth,figure=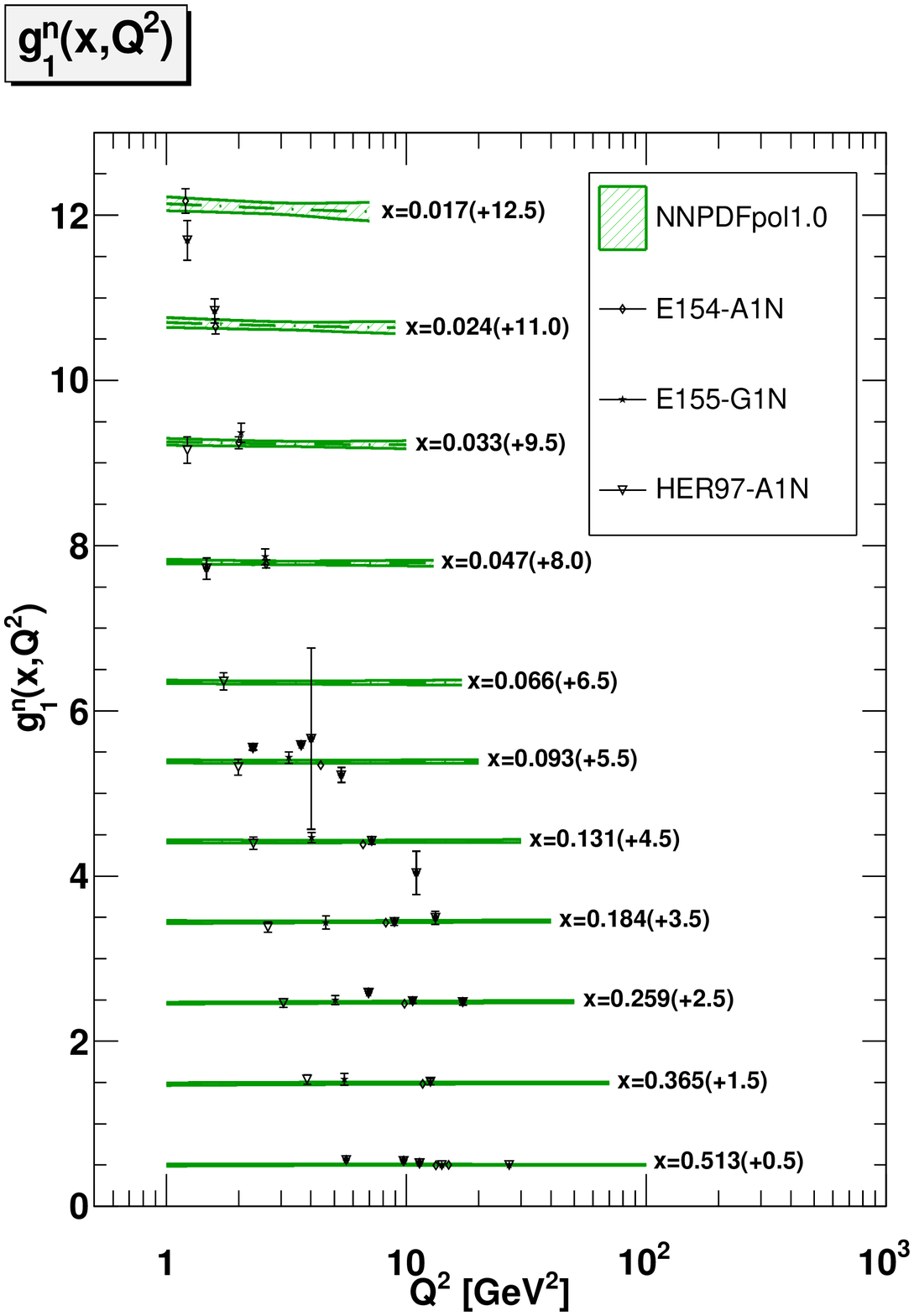}
\caption{\small The proton, neutron and deuteron 
polarized structure function $g_1(x,Q^2)$ as functions of $Q^2$ in 
different bins of $x$ compared to experimental data. 
Experimental data are grouped in bins of $x$, while
\texttt{NNPDFpol1.0} results are given at the center of each bin,
whose value is given next to each curve. In order to improve
legibility,
the values of $g_1(x,Q^2)$ 
have been shifted by the
amount given next to each curve.
\label{fig:g1}} 
\end{center}
\end{figure}

Finally, in Fig.~\ref{fig:g1} we compare 
 the 
structure function $g_1(x,Q^2)$  for proton,
deuteron and neutron, computed using {\tt NNPDFpol1.0} (with
its one-$\sigma$ uncertainty band) to the experimental data included in
the fit. Experimental data are grouped in bins of $x$
with a logarithmic spacing, while the NNPDF
prediction and its uncertainty are computed at the central value of
each bin.

The uncertainty band in the {\tt NNPDFpol1.0} result is typically
smaller than the experimental errors, except at small-$x$ where a much
more restricted dataset is available; in that region, the
uncertainties are comparable.  Scaling violations of the polarized
structure functions are clearly visible, especially for $g_1^p$,
despite the limited range in $Q^2$.

\subsection{Stability of the results}
\label{sec:stab}

Our results have been obtained with a number of theoretical
and methodological assumptions, discussed in
Sects.~\ref{sec:polpdfs}-\ref{sec:minim}. We will now test their
upon variation of these assumptions.

\subsubsection{Target-mass corrections and $g_2$.}
\label{sec:tmcres}

We have consistently included in our determination of $g_1$
corrections suppressed by powers of the nucleon mass which are of
kinematic origin. Thus in particular, 
as explained in Sec.~\ref{sec:tmc}, we have included
target-mass corrections (TMCs) up to first order in
${m^2}/{Q^2}$. Furthermore, both TMCs and the relation between the
measured asymmetries and the structure function $g_1$ involve
contributions to the structure
function $g_2$ proportional to powers of ${m^2}/{Q^2}$ which we
include according to Eq.~(\ref{eq:g1tog2}) or Eq.~(\ref{eq:g1tog2p})
(see the discussion in Sect.~\ref{sec:datasetl}). 
Our default PDF set is obtained
assuming that $g_2$ is given by the Wandzura-Wilczek relation,
Eq.~(\ref{eq:wwrel}).

In order to assess the impact of these assumptions on our results, we
have performed two more PDF determinations. In the first, we set $m=0$
consistently everywhere, both in the extraction of the structure functions 
from the
asymmetry data and in our computation of structure functions. 
This thus removes TMCs, and also contributions
proportional to $g_2$.  In the second, we retain
mass effects,  but we  assume $g_2=0$.
  
The statistical
estimators for each of these three fits over the full dataset are
shown in Tab.~\ref{tab:tmc_estimators}. Clearly, all fits
are of comparable quality.
\begin{table}[t]
 \centering
 \small
 \begin{tabular}{|c|c|c|c|}
  \hline
  Fit & {\tt NNPDFpol1.0} $g_2=g_2^{\mbox{\tiny WW}}$ & {\tt NNPDFpol1.0} $m=0$ & {\tt NNPDFpol1.0} $g_2=0$ \\
  \hline
  \hline
  $\chi^{2}_{\tot}$                            & 0.77 & 0.78             & 0.75             \\
  $\la E \ra \pm \sigma_{E}$                  & 1.82 $\pm$ 0.18   &  1.81 $\pm$ 0.16 & 1.83 $\pm$ 0.15  \\
  $\la E_{\rm tr} \ra \pm \sigma_{E_{\rm tr}}$   & 1.66 $\pm$ 0.49    & 1.62 $\pm$ 0.50  & 1.70 $\pm$ 0.38\\
  $\la E_{\rm val} \ra \pm \sigma_{E_{\rm val}}$ & 1.88 $\pm$ 0.67   & 1.84 $\pm$ 0.70 & 1.96 $\pm$ 0.56\\
  \hline
  $\la \chi^{2(k)} \ra \pm \sigma_{\chi^{2}}$  & 0.91 $\pm$ 0.12 & 0.90 $\pm$ 0.09 & 0.86 $\pm$ 0.09 \\
  \hline
 \end{tabular}
 \caption{\small The statistical estimators of Tab.~\ref{tab:chi2tab1}
   (obtained
assuming $g_2=g_2^{\mbox{\tiny WW}}$) compared to a fit with $m=0$ or
  with $g_2=0$.}
 \label{tab:tmc_estimators}
\end{table}

\begin{figure}[t]
\begin{center}
\epsfig{width=0.43\textwidth,figure=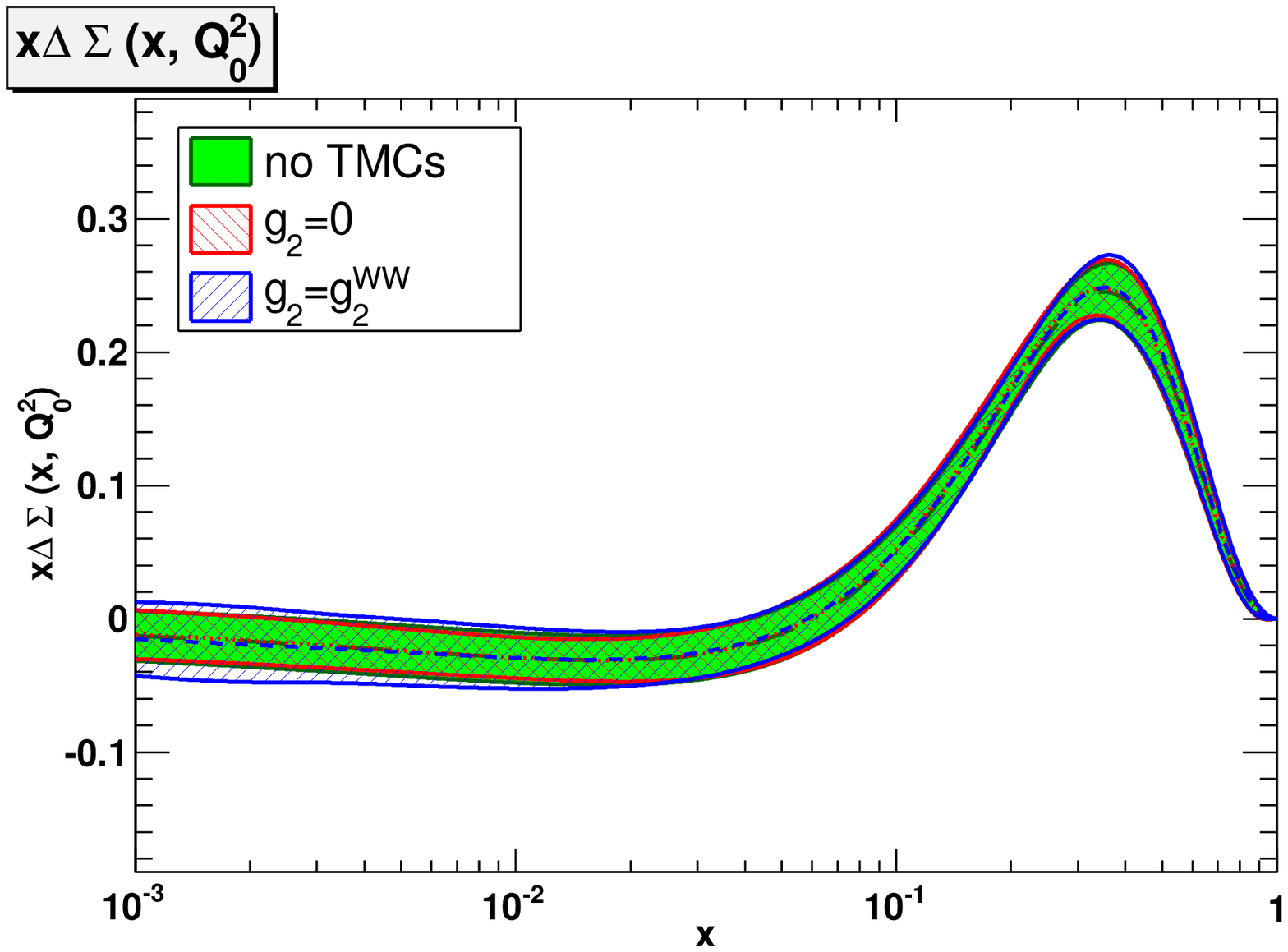}
\epsfig{width=0.43\textwidth,figure=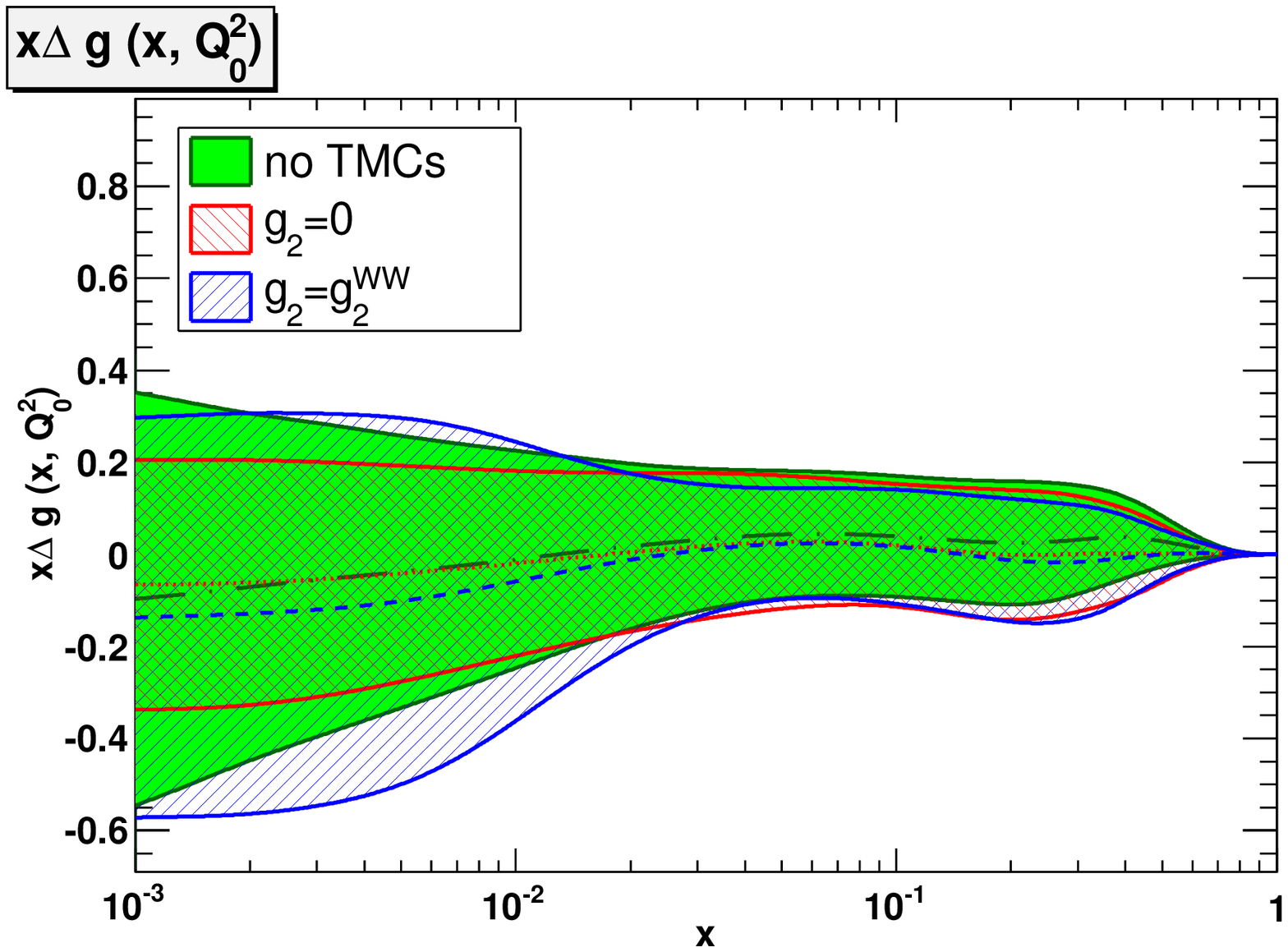}
\epsfig{width=0.43\textwidth,figure=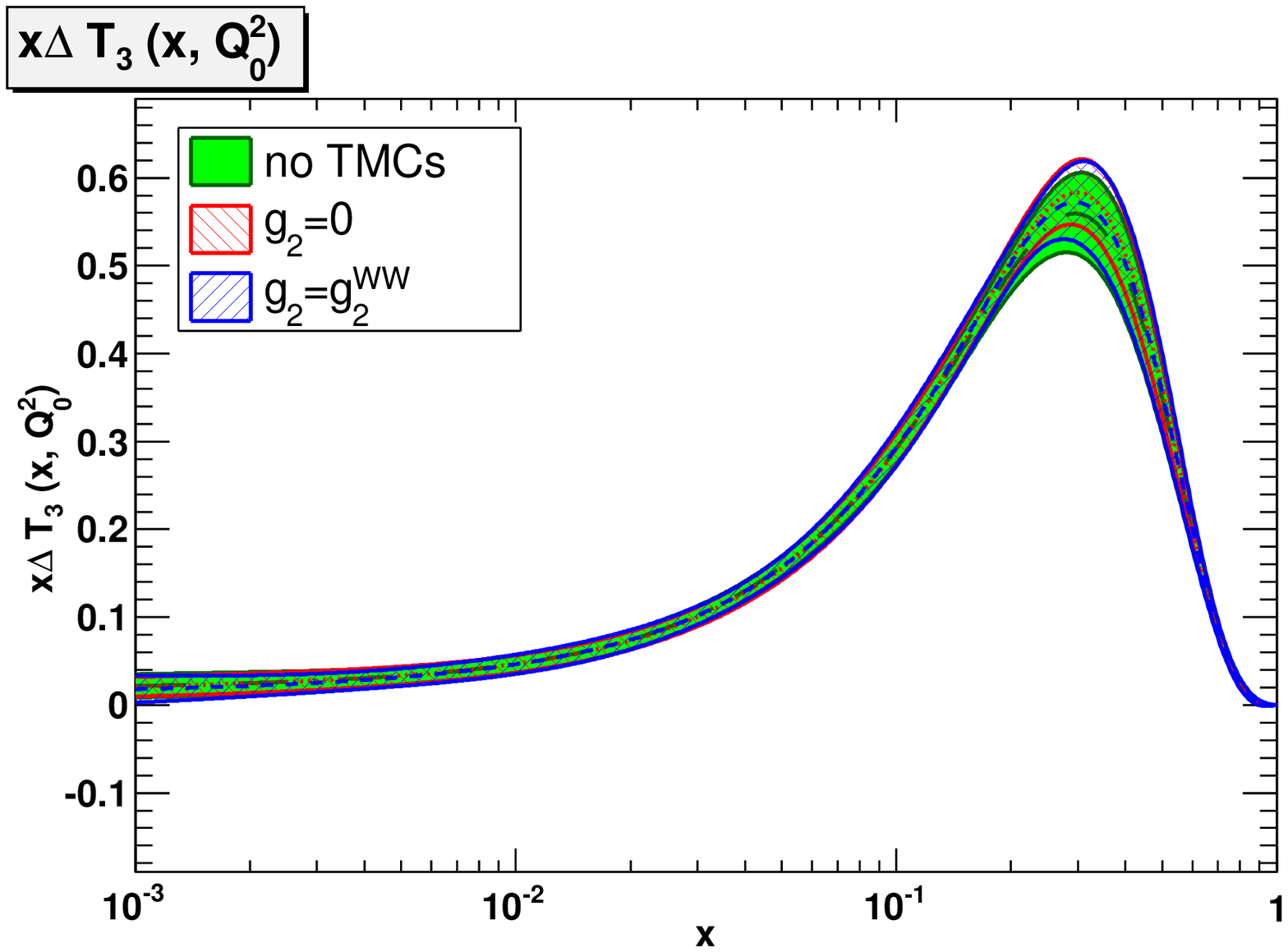}
\epsfig{width=0.43\textwidth,figure=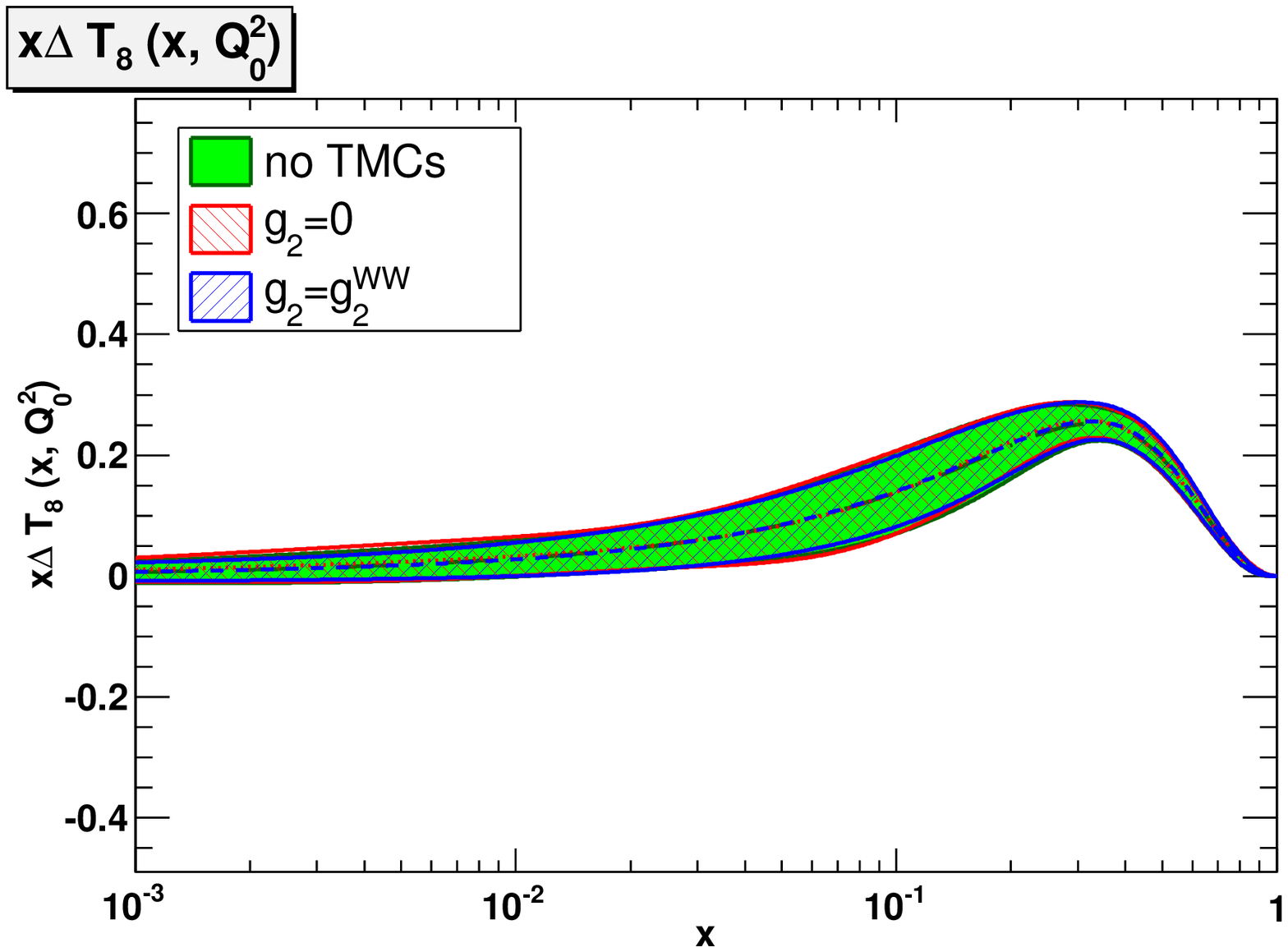}
\caption{\small Comparison between the default \texttt{NNPDFpol1.0}
  PDFs  (labeled as $g_2=g_2^{\mbox{\tiny WW}}$ in the plot), 
PDFs with $m=0$ (labeled as noTMCs in the plot) and PDFs with $g_2=0$;
each corresponds to the statistical estimators of Tab.~\ref{tab:tmc_estimators}.}
\label{fig:TMC_comparison}
\end{center}
\end{figure}

Furthermore,  in
Fig.~\ref{fig:TMC_comparison} we compare the PDFs at
the initial scale $Q_0^2$ determined in these fits to our default set:
differences are hardly visible.
This comparison can be made more quantitative by using the distance
$d(x,Q^2)$ between different fits, as defined in Appendix A of
Ref.~\cite{Ball:2010de}.  The distance  is defined in such a way that if we
compare two different samples of $N_{\rm rep}$ replicas each extracted
from the same distribution, then on average $d=1$, while if the two
samples are extracted from two distributions which differ by one
standard deviation, then on average $d=\sqrt{N_{\rm rep}}$ (the
difference being due to the fact that the standard deviation of the
mean scales as $1/\sqrt{N_{\rm rep}}$). 

\begin{figure}[p]
\begin{center}
\epsfig{width=0.80\textwidth,figure=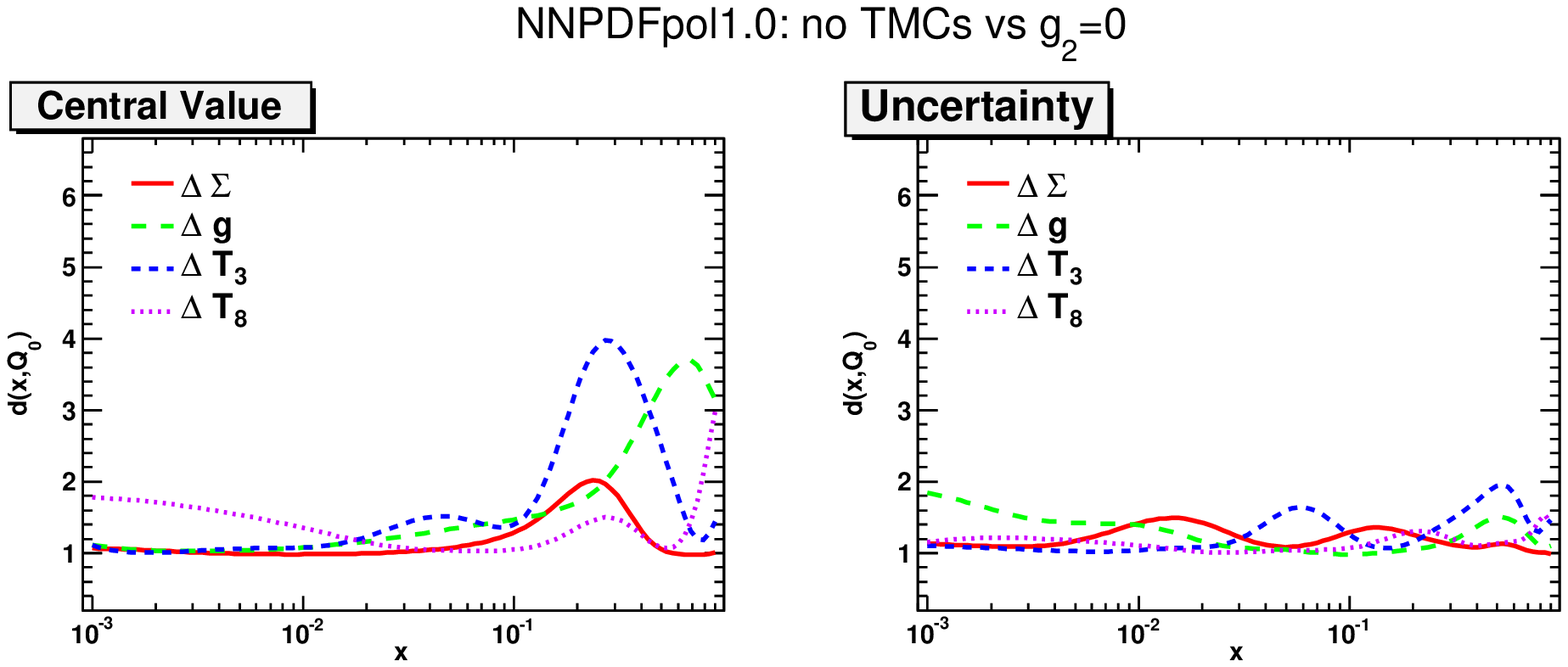}\\
\epsfig{width=0.80\textwidth,figure=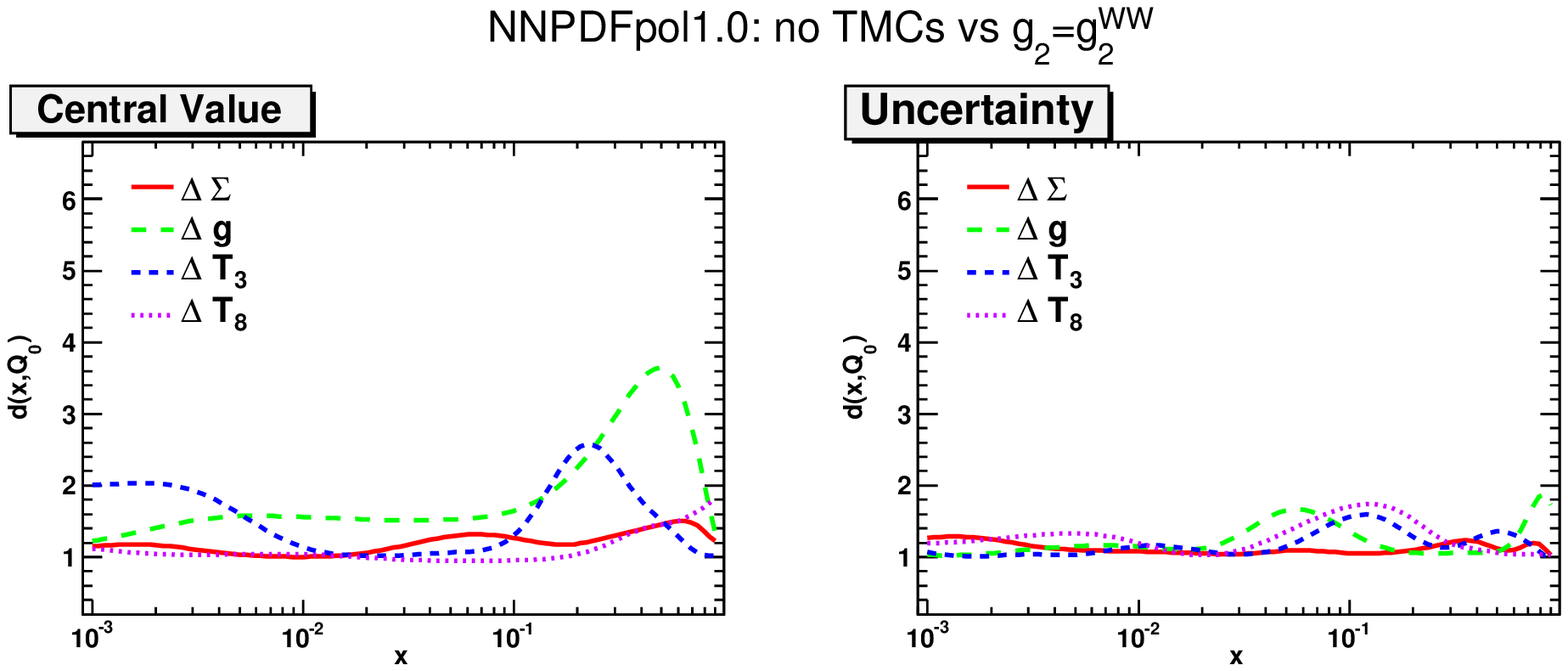}\\
\epsfig{width=0.80\textwidth,figure=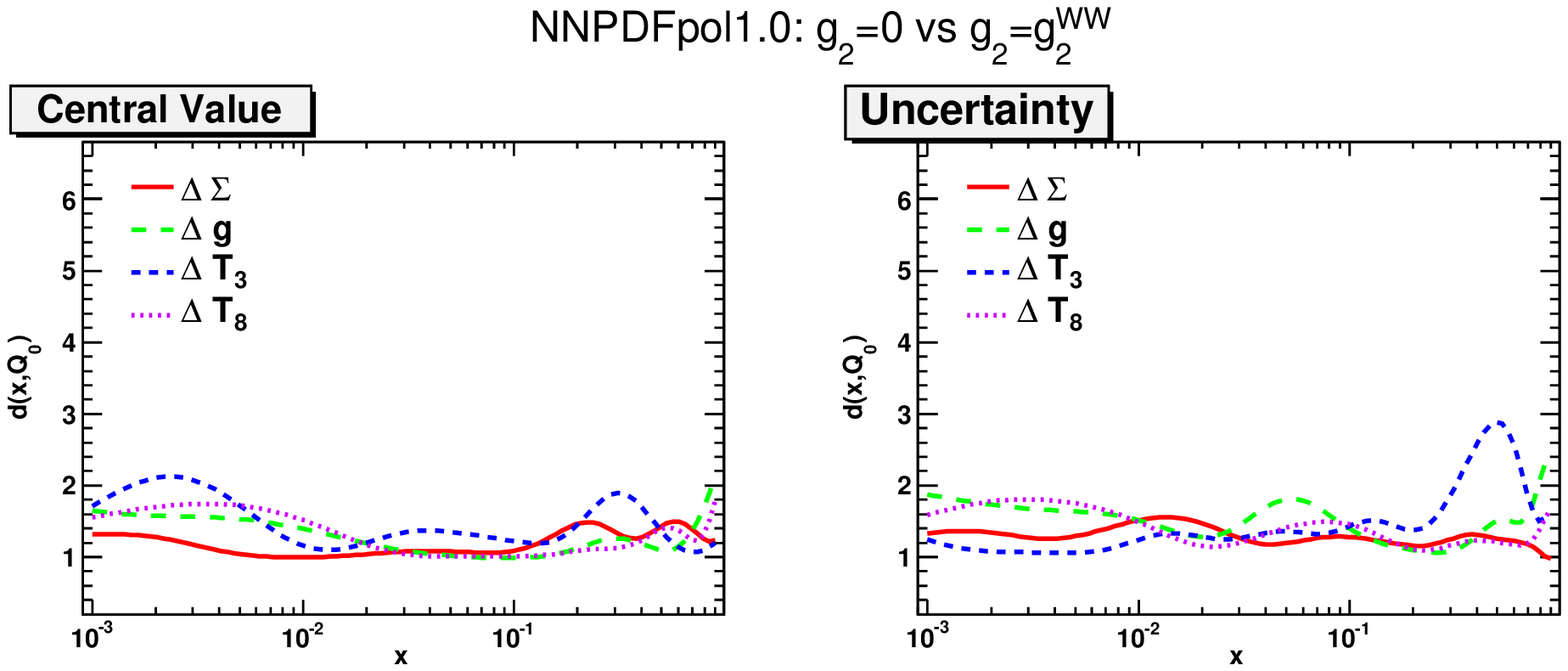}
\caption{\small Distances between each pair of the three sets  of PDFs
shown in Fig.~\ref{fig:TMC_comparison}.}
\label{fig:distances_noTMCs}
\end{center}
\end{figure}

The distances $d(x,Q^2)$
between central values and uncertainties of the three fits
of Tab.~\ref{tab:tmc_estimators} are shown in
Fig.~\ref{fig:distances_noTMCs}. They never exceed $d=4$, which means less
than half a standard deviation for $N_{\rm rep}=100$.
It is interesting to observe that distances tend to be larger in the
large-$x$ region, where the expansion in powers of $m^2/Q^2$ is less
accurate, and the effects of dynamical higher twists can become relevant.
It is reassuring that even in this region the distances are
reasonably small.

We conclude that inclusive DIS data, with our kinematic cuts, do not show 
sensitivity to finite nucleon mass effects, neither in terms of fit quality, nor in 
terms of the effect on PDFs.


\subsubsection{Sum rules}
\label{sec:srres}

Our default PDF fit is obtained by assuming that
the triplet axial charge $a_3$ is fixed to its value extracted from
$\beta$ decay, 
Eq.~(\ref{eq:a3}), and that the octet axial charge $a_8$ is fixed to 
the value of $a_8$ determined from baryon octet
decays, but with an inflated uncertainty in order to allow for SU(3) violation,
Eq.~(\ref{eq:a8p}). As discussed after Eq.~(\ref{eq:sumrules1})
uncertainties on them are included by randomizing their values among replicas.

In order to test the impact of these assumptions, we have produced two
more PDF determinations. In the first, we have not imposed the triplet
sum rule Eq.~(\ref{eq:t3sr}), so in particular $a_3$ is free and
determined by the data, 
instead of being fixed  to the value Eq.~(\ref{eq:a3}). In the
second, we have assumed that the uncertainty on $a_8$ is given by
the much smaller value of Eq.~(\ref{eq:a8}). 

\begin{table}[t]
 \centering
 \small
 \begin{tabular}{|c|c|c|}
  \hline
  Fit & free $a_3$  & $a_8$ Eq.~(\ref{eq:a8}) \\
  \hline
  \hline
  $\chi^{2}_{\tot}$ & 0.79 & 0.77 \\
  $\la E \ra \pm \sigma_{E}$ & 1.84 $\pm$ 0.19 & 1.86 $\pm$ 0.19  \\
  $\la E_{\rm tr} \ra \pm \sigma_{E_{\rm tr}}$ & 1.73 $\pm$ 0.41 & 1.66 $\pm$ 0.53  \\
  $\la E_{\rm val} \ra \pm \sigma_{E_{\rm val}}$ & 1.93 $\pm$ 0.58 & 1.87 $\pm$ 0.71 \\
  \hline
  $\la \chi^{2(k)} \ra \pm \sigma_{\chi^{2}}$ & 0.93 $\pm$ 0.12 & 0.92 $\pm$ 0.15 \\
  \hline
 \end{tabular}
 \caption{\small The statistical estimators of Tab.~\ref{tab:chi2tab1}, but for
   fits in which  the triplet sum rule is not imposed (free $a_3$) or
   in which the octet sum rule is imposed with the smaller uncertainty
   Eq.~(\ref{eq:a8}).}
 \label{tab:sr_estimators}
\end{table}

The statistical estimators for the total dataset  for each of these fits
are shown in Tab.~\ref{tab:sr_estimators}. Here too, there is no
significant difference in fit quality between these fits and the default.

\begin{figure}[t]
\begin{center}
\epsfig{width=0.80\textwidth,figure=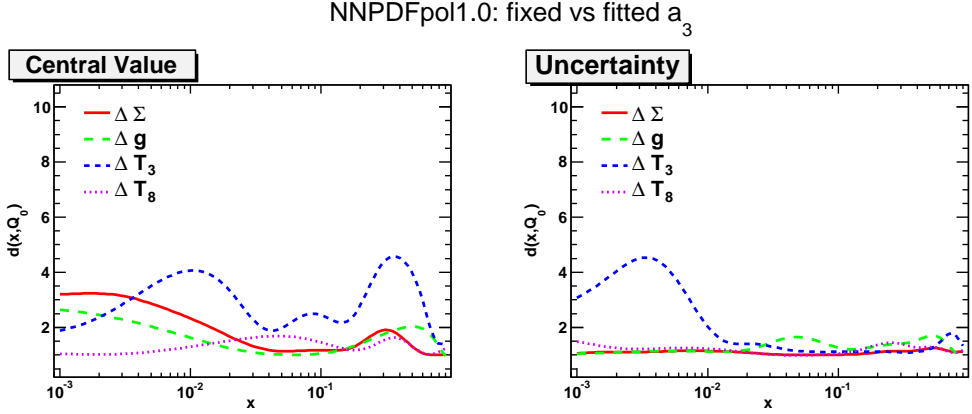}
\caption{\small Distances between PDFs (central values and
  uncertainties) for the
  default fit, with $a_3$ fixed,  and the fit with free $a_3$,
 computed using  $N_{\mbox{\tiny rep}}=100$
  replicas from each set.}
\label{fig:distances_a3}
\end{center}
\end{figure}

\begin{figure}[t]
\begin{center}
\epsfig{width=0.43\textwidth,figure=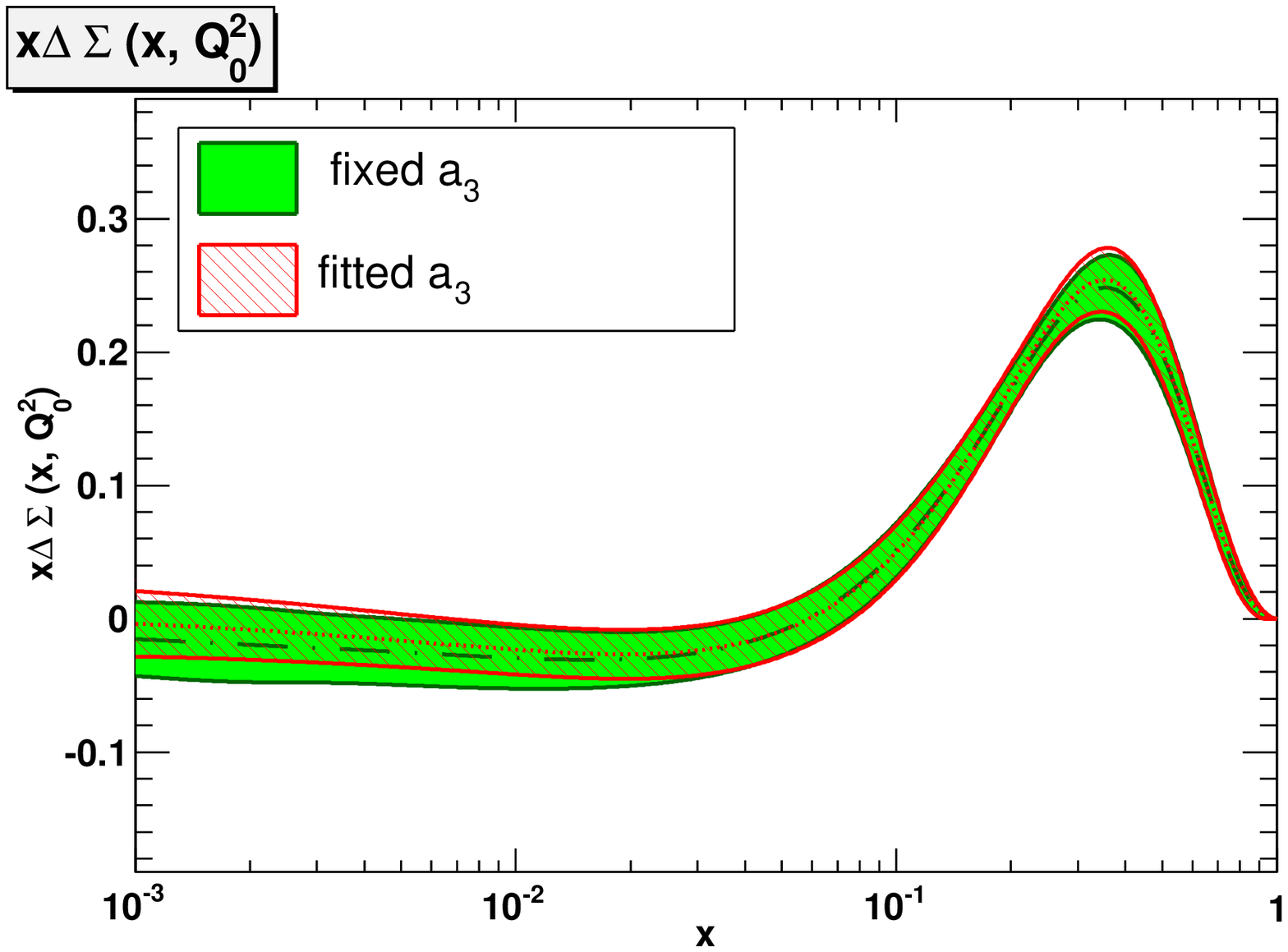}
\epsfig{width=0.43\textwidth,figure=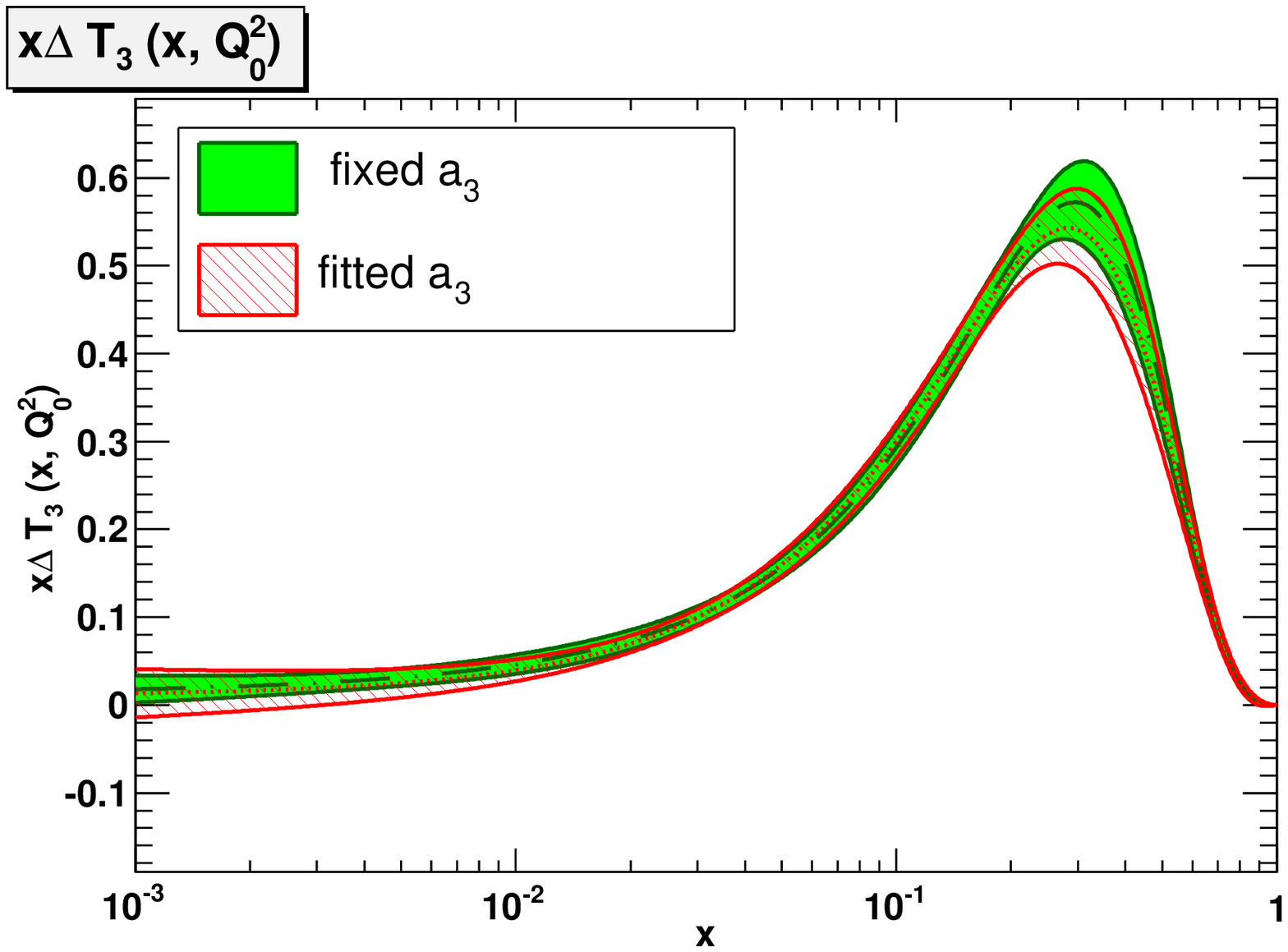}
\caption{\small Comparison of the singlet and triplet PDFs for the
  default fit, with $a_3$ fixed,  and the fit with free $a_3$.}
\label{fig:fit_a3}
\end{center}
\end{figure}

The distances between PDFs in the default and the free $a_3$
fits are displayed in Fig.~\ref{fig:distances_a3}. As one may expect,
only the triplet is affected significantly:
the central value is shifted by about $d \sim
5$, i.e. about half-$\sigma$, in the region $x\sim
0.3$, where $x\Delta T_3$ has a maximum, and also around $x\sim
0.01$. The uncertainties on the PDFs are very similar in both cases for
all PDFs, except $\Delta T_3$ at small-$x$: in this case, removing the
$a_3$ sum rule results in a moderate increase of the uncertainties;
the effect of removing $a_3$ is otherwise negligible.
The singlet and triplet PDFs for these two fits are compared
in Fig.~\ref{fig:fit_a3}.

\begin{figure}[t]
\begin{center}
\epsfig{width=0.80\textwidth,figure=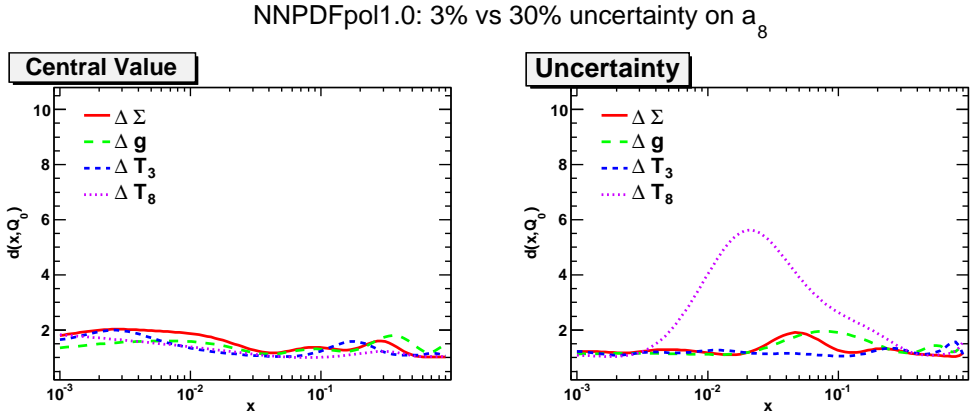}
\caption{\small Distances between PDFs (central values and
  uncertainties) for the default fit, with $a_8$ Eq.~(\ref{eq:a8p}),
  and the fit with the value of $a_8$ with smaller uncertainty,
  Eq.~(\ref{eq:a8}).}
\label{fig:distances_a8}
\end{center}
\end{figure}

\begin{figure}[t]
\begin{center}
\epsfig{width=0.43\textwidth,figure=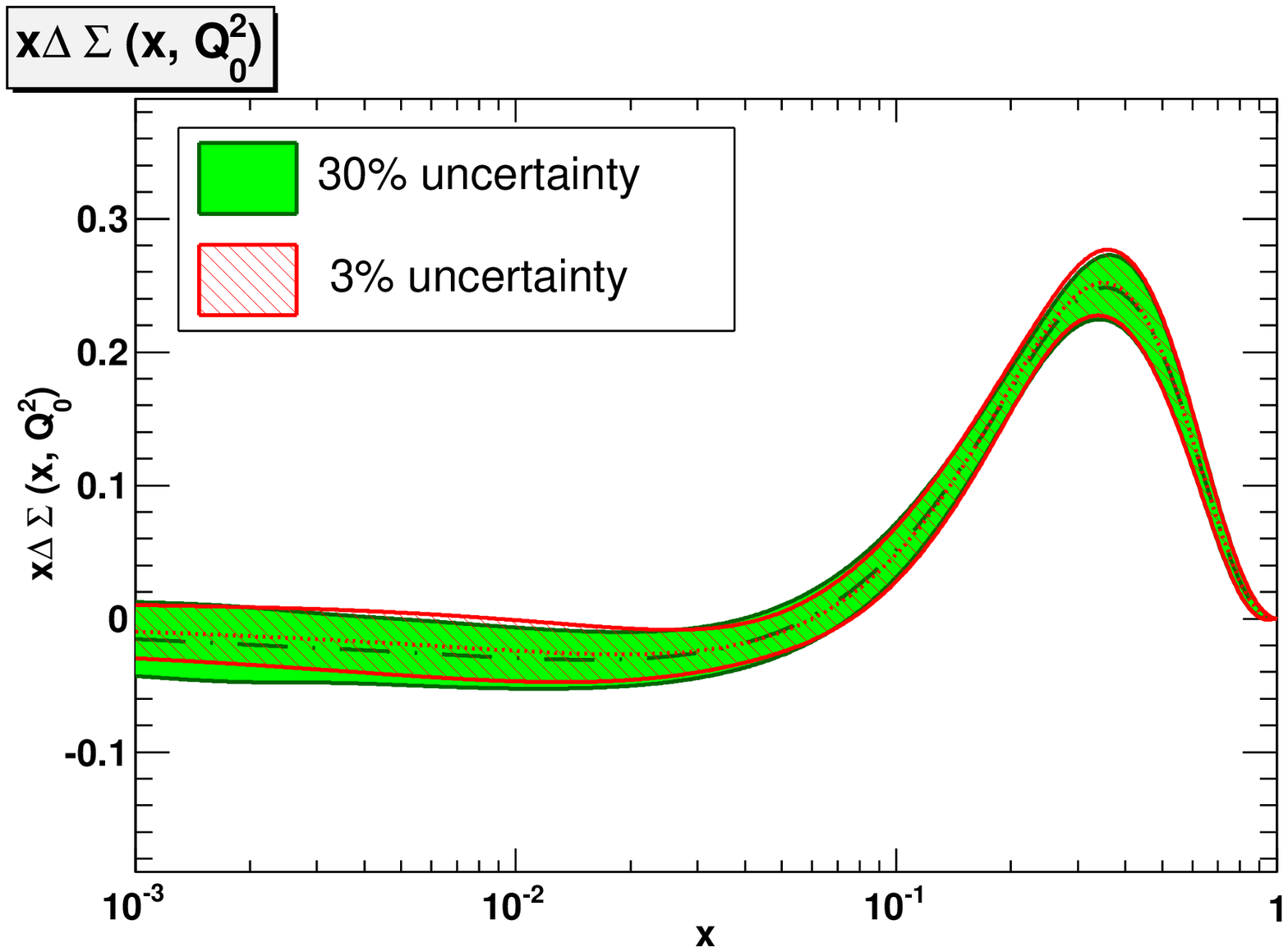}
\epsfig{width=0.43\textwidth,figure=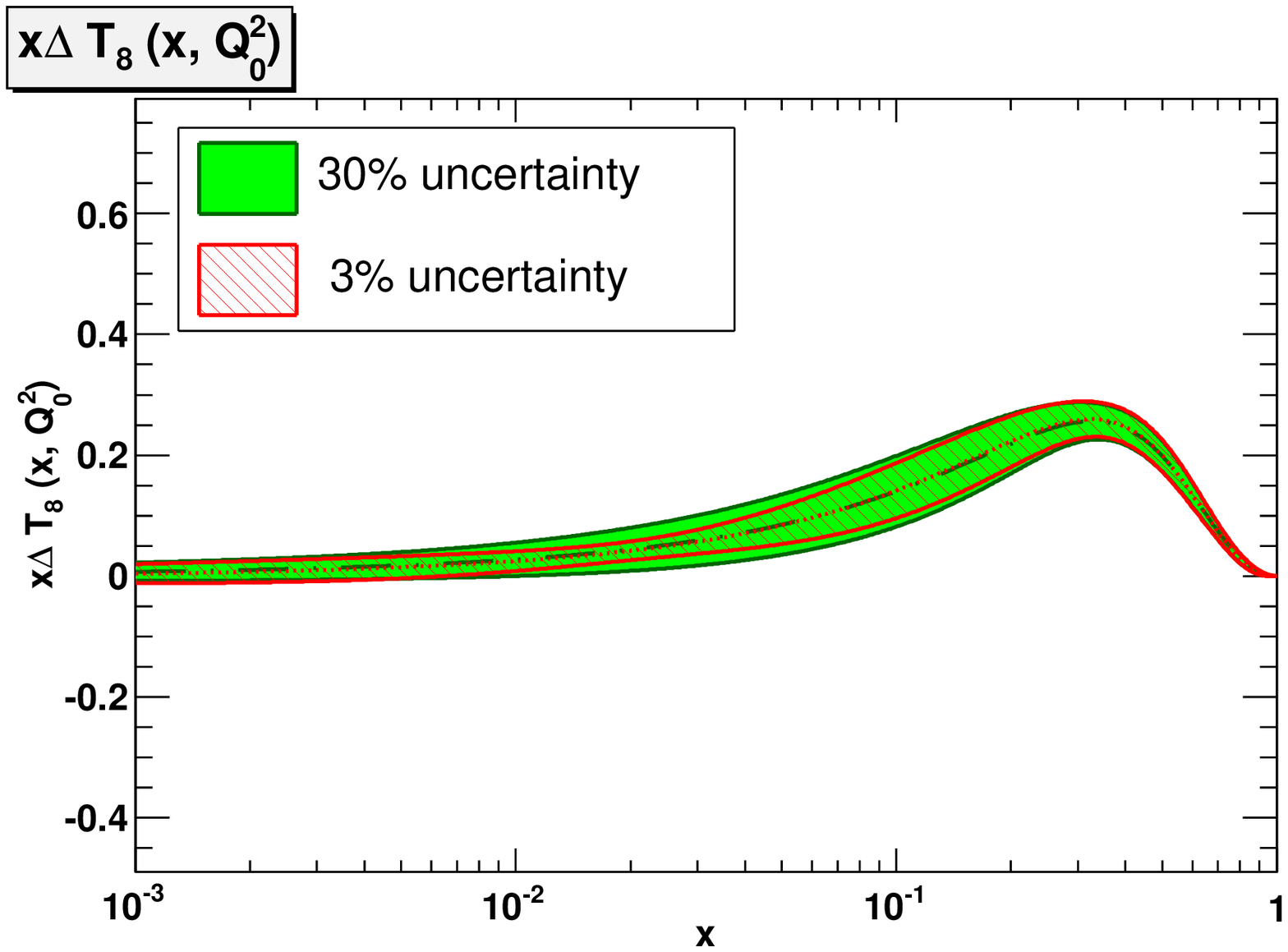}
\caption{\small Comparison of the singlet and octet PDFs for the
  default fit, with $a_8$ Eq.~(\ref{eq:a8p}), and the fit with the
  value of $a_8$ with smaller uncertainty, Eq.~(\ref{eq:a8}).}
\label{fig:fit_a8}
\end{center}
\end{figure}

The
distances between the default and the fit with the smaller uncertainty
on $a_8$, Eq.~(\ref{eq:a8}),
are shown in Fig.~\ref{fig:distances_a8}. In this case, again as
expected, the only effect is on the $\Delta T_8$ uncertainty, which changes
in the region $10^{-2}\lesssim x \lesssim 10^{-1}$
by up to $d\sim 6$ (about half a standard deviation): if a more
accurate value of $a_8$ is assumed, the determined $\Delta T_8$
is correspondingly more accurate. Central values are unaffected.
The singlet and octet PDFs for this fit are compared to the default 
in Fig.~\ref{fig:fit_a8}. We conclude that the size of the uncertainty
on  $\Delta T_8$ has a moderate effect on our fit; on the other hand
it is clear that if the octet sum rule were not imposed at all, the
uncertainty on the octet and thus on strangeness would increase very
significantly, as we have checked explicitly.


We conclude that our fit results are quite stable upon variations of our treatment
of both the triplet and the octet sum rules.

\subsection{Positivity}
\label{sec:thconstraints} 

As discussed in Sect.~\ref{sec:minim}, positivity of the individual
cross sections entering the polarized asymmetries
Eq.~(\ref{eq:xsecasy}) has been imposed at leading order according to
Eq.~(\ref{eq:possigma}), using the \texttt{NNPDF2.1 NNLO} PDF set~\cite{Ball:2011mu}, separately for the lightest polarized quark
PDF combinations $\Delta u + \Delta\bar{u}$, $\Delta d +
\Delta\bar{d}$, $\Delta s + \Delta\bar{s}$ and for the polarized gluon
PDF, by means of a Lagrange multiplier Eq.~(\ref{eq:lagrmult}). After
stopping, positivity is checked a posteriori and replicas which do not
satisfy it are discarded and retrained. 

\begin{figure}[t]
\begin{center}
\epsfig{width=0.43\textwidth,figure=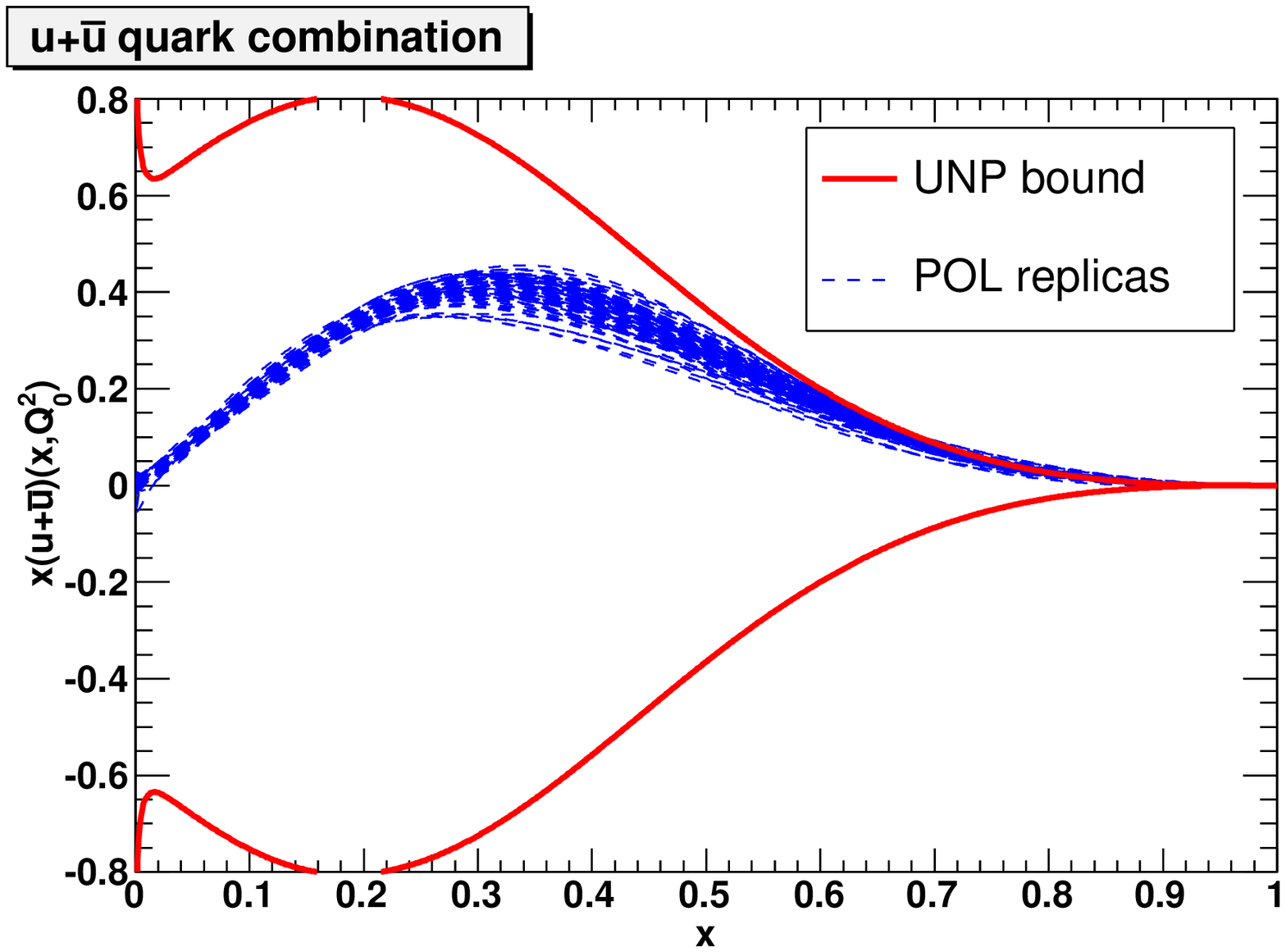}
\epsfig{width=0.43\textwidth,figure=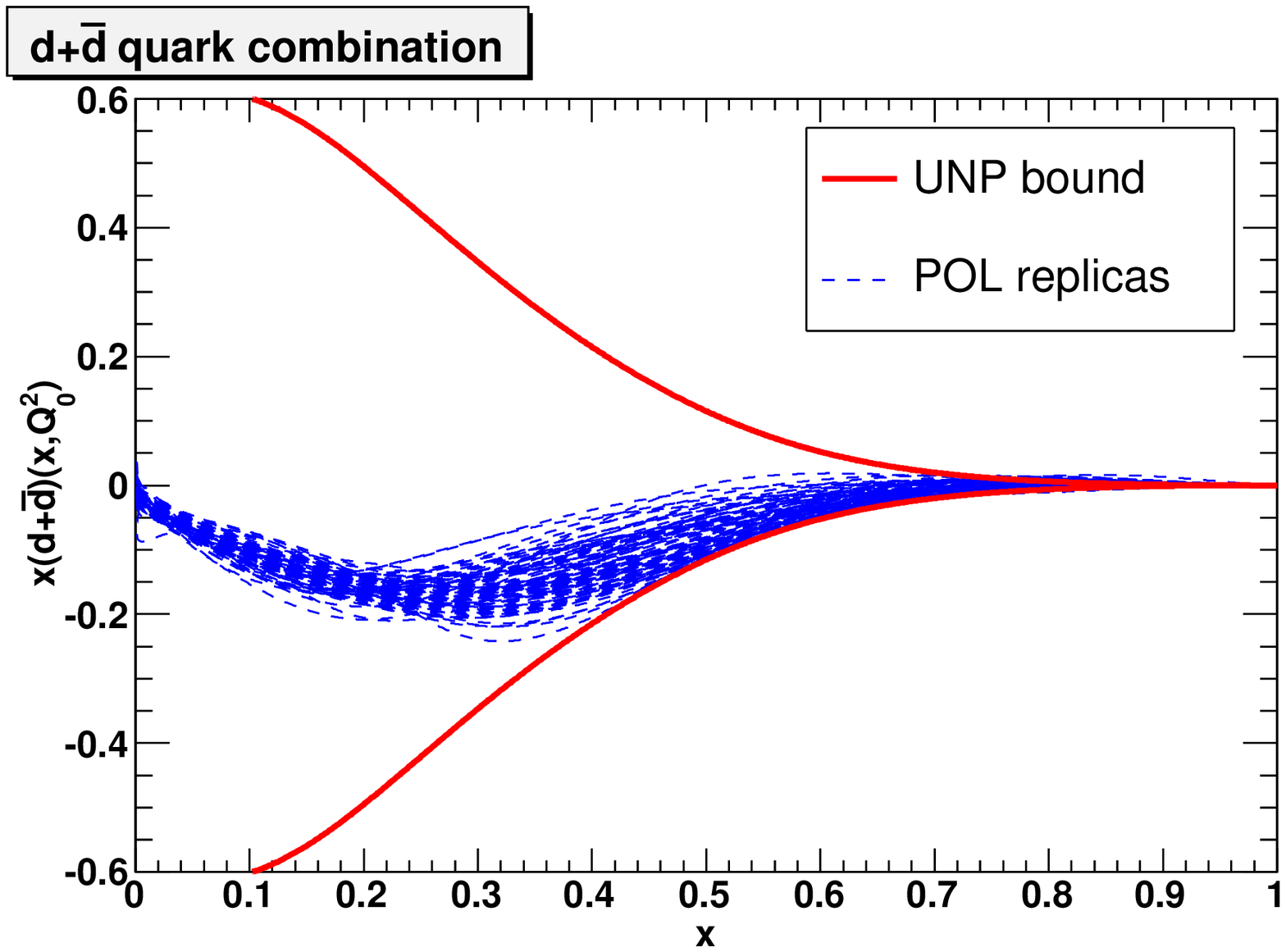}
\epsfig{width=0.43\textwidth,figure=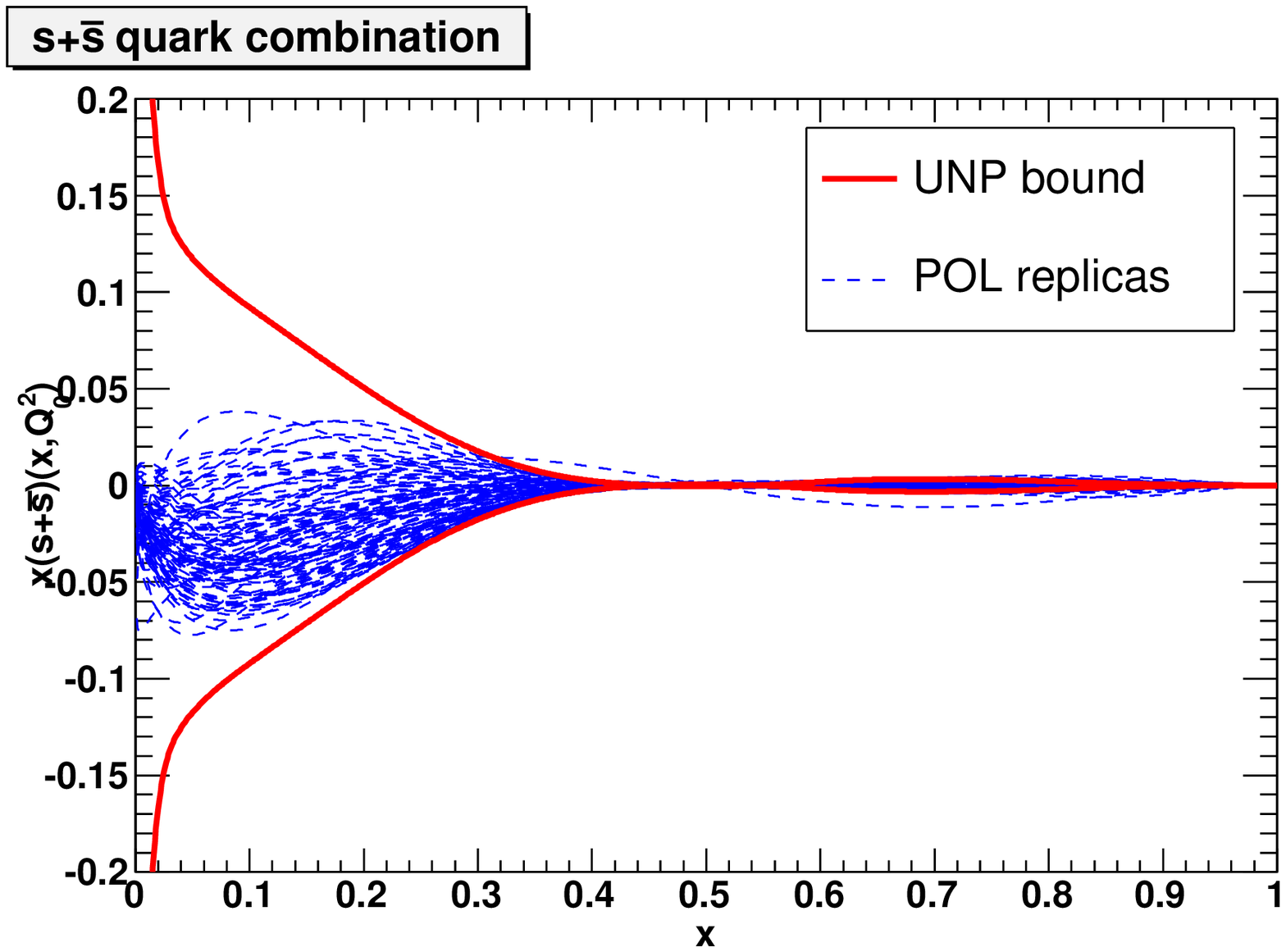}
\epsfig{width=0.43\textwidth,figure=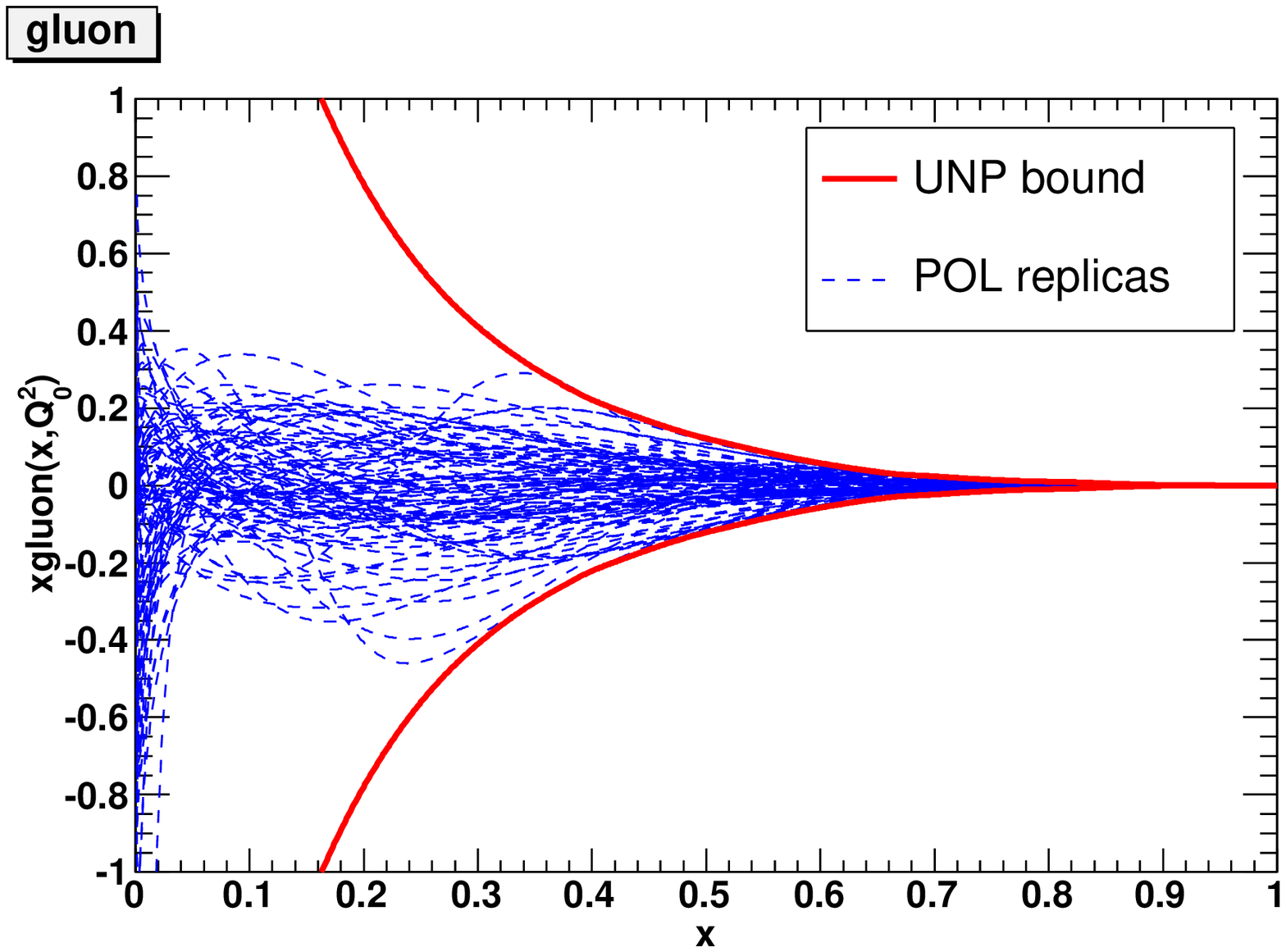}
\caption{\small The positivity bound Eq.~(\ref{eq:possigma}), compared
  to a set of $N_{\mbox{\tiny rep}}=100$  replicas  (dashed
  lines).}
\label{fig:pdfposconstr}
\end{center}
\end{figure}

In Fig.~\ref{fig:pdfposconstr} we compare  to the positivity bound for
the up, down, strange PDF combinations and
gluon PDF a set of $N_{\mbox{\tiny rep}}=100$  replicas obtained  by enforcing positivity through a
Lagrange multiplier, but before the final, \textit{a posteriori} check. 
Almost all replicas satisfy the constraint, but at least one replica
which clearly violates it for 
the
$s+\bar{s}$ combination (and thus will be discarded) is seen.

In order to assess the effect of the positivity
constraints  we have performed a
fit without imposing positivity. Because positivity significantly
affects PDFs in the region where no data are available, and 
thus in particular their large $x$ behaviour, preprocessing exponents for this
PDF determination had to be determined again using the procedure
described in Sect.~\ref{sec:net-param}. The values of the large $x$
preprocessing exponents used in the fit without positivity are shown
in Tab.~\ref{tab:prepexpsnopos}. The small $x$
 exponents are the same as in the
      baseline fit, Tab.~\ref{tab:prepexps}.
\begin{table}[h!]
  \begin{center}
 \begin{tabular}{c|c}
 \hline PDF & $m$ \\
 \hline
 $\Delta\Sigma(x, Q_0^2)$  & $\lc 0.5, 5.0 \rc$ \\
 \hline
 $\Delta g(x, Q_0^2)$          & $\lc 0.5, 5.0 \rc$\\
 \hline
 $\Delta T_3(x, Q_0^2)$      & $\lc 0.5, 4.0 \rc$\\
 \hline
 $\Delta T_8(x, Q_0^2)$      & $\lc 0.5, 6.0 \rc$ \\
  \hline
  \end{tabular}
    \caption{\small \label{tab:prepexpsnopos} Ranges for the
      large $x$
 preprocessing exponents Eq.~(\ref{eq:PDFbasisnets})
      for the fit in which no positivity
      is imposed. The small $x$ exponents are the same as in the
      baseline fit Tab.~\ref{tab:prepexps}.}
  \end{center}
\end{table}

\begin{figure}[p]
\begin{center}
\epsfig{width=0.43\textwidth,figure=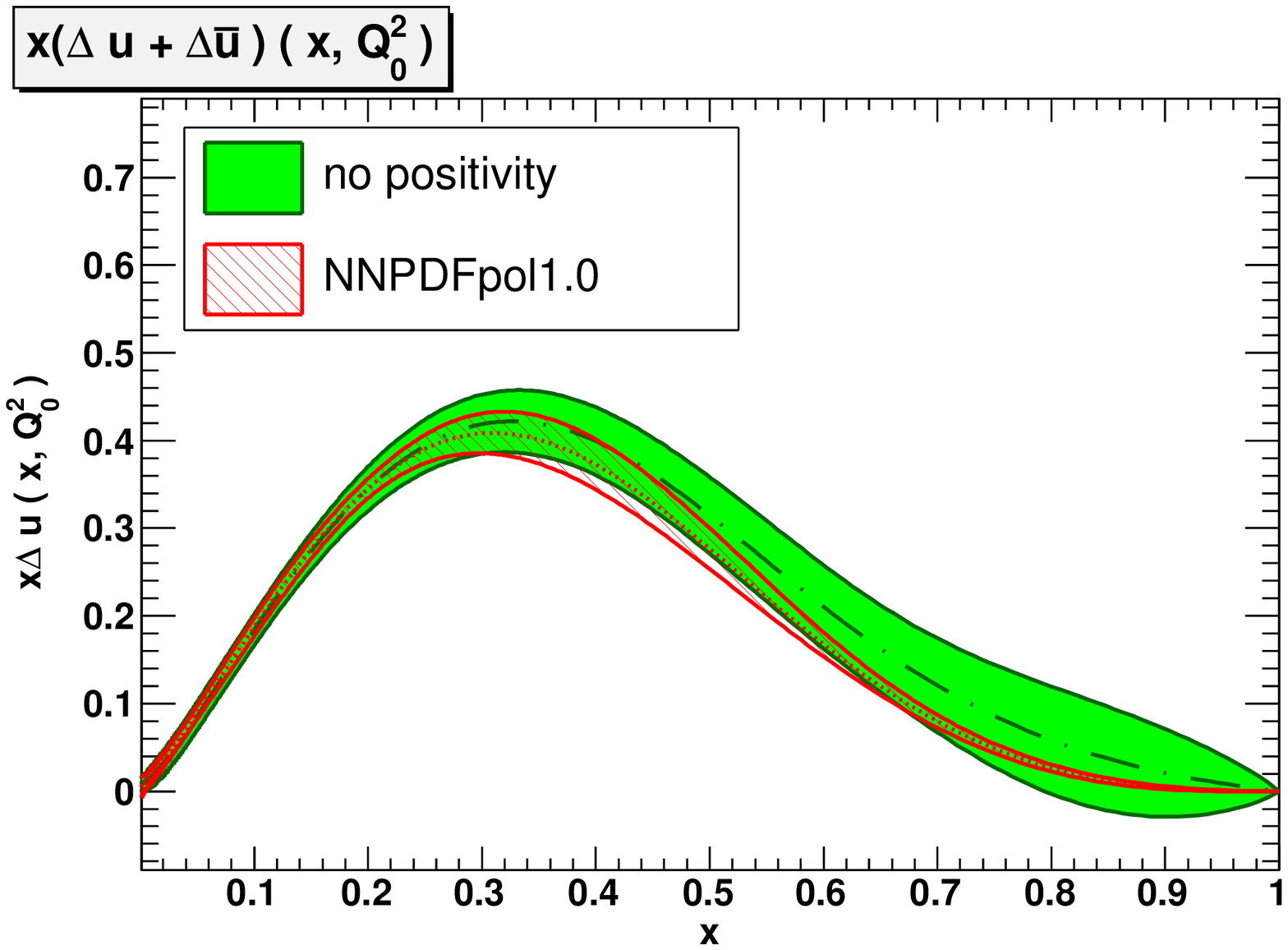}
\epsfig{width=0.43\textwidth,figure=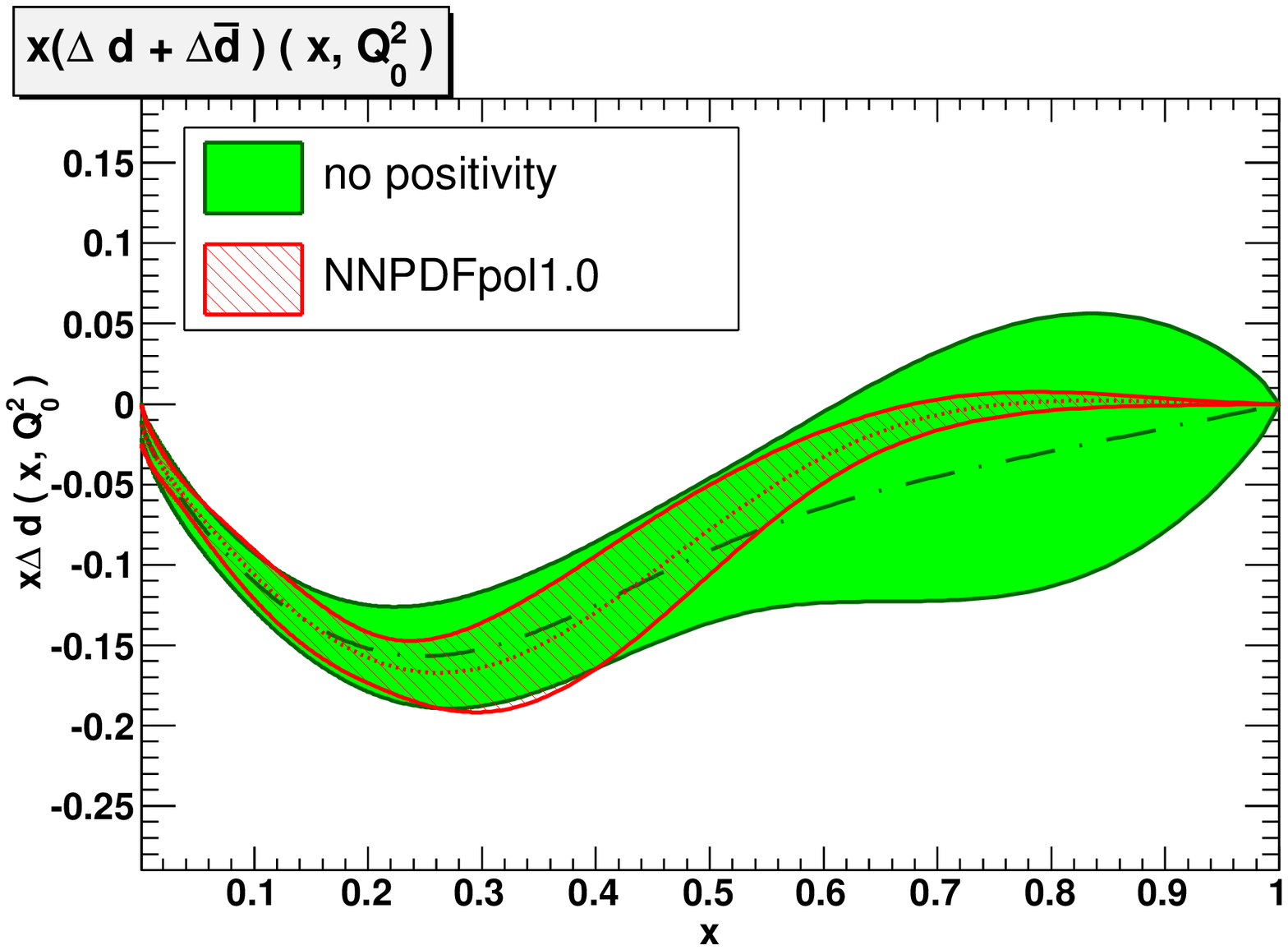}
\epsfig{width=0.43\textwidth,figure=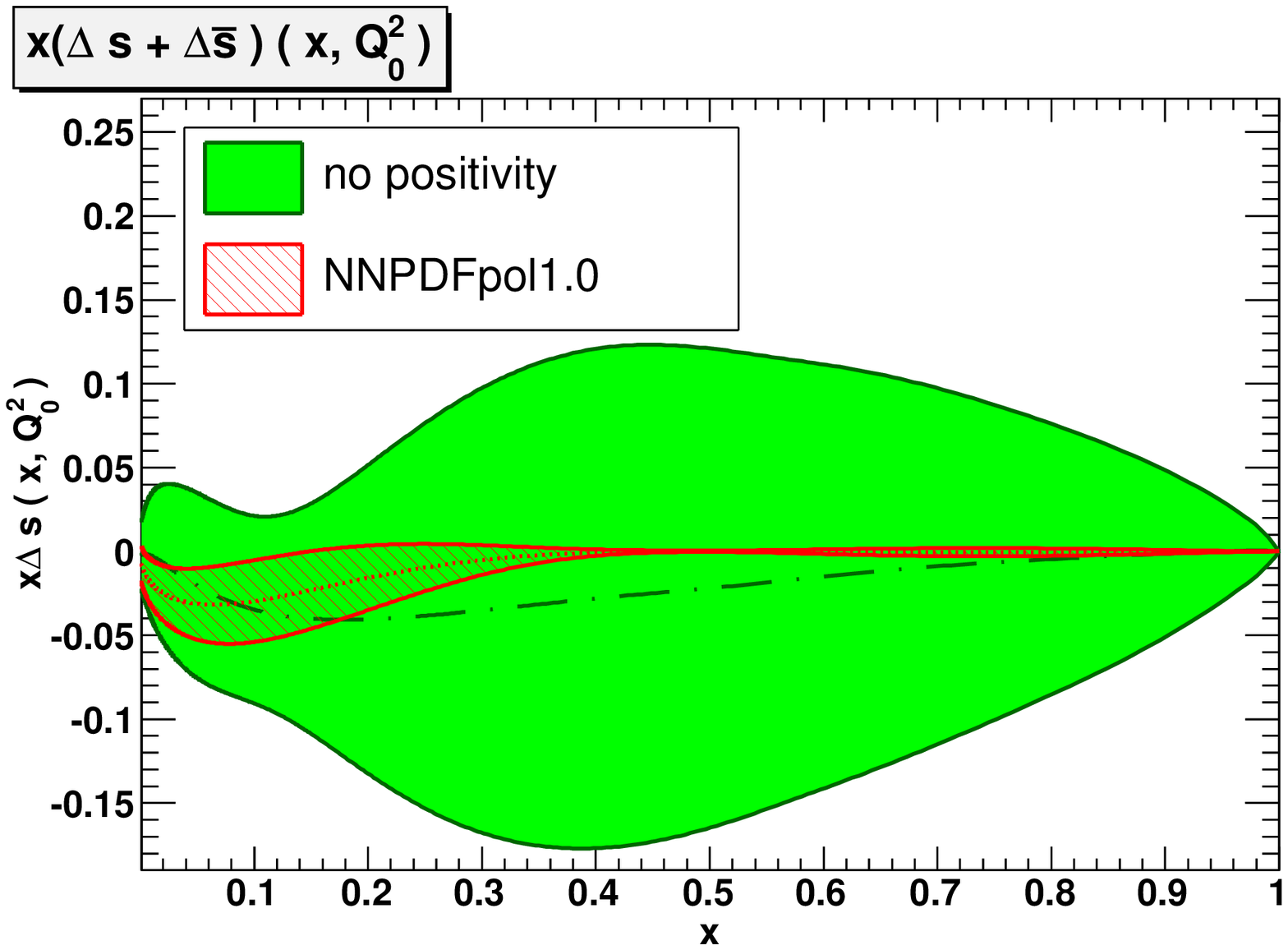}
\epsfig{width=0.43\textwidth,figure=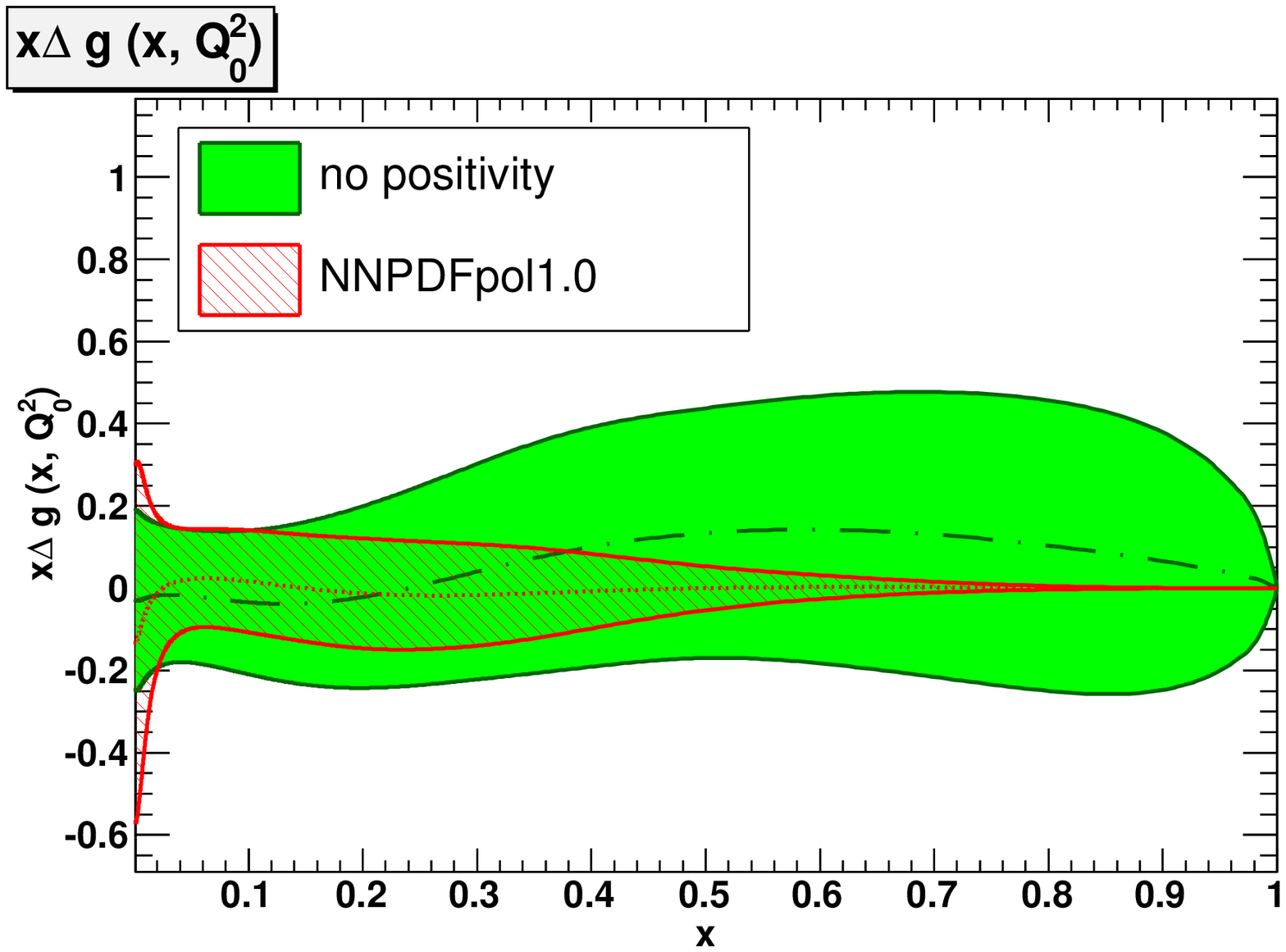}
\caption{\small The \texttt{NNPDFpol1.0} PDFs with and without
  positivity constraints compared at the initial parametrization scale
  $Q_0^2=1$ GeV$^2$ in the flavor basis. \label{fig:pdfposbench}}
\end{center}
\end{figure}

The corresponding estimators are
shown in Tab.~\ref{tab:pos_estimators}. Also in this case, we see no
significant change in fit quality, with only a slight improvement in 
$\chi^2_{\rm tot}$ when the constraint is removed. This shows that
our PDF parametrization is flexible enough to easily accommodate positivity.
On the other hand, clearly the positivity bound has a significant
impact on PDFs, especially in the large $x$ region, as shown in 
Fig.~\ref{fig:pdfposbench}, where PDFs obtained from this fit are
compared to the baseline.
At small $x$, instead, the impact of positivity is moderate, because,
as discussed in Sect.~\ref{sec:thconst}, $g_1/F_1\sim x$ as
$x\to0$~\cite{Ball:1995ye} so there is no constraint in the
limit. This in particular implies that there is no significant loss of
accuracy in imposing the LO positivity bound, because 
in the small $x\lesssim10^{-2}$  region, where the LO and NLO 
positivity bounds differ
significantly~\cite{Forte:1998kd} the bound is not significant.
\begin{table}[t]
 \centering
 \small
 \begin{tabular}{|c|c|}
  \hline
  Fit & {\tt NNPDFpol1.0} no positivity  \\
  \hline
  \hline
  $\chi^{2}_{\tot}$ &  0.72\\
  $\la E \ra \pm \sigma_{E}$ &  1.84 $\pm$ 0.22\\
  $\la E_{\rm tr} \ra \pm \sigma_{E_{\rm tr}}$ &  1.60 $\pm$ 0.20\\
  $\la E_{\rm val} \ra \pm \sigma_{E_{\rm val}}$ &  2.07 $\pm$ 0.39\\
  \hline
  $\la \chi^{2(k)} \ra \pm \sigma_{\chi^{2}}$ &  0.95 $\pm$ 0.16 \\
  \hline
\end{tabular}
\caption{\small The statistical estimators of Tab.~\ref{tab:chi2tab1}
     for a fit without positivity constraints.}
 \label{tab:pos_estimators}
\end{table}

\subsection{Small- and large-$x$ behaviour and preprocessing}
\label{sec:prepexp}

The asymptotic behavior of both polarized and unpolarized
PDFs for $x$ close to 0 or 1 is not controlled by perturbation theory,
because powers of $\ln\frac{1}{x}$ and $\ln(1-x)$
respectively appear in the perturbative coefficients, thereby spoiling
the reliability of the perturbative expansion close to the endpoints. 
Non-perturbative effects are also expected to set in eventually (see
e.g.~\cite{Roberts:1990ww,Ball:1995ye}). For this reason, 
our fitting procedure makes no assumptions on the large- and small-$x$
behaviors of PDFs, apart from the positivity and integrability constraints
discussed in the previous Section.

\begin{figure}[p]
\begin{center}
\epsfig{width=0.41\textwidth,figure=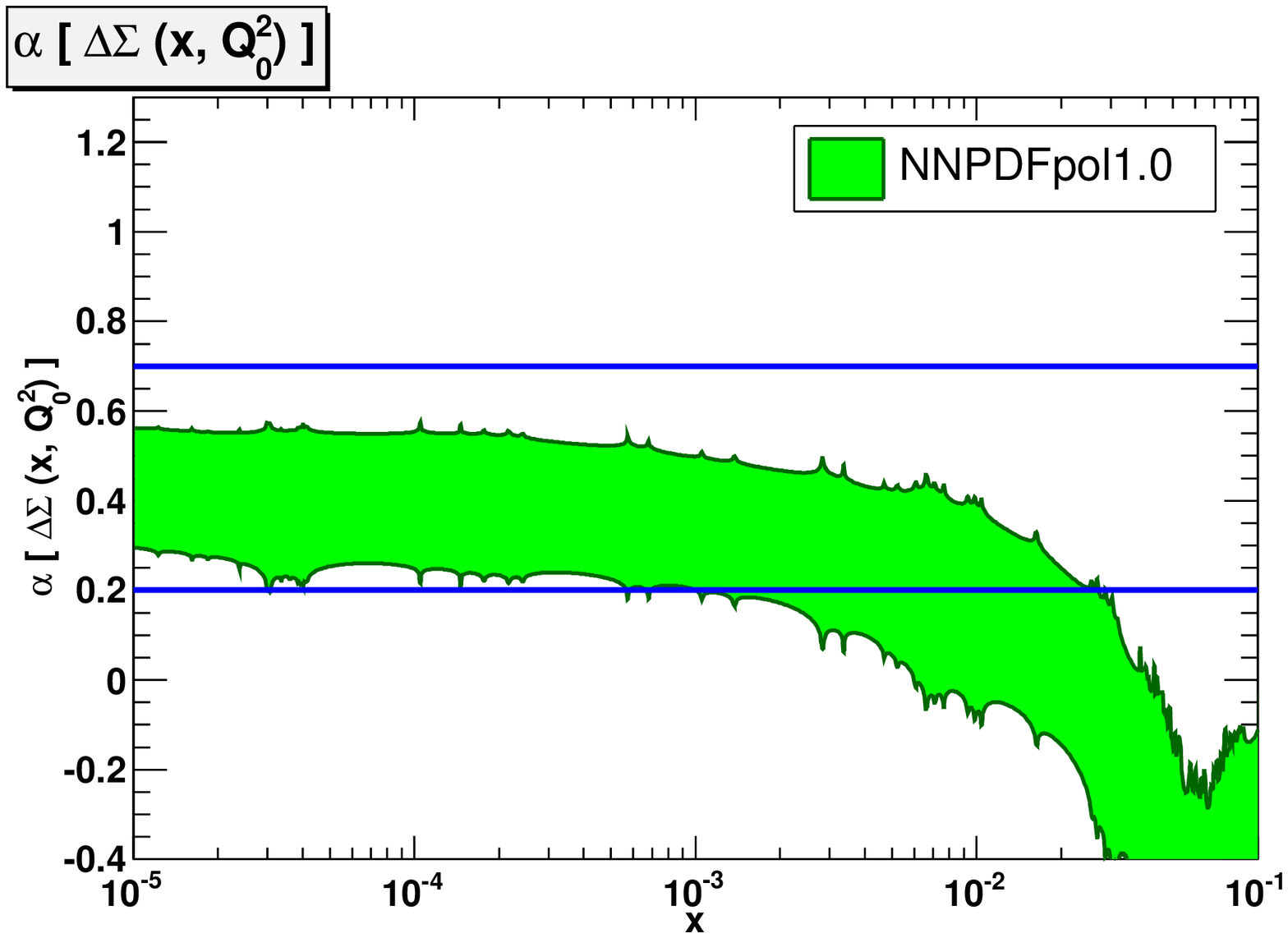}
\epsfig{width=0.41\textwidth,figure=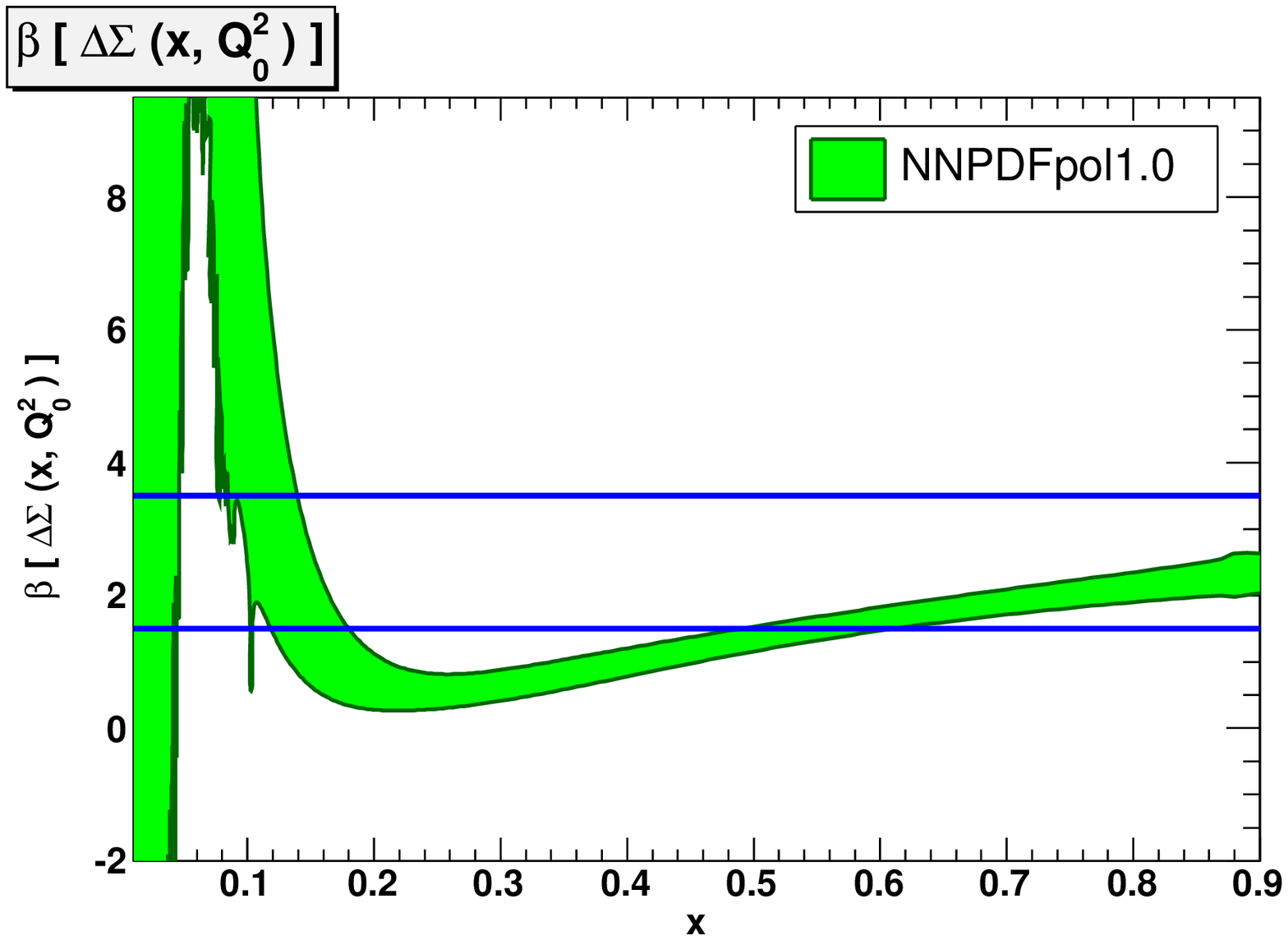}
\epsfig{width=0.41\textwidth,figure=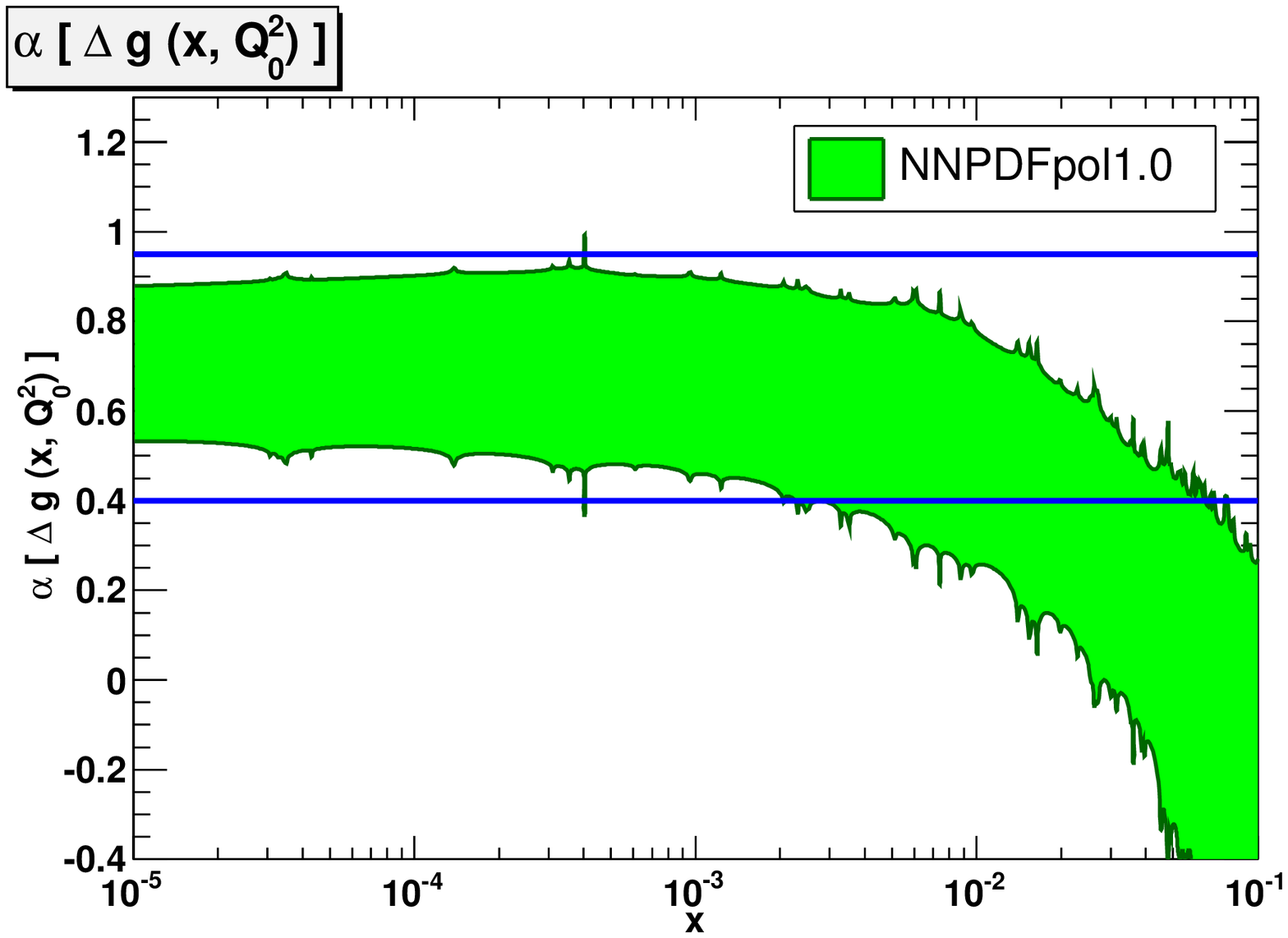}
\epsfig{width=0.41\textwidth,figure=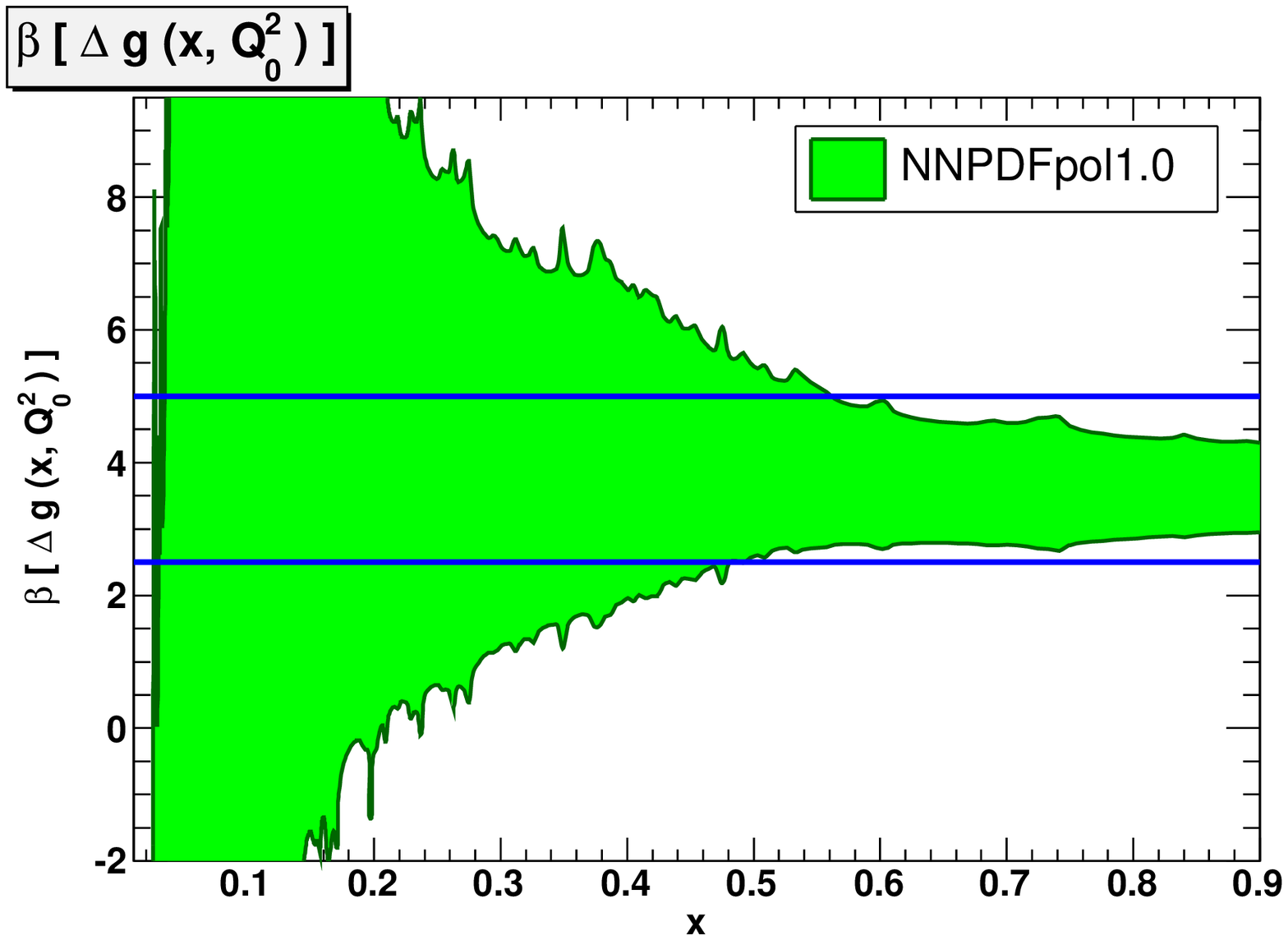}
\epsfig{width=0.41\textwidth,figure=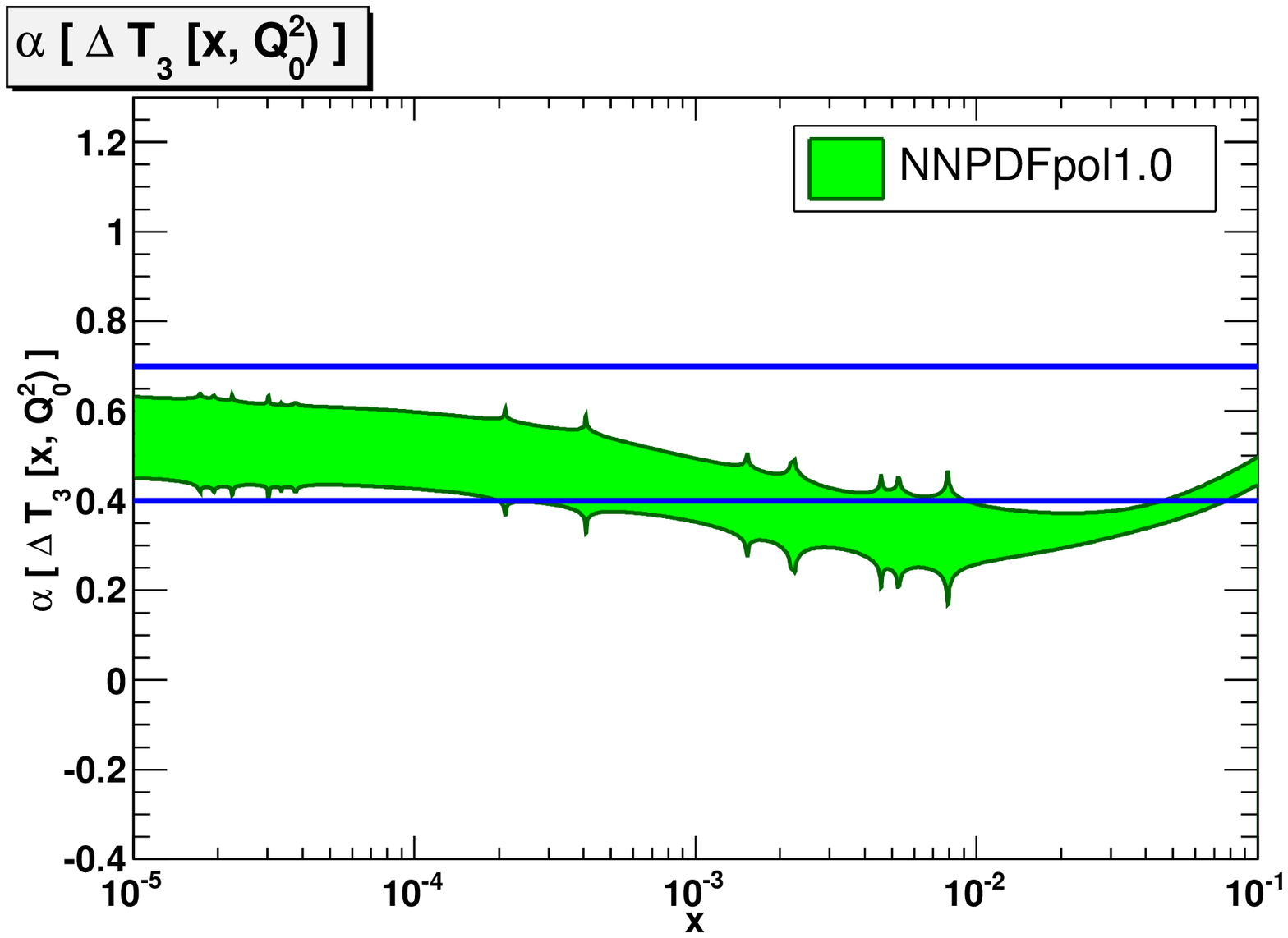}
\epsfig{width=0.41\textwidth,figure=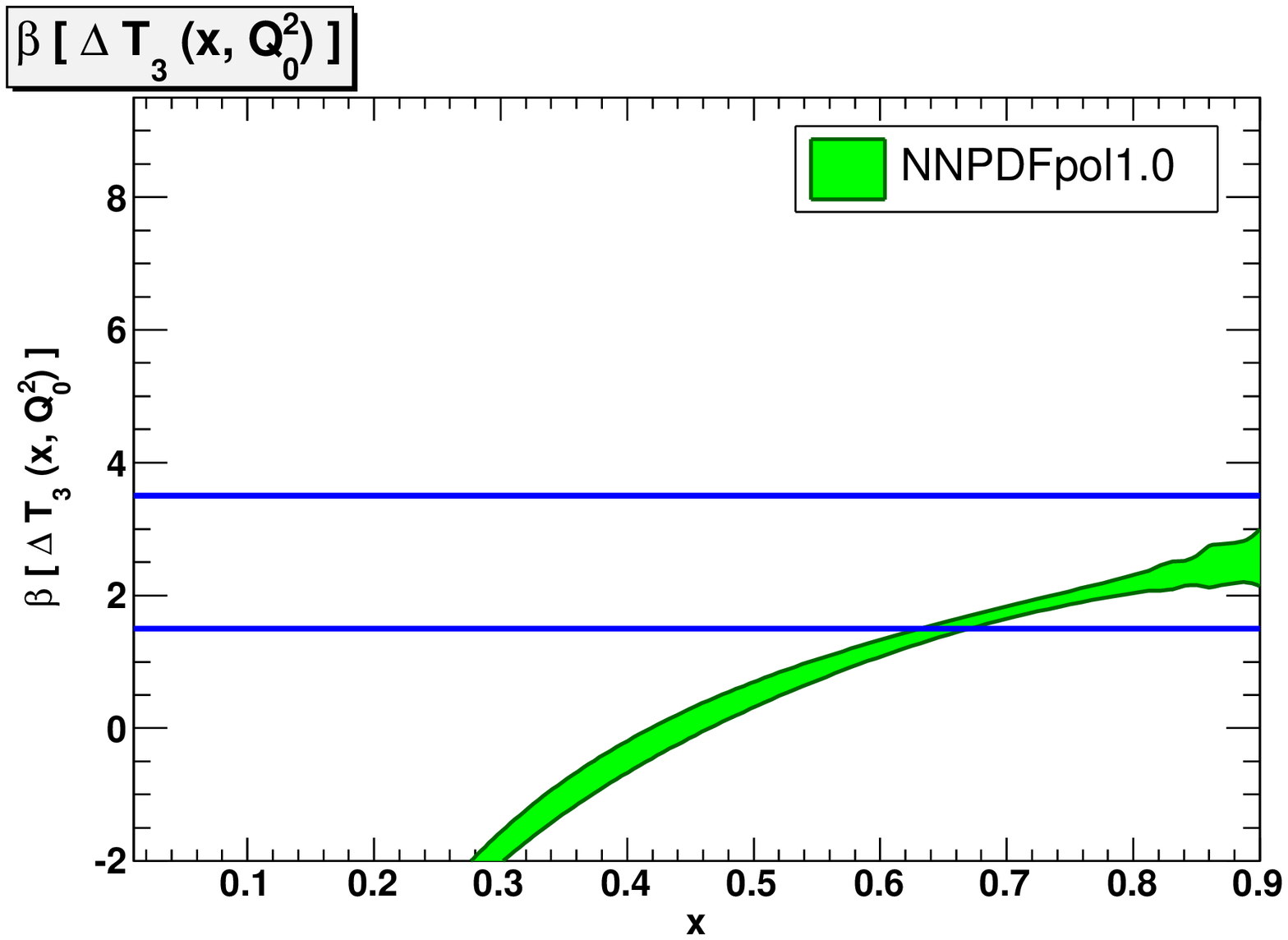}
\epsfig{width=0.41\textwidth,figure=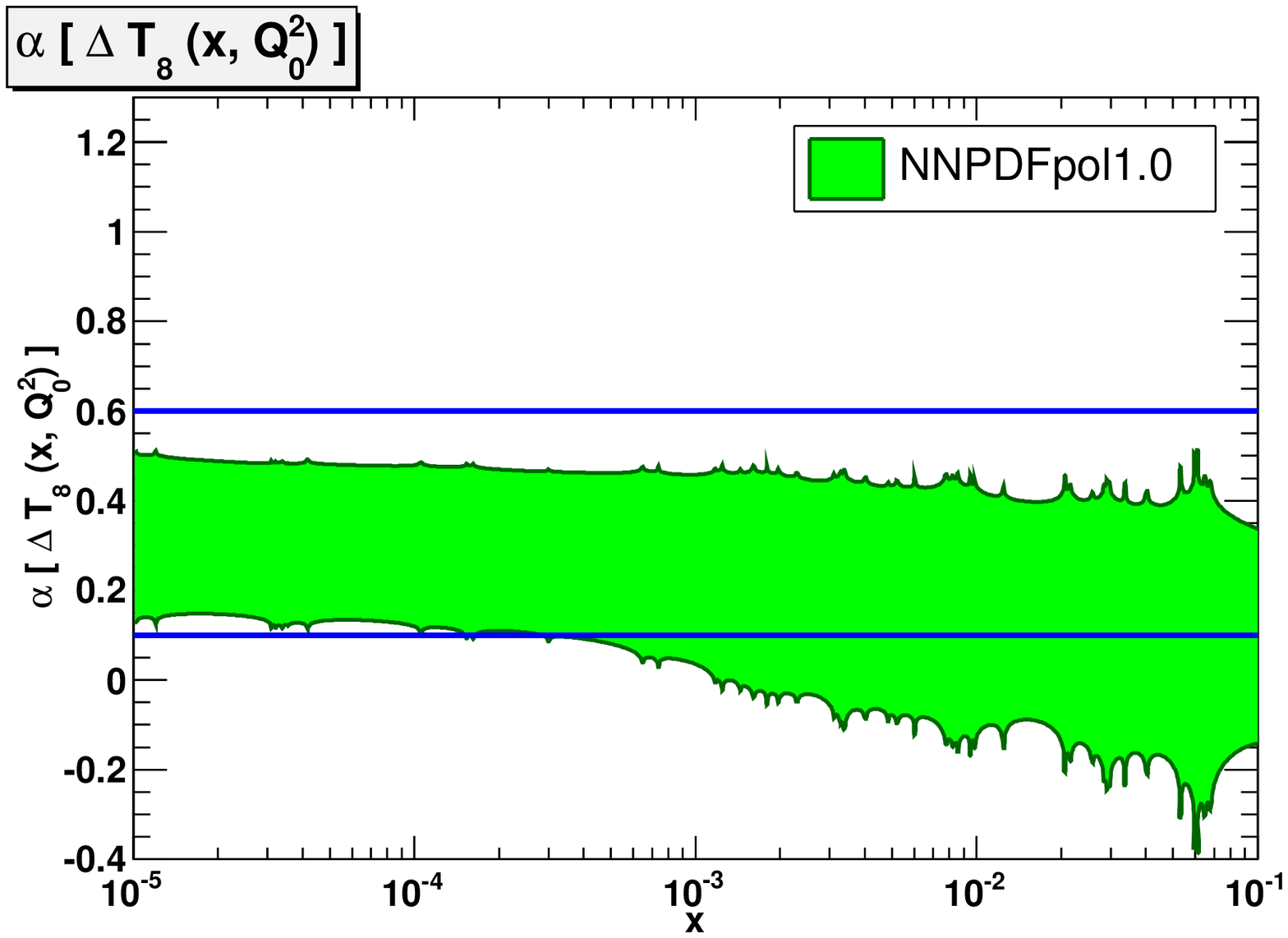}
\epsfig{width=0.41\textwidth,figure=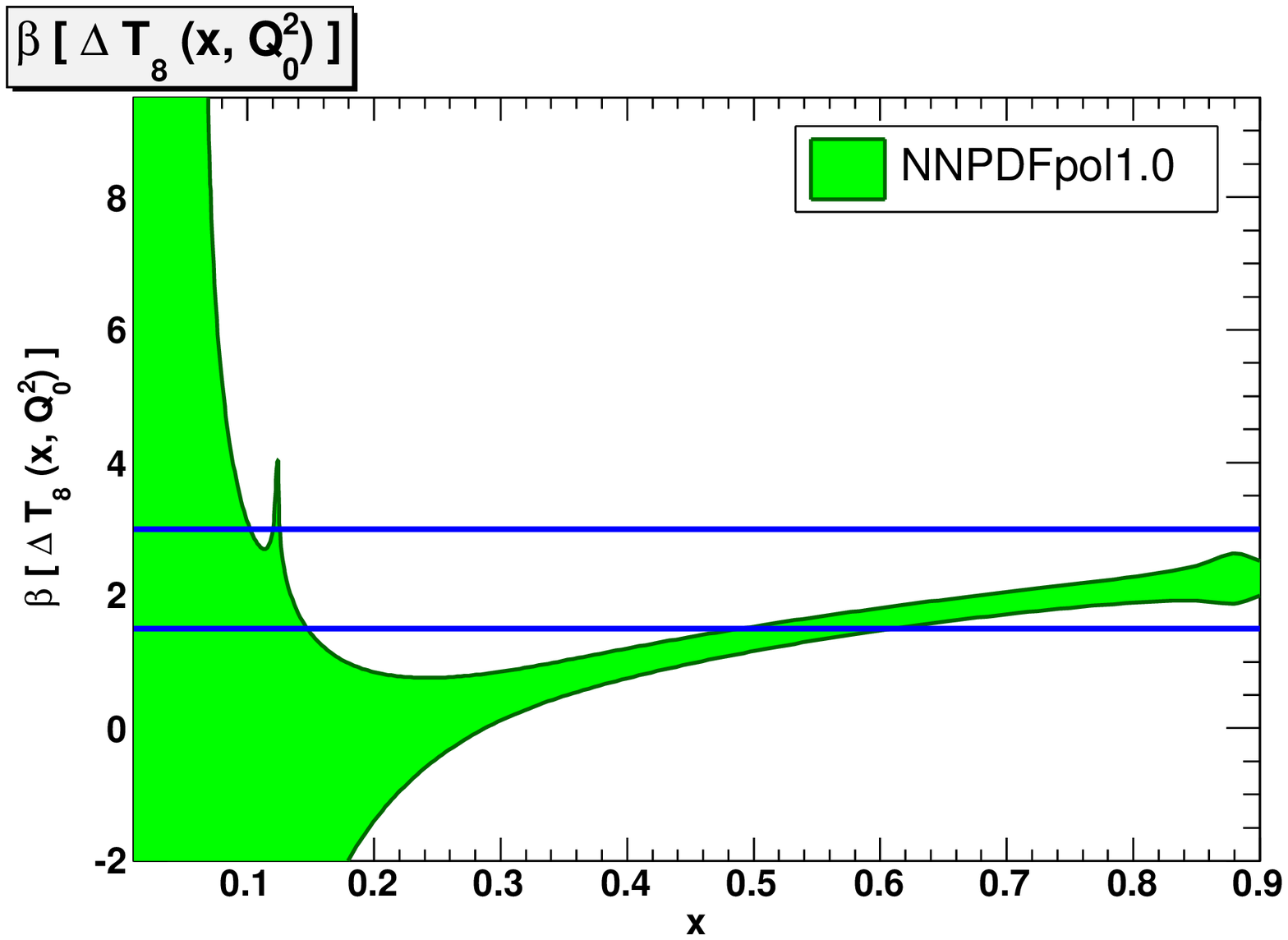}
\caption{\small The 68\% confidence level of the distribution of
  effective small- and large-$x$ exponents
  Eqs.~(\ref{eq:exp2}-\ref{eq:exp1}) for the default $N_{\rm
    rep}=100$ replica {\tt NNPDFpol1.0} set at $Q_0^2=1$~GeV$^2$, plotted as
  functions of $x$. The range of variation of the preprocessing
  exponents of Tab.~\ref{tab:prepexps} is also shown in each case
  (solid lines).}
\label{fig:prep}
\end{center}
\end{figure}

It is however necessary to check that no bias is introduced by the
preprocessing. We do this following the iterative method
described in Sect.~\ref{sec:net-param}. The outcome of the procedure
is the set of exponents Eq.~(\ref{eq:PDFbasisnets}), listed in
Tab.~\ref{tab:prepexps}. The lack of bias with these choices is
explicitly demonstrated  in  Fig.~\ref{fig:prep}, where we plot the 68\%
confidence level of the distribution of 
\begin{align}
&\alpha[\Delta q(x,Q^2)]=\frac{\ln \Delta q(x,Q^2)}{\ln\frac{1}{x}}\mbox{ ,}
\label{eq:exp2}
\\
&\beta[\Delta q(x,Q^2)]=
\frac{ \ln \Delta q(x,Q^2) }{\ln(1-x)}\mbox{ ,}
\label{eq:exp1}
\end{align}
$\Delta q=\Delta\Sigma\mbox{, }\Delta g\mbox{, }\Delta T_3\mbox{, }\Delta T_8$, for the default \texttt{NNPDFpol1.0} $N_{\rm rep}=100$ replica
 set, at $Q^2=Q_0^2=1$~GeV$^2$, and compare them to the ranges of
Tab.~\ref{tab:prepexps}.
It is apparent that  as the endpoints 
$x=0$ and $x=1$ are approached, the uncertainties on both
the small-$x$ and the large-$x$ exponents lie well within the range
of the preprocessing exponents for all PDFs, thus confirming
that the latter do not introduce any bias.

\clearpage

\section{Polarized nucleon structure}
\label{sec:phenoimplications}

The \texttt{NNPDFpol1.0} PDF set may be used for a determination of the first
moments of polarized parton distributions. As briefly summarized in
the introduction, these are the quantities of greatest physical
interest in that they are directly related to the spin structure of
the nucleon, and indeed their determination, in particular the
determination of the first moments of the quark and gluon
distributions, has been the main motivation for the experimental
campaign of $g_1$ measurements. The determination of the isotriplet
first moment, because of the Bjorken sum rule, provides a potentially
accurate and unbiased handle on the strong coupling $\alpha_s$.

\subsection{First moments}

\begin{figure}[t]
\begin{center}
\epsfig{width=0.47\textwidth,figure=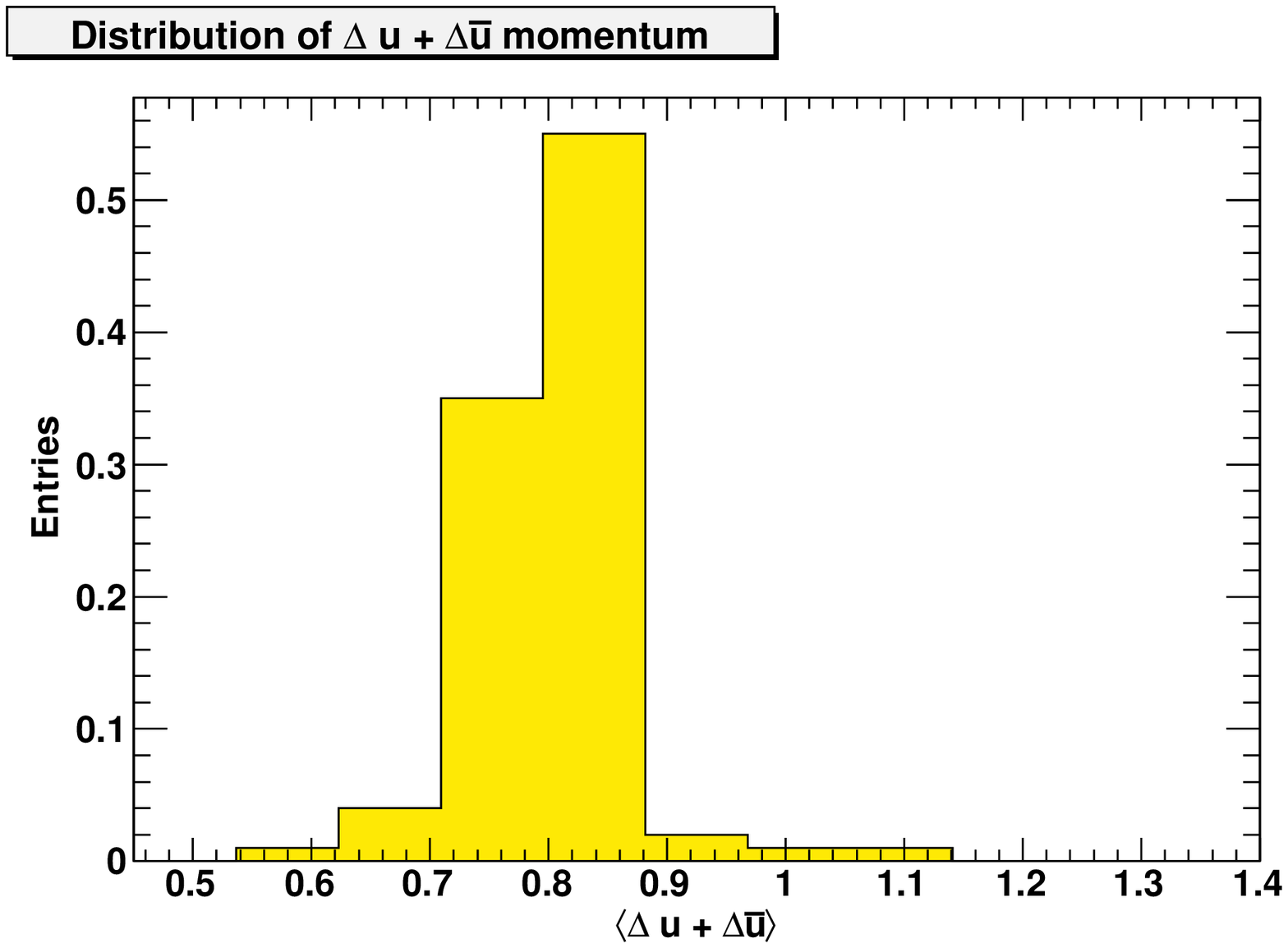}
\epsfig{width=0.47\textwidth,figure=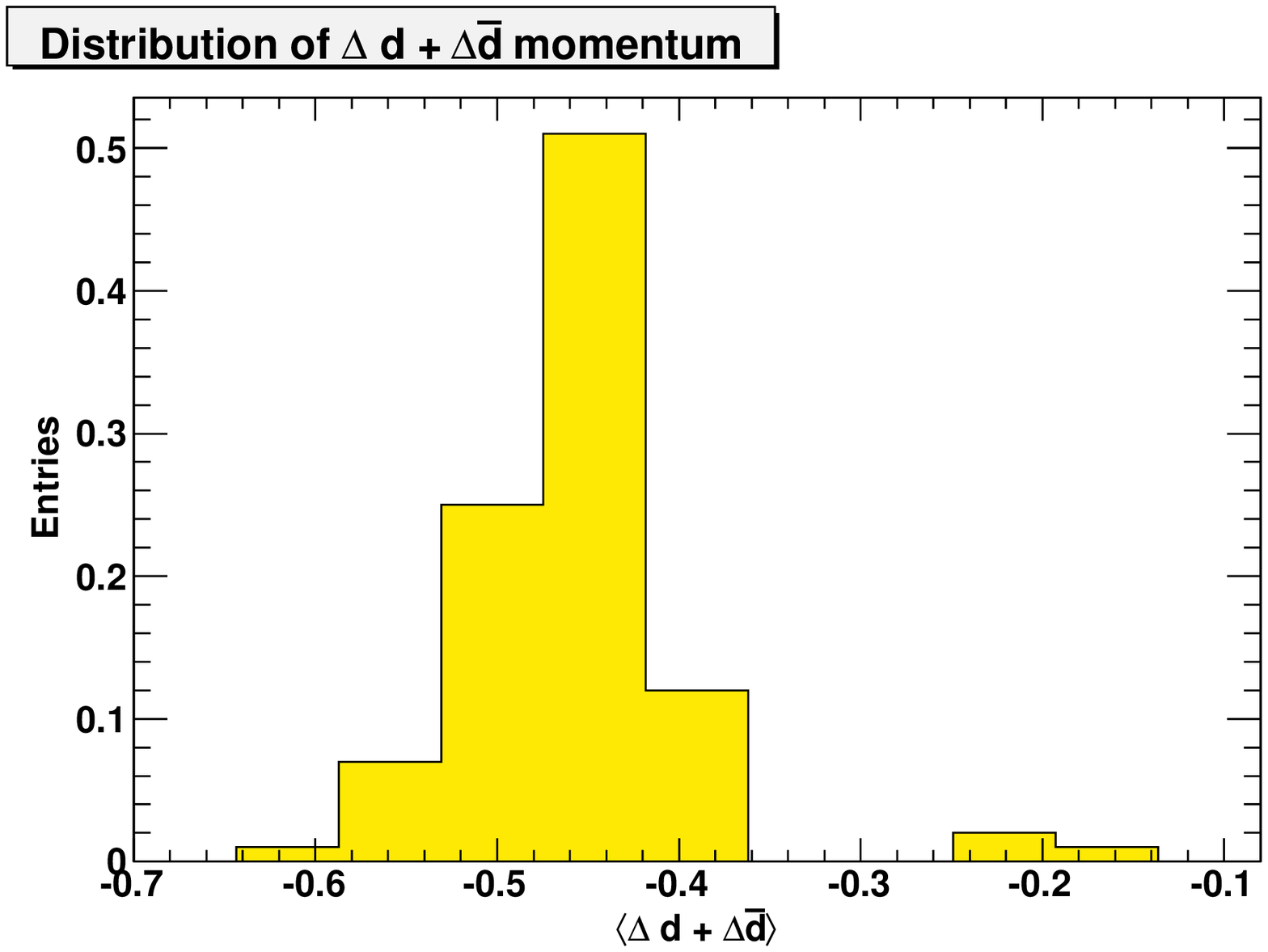}
\epsfig{width=0.47\textwidth,figure=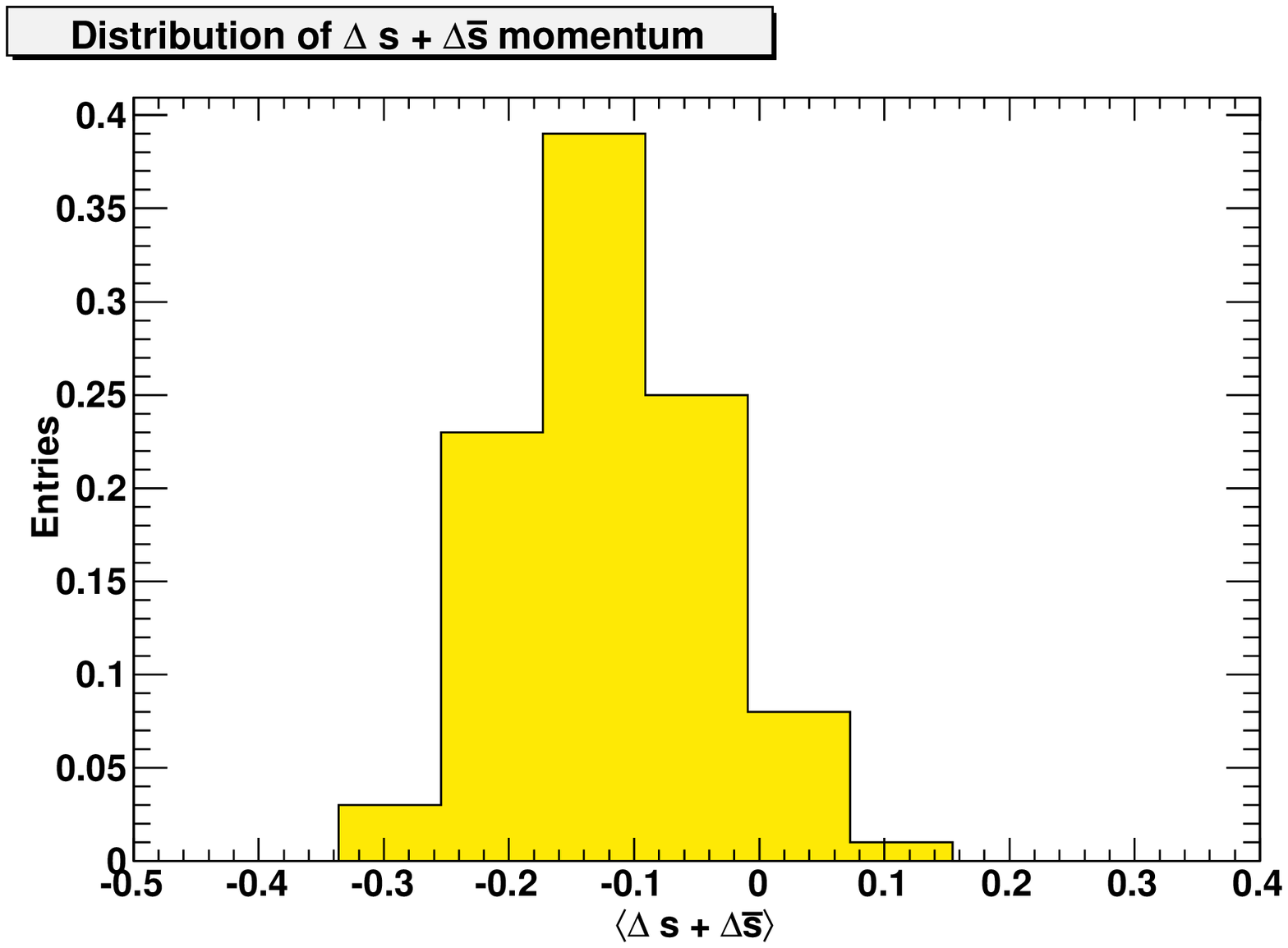}
\epsfig{width=0.47\textwidth,figure=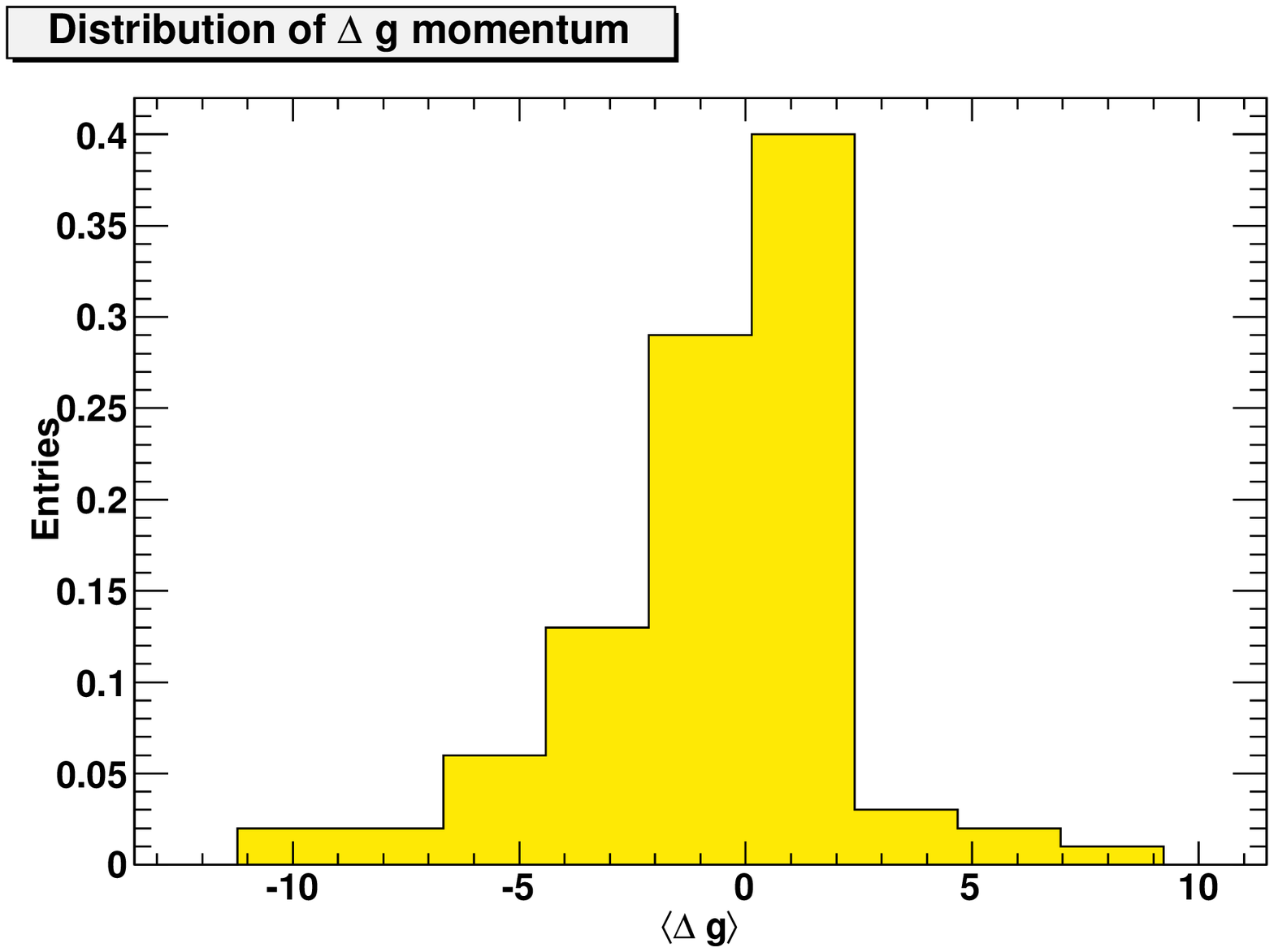}
\caption{\small Distribution of the first moments of 
$\Delta u + \Delta\bar{u}$ (top left), $\Delta d + \Delta\bar{d}$ (top right),
$\Delta s + \Delta\bar{s}$ (bottom left) over a set of
$N_\mathrm{rep}=100$  \texttt{NNPDFpol1.0} 
PDF replicas.
\label{fig:mom_distr}} 
\end{center}
\end{figure}

We have computed the first moments 
\be
\langle \Delta f(Q^2) \rangle \equiv
\int_0^1 dx \, \Delta f(x,Q^2)
\label{eq:moments}
\ee 
of each light polarized quark-antiquark and gluon
distribution using a sample of
$N_\mathrm{rep}=100$  \texttt{NNPDFpol1.0} 
PDF replicas.
 The histogram  of the distribution of first moments over the replica
 sample at $Q_0^2=1~{\rm
  GeV^2}$ are  displayed  in Fig.~\ref{fig:mom_distr}: they appear to
 be reasonably  approximated by a
Gaussian.

The central value and one-$\sigma$ uncertainties of the quark first
moments are listed in Tab.~\ref{tab:spin2}, while those of the singlet
quark combination Eq.~(\ref{2}) and the gluon are given in
Tab.~\ref{tab:spin1}.
Results are compared to those from other
parton sets, namely ABFR98~\cite{Altarelli:1998nb},
DSSV10~\cite{deFlorian:2009vb}, AAC08~\cite{Hirai:2008aj},
BB10~\cite{Blumlein:2010rn} and
LSS10~\cite{Leader:2010rb}. Results from other PDF sets are not
available for all combinations and scales, because public codes only
allow for the computation of first moments in a limited $x$ range, in
particular down to a minimum value of $x$: hence we must rely on
published values for the first moments. In particular, the
DSSV and AAC results are shown at $Q_0^2=1~{\rm
  GeV^2}$, while the BB and LSS results are shown at  $Q^2=4~{\rm
  GeV^2}$. For ease of reference, the NNPDF values for both scales are shown
in Tab.~\ref{tab:spin1}.

\begin{table}[h!]
\begin {center}
\small
\begin{tabular}{l||r@{.}l|r@{.}l|r@{.}l|r@{.}l||r@{.}l|r@{.}l|r@{.}l|r@{.}l||r@{.}l|r@{.}l|r@{.}l|r@{.}l}
\hline 
             & \multicolumn{8}{c||}{$\langle \Delta u +\Delta \bar u  \rangle$} 
             & \multicolumn{8}{c||}{$\langle \Delta d +\Delta \bar d  \rangle$} 
             & \multicolumn{8}{c}{$\langle   \Delta s +\Delta \bar s  \rangle$}\\
\hline 
\hline 
             & \multicolumn{2}{c|}{\textsl{\rm cv}} & \multicolumn{2}{c|}{\textsl{\rm exp}} 
             & \multicolumn{2}{c|}{\textsl{\rm th}} & \multicolumn{2}{c||}{\textsl{\rm tot}}
             & \multicolumn{2}{c|}{\textsl{\rm cv}} & \multicolumn{2}{c|}{\textsl{\rm exp}} 
             & \multicolumn{2}{c|}{\textsl{\rm th}} & \multicolumn{2}{c||}{\textsl{\rm tot}}  
             & \multicolumn{2}{c|}{\textsl{\rm cv}} & \multicolumn{2}{c|}{\textsl{\rm exp}} 
             & \multicolumn{2}{c|}{\textsl{\rm th}} & \multicolumn{2}{c}{\textsl{\rm tot}} \\
\hline
NNPDFpol1.0  
             &  0 & 80 & 0 & 08 & \multicolumn{2}{c|}{---} & 0 & 08
             & -0 & 46 & 0 & 08 & \multicolumn{2}{c|}{---} & 0 & 08
             & -0 & 13 & 0 & 09 & \multicolumn{2}{c|}{---} & 0 & 09 \\
DSSV08~\cite{deFlorian:2008mr}
             &  0 & 817 & 0 & 013 & 0 & 008 & 0 & 015
             & -0 & 453 & 0 & 011 & 0 & 036 & 0 & 038
             & -0 & 110 & 0 & 023 & 0 & 098 & 0 & 101 \\
\hline
\end{tabular}
\end {center}
\caption{\small
  First moments of the polarized quark distributions at   $Q_0^2=1$
  GeV$^2$; cv denotes the central value, whil exp and th denote
  uncertainties (see text) whose sum in quadrature is given by tot.}
\label{tab:spin2}
\end{table}

In order to compare the results for first moments shown in
Tabs.~\ref{tab:spin2}-\ref{tab:spin1}, it should be understood that
the uncertainties shown, and sometimes also the central values, 
have somewhat different meanings. In particular:
\begin{itemize}
\item For \texttt{NNPDFpol1.0} the {\textit exp} uncertainty, determined as 
the standard deviation of the  replica sample, is a pure PDF
uncertainty: it  includes the
propagation of the experimental data uncertainties and the uncertainty
due to the interpolation and extrapolation.
\item In the ABFR98 study, the central values were obtained in the
  so-called AB factorization scheme~\cite{Ball:1995td}. While the
  gluon in this scheme coincides with the gluon in the $\overline{\rm
    MS}$ scheme used here (and thus the value from
  Ref.~\cite{Altarelli:1998nb} for the gluon is shown in
  Tab.~\ref{tab:spin1}), the quark singlet differs from it. However,
  in Ref.~\cite{Altarelli:1998nb} a value of the singlet axial charge
  $a_0$ in the limit of infinite $Q^2$ was also given. 
  In the $\overline{\rm MS}$,
  the singlet axial charge and the first moment of $\Delta\Sigma$
  coincide~\cite{Ball:1995td}, hence we have determined $\la\Delta\Sigma\ra$
  for ABFR98 by evolving down to $Q^2=1$~GeV$^2$ the value of
  $a_0(\infty)$ given in Ref.~\cite{Altarelli:1998nb}, at NLO and with
  $\alpha_s(M_z)=0.118$~\cite{Beringer:1900zz} (the impact of the
  $\alpha_s$ uncertainty is negligible). We have checked that
  the same result is obtained if $a_0$ is computed as the appropriate
  linear combination of $\la \Delta\Sigma\ra$ in the AB scheme and the
  first moment of $\Delta g$.
  In the ABFR98 study, the
  \textit{exp} uncertainty is the Hessian uncertainty on the best fit,
  and it thus includes the propagated data uncertainty. The
  \textit{th} uncertainty includes the uncertainty originated by
  neglected higher orders (estimated by renormalization and
  factorization scale variations), higher twists, position of heavy
  quark thresholds, value of the strong coupling, violation of SU(3)
  (uncertainty on $a_8$ Eq.~(\ref{eq:t8sr})), and finally
  uncertainties related to the choice of functional form, estimated by
  varying the functional form. This latter source of theoretical
  uncertainty corresponds to interpolation and extrapolation
  uncertainties which are included in the \textit{exp} for NNPDF.
\item For DSSV08 and BB10 PDFs, the central value is obtained by
  computing the first moment integral of the best-fit with a fixed
  functional form restricted to the data region, and then
  supplementing it with a contribution due to the extrapolation in the
  unmeasured (small $x$) region. The \textit{exp} uncertainty in the
  table is the Hessian uncertainty given by DSSV08 or BB10 on the
  moment in the measured region, and it thus includes the propagated
  data uncertainty. In both cases, we have determined the \textit{th}
  uncertainty shown in the table as the difference between the full first
  moment quoted by DSSV08 or BB10, and the first moment in the
  measured region. It is thus the contribution from the extrapolation
  region, which we assume to be $100\%$ uncertain. In both cases,
  we have computed  the truncated first moment in the measured region
  using publicly available codes, and checked that it coincides with
  the values quoted by  DSSV08 and BB10.
\item For AAC08, the central value is obtained by computing the first moment
  integral of the best-fit with a fixed functional form, and the
  \textit{exp} uncertainty is the Hessian uncertainty on it. However,
  AAC08 uses a so-called tolerance~\cite{Pumplin:2002vw} criterion for
  the determination of Hessian uncertainties, which rescales the
  $\Delta\chi^2=1$ region by a suitable factor, in order to
  effectively keep into
  account also interpolation errors. Hence, the
  \textit{exp} uncertainties include propagated data uncertainties, as
  well as uncertainties on the PDF shape.
\item For LSS10, the central value is obtained by computing the first
  moment integral of the
  best-fit with a fixed functional form, and the \textit{exp} uncertainty is
  the Hessian uncertainty on it. Hence it includes the
  propagated data uncertainty.
\end{itemize}
\begin{table}[t]
\begin{center}
\begin{tabular}{ll||r@{.}l|r@{.}l|r@{.}l|r@{.}l||r@{.}l|r@{.}l|r@{.}l|r@{.}l}
\hline 
        & & \multicolumn{8}{c||}{$\langle  \Delta\Sigma  \rangle$} 
          & \multicolumn{8}{c}{$\langle \Delta g  \rangle$} \\
\hline 
\hline 
        & & \multicolumn{2}{c|}{\textsl{\rm cv}} & \multicolumn{2}{c|}{\textsl{\rm exp}} 
          & \multicolumn{2}{c|}{\textsl{\rm th}} & \multicolumn{2}{c||}{\textsl{\rm tot}} 
          & \multicolumn{2}{c|}{\textsl{\rm cv}} & \multicolumn{2}{c|}{\textsl{\rm exp}} 
          & \multicolumn{2}{c|}{\textsl{\rm th}} & \multicolumn{2}{c}{\textsl{\rm tot}} \\
\hline
NNPDFpol1.0 (1GeV$^2$)     
                  & &  0 & 22    & 0 & 20    & \multicolumn{2}{c|}{---}                & 0 & 20 
                                 & -1 & 2     & 4 & 2   & \multicolumn{2}{c|}{---}                 & 4 & 2 \\
NNPDFpol1.0 (4GeV$^2$)     
                  & &  0 & 18   & 0 & 20    & \multicolumn{2}{c|}{---}                & 0 & 20 
                                 & -0 & 9     & 3 & 9   &
\multicolumn{2}{c|}{---}                 & 4 & 2 \\
\hline
ABFR98~\cite{Altarelli:1998nb} & &  0 & 12    & 0 & 05   & \multicolumn{2}{c|}{$^{+0.19}_{-0.12}$} & \multicolumn{2}{c||}{$^{+0.19}_{-0.13}$} 
                                 &  1 & 6     & 0 & 4    & 0 & 8                                   & 0 & 9 \\
DSSV08~\cite{deFlorian:2008mr} & &  0 & 255   & 0 & 019  & 0 & 126                                 & 0 & 127
                                 & -0 & 12   & 0 & 12  & 0 & 06                                 & 0 & 13 \\
\multirow{2}*{AAC08~\cite{Hirai:2008aj}}      & (\textsl{positive})      
                                 &  0 & 26    & 0 & 06   & \multicolumn{2}{c|}{---}                & 0 & 06 
                                 &  0 & 40    & 0 & 28   & \multicolumn{2}{c|}{---}                & 0 & 28 \\
                                              & (\textsl{node})                
                                 &  0 & 25    & 0 & 07   & \multicolumn{2}{c|}{---}                & 0 & 07 
                                 & -0 & 12    & 1 & 78   & \multicolumn{2}{c|}{---}                & 1 & 78  \\
BB10~\cite{Blumlein:2010rn}    & &  0 & 19   & 0 & 08  & 0 & 23                                 & 0 & 24 
                                 &  0 & 46   & 0 & 43  & 0 & 004                                 & 0 & 43 \\
\multirow{2}*{LSS10~\cite{Leader:2010rb}}     & (\textsl{positive})      
                                 &  0 & 207   &  0 & 034 & \multicolumn{2}{c|}{---}                & 0 & 034 
                                 &  0 & 316   &  0 & 190 & \multicolumn{2}{c|}{---}                & 0 & 190 \\
                                              & (\textsl{node})                
                                 &  0 & 254   &  0 & 042 & \multicolumn{2}{c|}{---}                & 0 & 042 
                                 & -0 & 34   &  0 & 46 & \multicolumn{2}{c|}{---}                & 0 & 46 \\
\hline 
\end{tabular}
\end{center}
\caption{\small Same as Tab.~\ref{tab:spin2}, but for the total
  singlet quark distribution and the gluon distribution. The NNPDF
  results are shown both at $Q_0^2=1$ GeV$^2$ and  $Q^2=4~{\rm GeV^2}$,  the
  ABFR, DSSV and AAC results are shown at $Q_0^2=1$ GeV$^2$, and the
  BB10 and LSS10 are shown at $Q^2=4~{\rm GeV^2}$).}
\label{tab:spin1}
\end{table}

In all cases, the total uncertainty is computed as the sum in quadrature of the
\textit{exp} and \textit{th} uncertainties. Roughly speaking, for
LSS10 this includes only the data uncertainties; for DSSV08, and BB10
it also includes extrapolation uncertainties; 
for AAC08 interpolation uncertainties;
for \texttt{NNPDFpol1.0} both extrapolation and
interpolation uncertainties; and for ABFR98 all of the
above, but also theoretical (QCD) uncertainties.  For LSS10 and AAC08,
we quote the results obtained from two different fits, both
assuming positive- or node-gluon PDF: their spread gives a feeling for
the missing uncertainty due to the choice of functional form. Note
that the AAC08 results correspond to their Set B which includes,
besides DIS data, also RHIC $\pi^0$ production data; the DSSV08 fit
also includes, on top of these, RHIC jet data and semi-inclusive DIS
data; LSS10 includes, beside DIS, also semi-inclusive DIS data. All other sets are
based on DIS data only.

Coming now to a comparison of results, we see that for the 
singlet first moment  $\langle
\Delta\Sigma \rangle$ the {\tt NNPDFpol1.0} result is consistent within
uncertainties with that of other groups. The uncertainty on the {\tt
  NNPDFpol1.0} result  is
comparable (if somewhat larger) to that found whenever the
extrapolation uncertainty has been included. 
For individual quark flavors
(Tab.~\ref{tab:spin2}) we find excellent agreement in the central
values obtained between {\tt NNPDFpol1.0} and DSSV08; the NNPDF
uncertainties are rather larger, but this could also be due to the
fact that the DSSV08 dataset is sensitive to flavour separation. 

For the gluon first moment $\langle \Delta g
\rangle$, the {\tt NNPDFpol1.0} result is characterized by an
uncertainty which is much larger than that of any other determination:
a factor of three or four larger than ABFR98 and AAC08, ten times larger
than BB10, and twenty times larger than DSSV08 and LSS10. It is
compatible with zero within this large uncertainty.
We have seen that for the quark singlet, the {\tt NNPDFpol1.0}
uncertainty is similar to that of groups which include an estimate of
extrapolation uncertainties. In order to assess the impact
of the extrapolation uncertainty for the  gluon, we have
computed the gluon first truncated moment in the region
$x\in[10^{-3},1]$:
\be
\int_{10^{-3}}^1dx\, \Delta g(x, Q^2=1~{\rm GeV^2}) = -0.26 \pm 1.19 \,, 
\ee
to be compared with the result of Tab.~\ref{tab:spin1}, which is
larger by almost a factor four. 

We must conclude that the experimental status of the gluon first
moment is still completely uncertain, unless one is willing to make
strong theoretical assumptions on the behaviour of the polarized gluon
at small $x$, and that previous different conclusions were affected by
a significant under-estimate of the impact of the bias in the choice
of functional form, in the data and especially in the extrapolation
region. Because of the large uncertainty related to the extrapolation
region, only low $x$ data can improve this situation, such as those
which could be collected at a high
energy electron-ion collider~\cite{Deshpande:2005wd,Boer:2011fh}.

\subsection{The Bjorken sum rule}
\label{sec:bjorken}

Perturbative factorization, expressed in this context
by Eq.~(\ref{1}) for the structure function $g_1(x,Q^2)$,
and the assumption of exact isospin symmetry, immediately
lead to the so-called Bjorken sum rule (originally
derived~\cite{Bjorken:1966jh,Bjorken:1969mm} using  current algebra):
\be
\label{eq:bjorken}
\Gamma_1^p\lp Q^2\rp - \Gamma_1^n\lp Q^2\rp = \frac{1}{6}
\Delta C_{\mathrm{NS}} (\alpha_s(Q^2)) a_3 \ ,
\ee
where
\be
\label{eq:g1pmomentum}
\Gamma_1^{p,n}(Q^2) \equiv \int_0^1 dx\, g_1^{p,n} (x,Q^2) \ ,
\ee
and $\Delta C_{\mathrm{NS}} (\alpha_s(Q^2))$
is the first moment of the non-singlet coefficient function,
while $a_3$ is defined in Eq.~\eqref{eq:t3sr}.

Because the first moment of the non-singlet coefficient function
$\Delta C_{\rm NS}$ is known up to three loops~\cite{Larin:1991tj} and
isospin symmetry is expected to hold to high accuracy, the Bjorken sum
rule Eq.~(\ref{eq:bjorken}) potentially provides a theoretically very
accurate handle on the strong coupling constant: in principle,
the truncated isotriplet first moment
\be
\label{eq:bjorken-cut}
\Gamma_1^{\rm NS}\lp Q^2,x_{\rm min}\rp \equiv
\int_{x_{\rm min}}^1 dx \lc g_1^p\lp x,Q^2 \rp - g_1^n\lp x,Q^2 \rp \rc
\ee
can be extracted from
the data without any theoretical assumption. Given a measurement of 
$\Gamma_1^{\rm NS}\lp Q^2,0\rp$ at one scale the strong coupling can then be
extracted from Eq.~(\ref{eq:bjorken}) using the value of $a_3$ from
$\beta$ decays, while given a measurement of 
$\Gamma_1^{\rm NS}\lp Q^2,0\rp$ at two scales both $a_3$ and the value
of $\alpha_s$ can be extracted simultaneously.

In Ref.~\cite{DelDebbio:2009sq}, $a_3$ and $\alpha_s$ where simultaneously determined
from a set of nonsinglet truncated moments (both the first and higher
moments),  by exploiting the scale
dependence of the latter~\cite{Forte:1998nw}, with the result
$g_A=1.04\pm0.13$ and $\alpha_s(M_z)=0.126^{+0.006}_{-0.014}$, where
the uncertainty is dominated by the data, interpolation and
extrapolation, but also includes theoretical (QCD)
uncertainties. In this reference, truncated moments were determined
from a neural network interpolation of existing data, sufficient for 
a computation of moments at any scale. However, because the small $x$
behaviour of the structure function is only weakly constrained by
data, the $x\to0$ extrapolation was done by assuming a powerlike
(Regge) behaviour~\cite{Close:1994he}.

\begin{figure}[t!]
\begin{center}
\epsfig{width=0.47\textwidth,figure=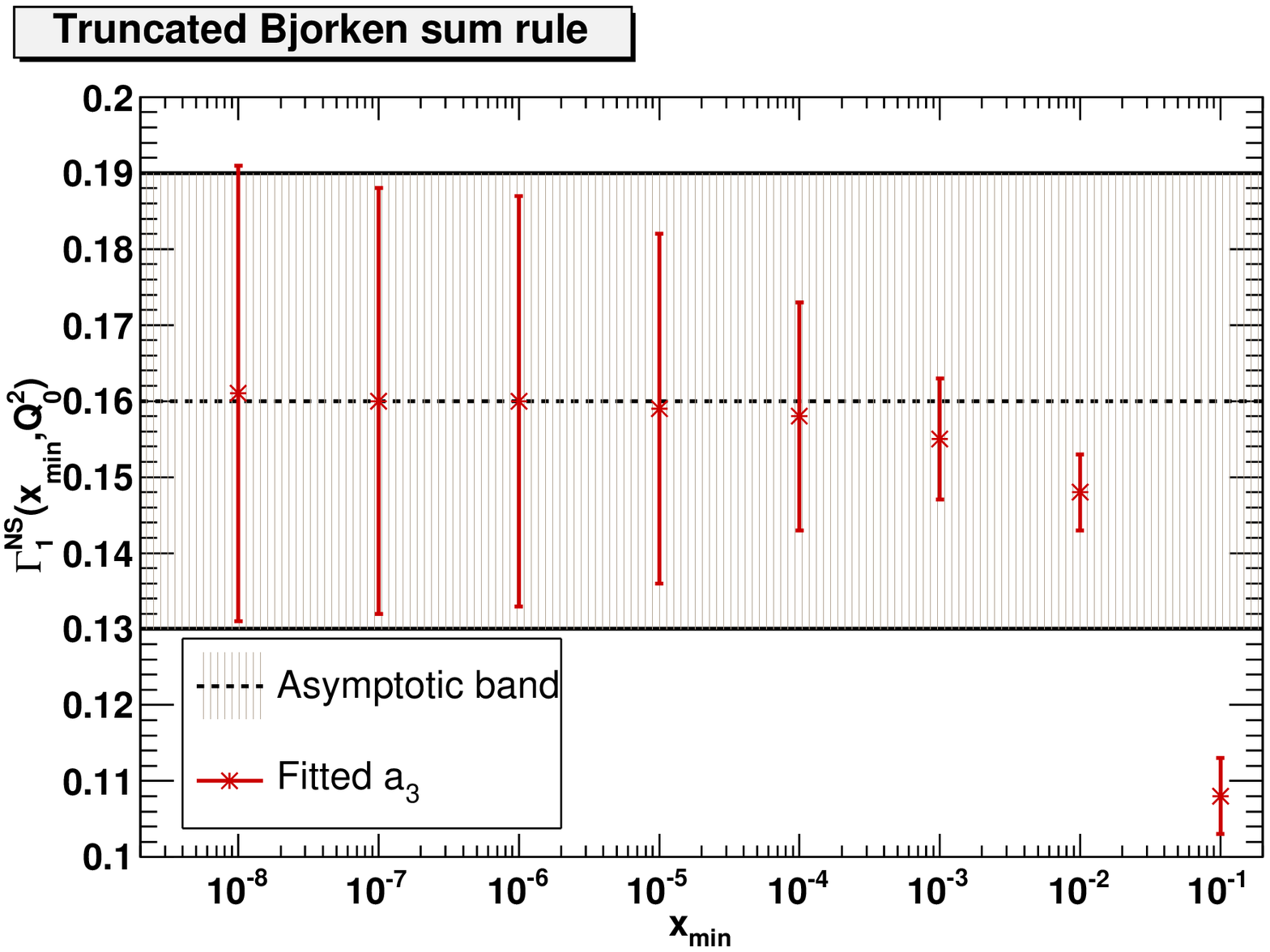}
\epsfig{width=0.47\textwidth,figure=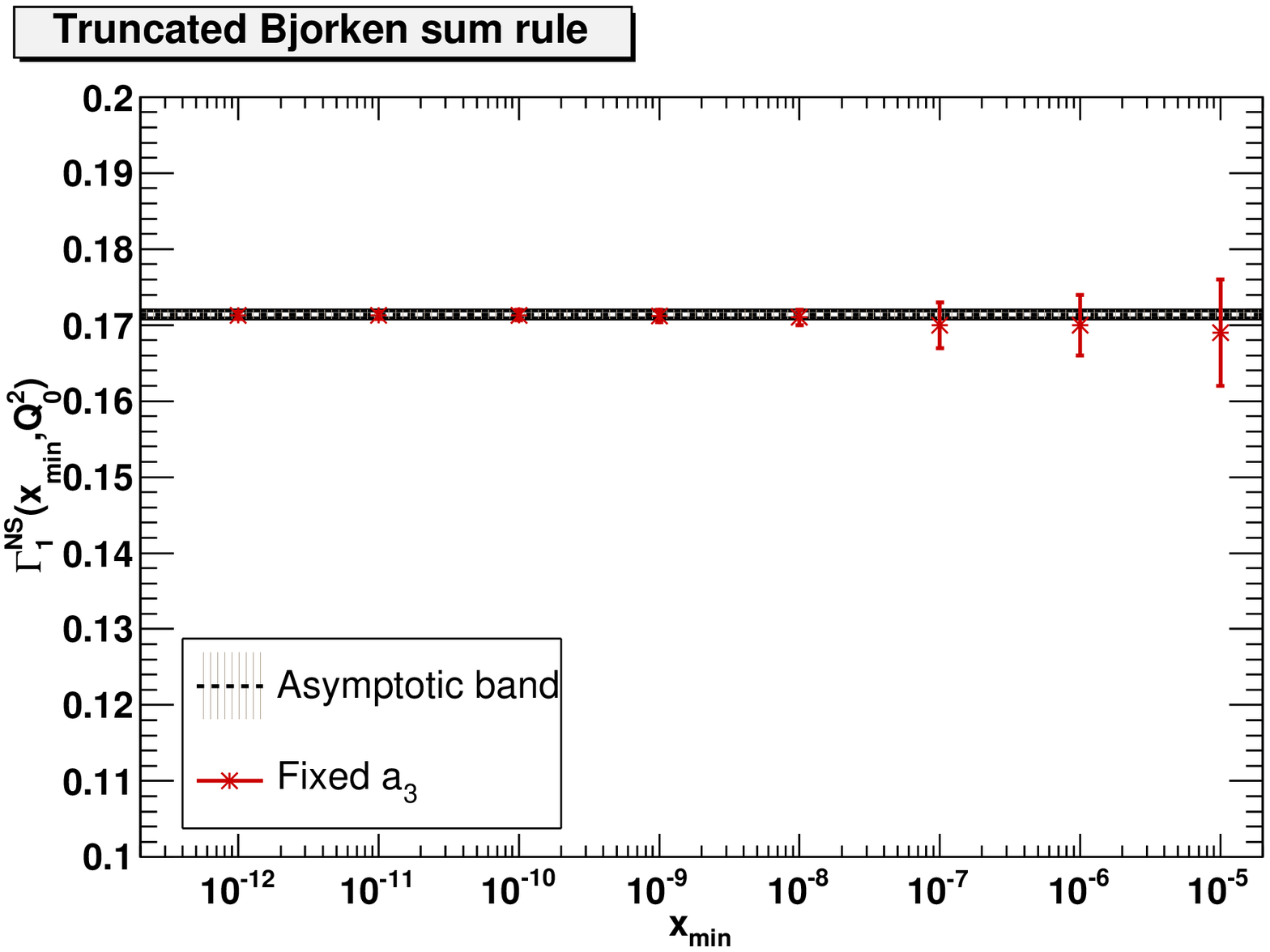}
\caption{\small The truncated 
Bjorken sum rule  $\Gamma_1^{\rm NS}\lp Q^2,x\rp$ Eq.~(\ref{eq:bjorken-cut})
plotted as a function of $x$ for $Q^2=1$~GeV$^2$,
 for the 
fit with free $a_3$  (left) and for the
reference fit with $a_3$ fixed to the value 
Eq.~(\ref{eq:a3}),
(right panel). In the left plot,
the shaded band corresponds to the 
asymptotic value of the truncated sum rule, Eq.~(\ref{eq:extriplet}),
while in the right plot it corresponds to the
experimental value Eq.~(\ref{eq:a3}).
\label{fig:bjsr}} 
\end{center}
\end{figure}

The situation within {\tt NNPDFpol1.0} can be understood by exploiting
the PDF
determination in which $a_3$ is not fixed by the triplet sum
rule, discussed in  Sect.~\ref{sec:srres}. Using the results of this
determination, we find 
\begin{equation}\label{eq:a3exp}
a_3=\int_0^1dx\,\Delta T_3 (x, Q^2) = 1.19 \pm 0.22\ .
\end{equation}
The uncertainty is about twice that of the determination of
Ref.~\cite{DelDebbio:2009sq}. As mentioned, the latter was obtained 
from a neural network parametrization of the data with no theoretical
assumptions, and based on a methodology which is quite close to that
of the  {\tt NNPDFpol1.0} PDF determination discussed here, the only
difference being the assumption of Regge behaviour in order to perform
the small $x$ extrapolation. This strongly suggests that, as in the
case of the gluon distribution discussed above, the uncertainty on the
value Eq.~(\ref{eq:a3exp}) is dominated by the small $x$
extrapolation.

To study this, in
Fig.~\ref{fig:bjsr} we plot the value of the truncated Bjorken sum
rule $\Gamma_1^{\rm NS}\lp Q^2,x_{\rm min}\rp$
Eq.~(\ref{eq:bjorken-cut}) as  a function 
of the lower limit of integration $x_{\rm min}$ at $Q_0^2=1$~GeV$^2$, along
with the asymptotic value 
\be\label{eq:extriplet}
\Gamma_1^{\rm NS}\lp 1~\hbox{GeV}^2,0 \rp= 0.16 \pm 0.03
\ee
which at NLO corresponds to the value of $a_3$ given by Eq.~(\ref{eq:a3exp}).
As a consistency check, we also show the same plot  for our baseline fit, 
in which $a_3$ is fixed by the sum rule to the value
Eq.~(\ref{eq:a3}). It is clear that indeed the uncertainty is
completely dominated by the small $x$ extrapolation.

This suggests that a determination of $\alpha_s$ from the Bjorken sum
rule is not competitive unless one is willing to make assumptions on
the small $x$ behaviour of the nonsinglet structure function in the
unmeasured region. Indeed, it is clear that a determination based on
{\tt NNPDFpol1.0} would be affected by an uncertainty which is
necessarily larger than that found in Ref.~\cite{DelDebbio:2009sq},
which is already not competitive. The fact that a determination
of $\alpha_s$ from the Bjorken sum rule is not competitive due to
small $x$ extrapolation ambiguities was already pointed out in
Ref.~\cite{Altarelli:1998nb}, where values of $a_3$ and $\alpha_s$
similar to those of Ref.~\cite{DelDebbio:2009sq} were obtained.

\clearpage

\section{Conclusions and outlook}
\label{sec:conclusions}

We have presented a first determination of polarized parton
distributions based on the NNPDF methodology: {\tt NNPDFpol1.0}.  We
have determined polarized PDFs from the most recent inclusive data on
proton, deuteron and neutron deep-inelastic polarized asymmetries and
structure functions.  Our main result is that the uncertainty in the
gluon distribution, and to a lesser extent the strange distribution,
and in the small $x$ extrapolation for all parton distributions, is
rather larger than in previous polarized PDF determinations. Also,
there seems to be some tension between strangeness determined in
deep-inelastic scattering and using sem-inclusive data.

In particular, we find that the role of the gluon distribution in the
spin structure of the nucleon is essentially unknown, as the first
moment of the gluon distribution is compatible with zero, but with an
uncertainty which is compatible with a very large positive or negative
gluon spin fraction. Likewise, the contribution from the small $x$
region to the Bjorken sum rule makes its use as a means to determine
$\alpha_s$ essentially impossible. Different conclusions can be
reached only if one is willing to make strong theoretical assumptions
on the small $x$ behaviour of polarized PDFs.

Future experiments, in particular open charm and hadron production in
fixed target experiments,~\cite{Adolph:2012ca,Adolph:2012vj}
inclusive jet production~\cite{Adare:2010cc,Adamczyk:2012qj} and $W$
boson production~\cite{Aggarwal:2010xx,Adare:2010xx,Stevens:2013xx} 
from the RHIC collider may
improve the knowledge on individual polarized flavors and antiflavors
and on the gluon distribution in the valence region. However, only a
high-energy electron-ion collider~\cite{Deshpande:2005wd,Boer:2011fh}
might provide information on polarized PDFs at small $x$ and thus
reduce the uncertainty on first moments in a significant way.

\bigskip
\bigskip
\begin{center}
\rule{5cm}{.1pt}
\end{center}
\bigskip
\bigskip
The {\tt NNPDFpol1.0} polarized PDFs, with $N_{\rm rep}=100$ replicas,
are available from the
NNPDF HEPFORGE web site,
\begin{center}
{\bf \url{http://nnpdf.hepforge.org/}~}.
\end{center}
A  Mathematica driver code is also available from the same source.


\bigskip
\bigskip

{\bf\noindent  Acknowledgments \\}

We would like to thank C.~Aidala and G.~Altarelli for encouragement in
this project, D.~De~Florian, D.~Hasch, A.~Bacchetta and M.~Radici for
useful discussions on polarized PDFs.  We thank G. Salam for
assistance with the polarized PDF evolution Les Houches benchmark
tables.  The research of JR has been supported by a Marie Curie
Intra--European Fellowship of the European Community's 7th Framework
Programme under contract number PIEF-GA-2010-272515. SF, ERN and GR
are supported by a PRIN2010 grant.


\clearpage

\appendix

\section{Benchmarking of polarized PDF evolution}
\label{sec:apppdfevol}

We have benchmarked our implementation of the evolution of polarized
parton densities by
cross-checking against the
Les Houches polarized PDF evolution benchmark
tables~\cite{Dittmar:2005ed}. Note that in
  Ref.~\cite{Dittmar:2005ed}
the  polarized sea PDFs are given incorrectly, and should be \bea
  x\Delta \bar{u}=-0.045 x^{0.3} (1-x)^7 \nonumber\\
  x\Delta \bar{d}=-0.055 x^{0.3} (1-x)^7 \ .  \eea  
These tables were obtained from a
comparison of the {\tt HOPPET}~\cite{Salam:2008qg} and {\tt
  PEGASUS}~\cite{Vogt:2008yw} evolution codes, which are $x-$space and
$N-$space codes respectively. In order to perform a meaningful
comparison, we use the so-called iterated solution of the $N-$space evolution
equations and use the same initial PDFs and running coupling as
in~\cite{Dittmar:2005ed}. The
relative difference $\epsilon_{\rm rel}$ 
between our PDF evolution and the benchmark tables of
Refs.~\cite{Dittmar:2005ed} at NLO in the ZM-VFNS scheme
are tabulated in Tab.~\ref{tab:lhacc} for various combinations of
polarized PDFs: the accuracy of our code is $\mathcal{O}\lp
10^{-5}\rp$  for all relevant values of $x$, which is the
nominal accuracy of the agreement between {\tt HOPPET} and {\tt
  PEGASUS}.

\begin{table}[t]
\begin{center}
\vskip-0.1cm
\begin{tabular}{|c||c|c|c|c|}
\hline
$x$  &  $\epsilon_{\rm rel}\lp \Delta u_V\rp$ & 
 $\epsilon_{\rm rel}\lp \Delta d_V\rp$ &  $\epsilon_{\rm rel}\lp \Delta \Sigma\rp$ &
 $\epsilon_{\rm rel}\lp   \Delta g\rp$  \\
\hline
\hline   
$10^{-3}$ & $1.1\,10^{-4}$ & $9.2\,10^{-5}$ & $9.9\,10^{-5}$& $1.1\,10^{-4}$\\
$10^{-2}$  & $1.4\,10^{-4}$ & $1.9\,10^{-4}$ & $3.5\,10^{-4}$& $9.3\,10^{-5}$\\
$0.1$  & $1.2\,10^{-4}$ & $1.6\,10^{-4}$ & $5.4\,10^{-6}$& $1.7\,10^{-4}$\\
$0.3$  & $2.3\,10^{-6}$ & $1.1\,10^{-5}$ & $7.5\,10^{-6}$& $1.7\,10^{-5}$\\
$0.5$  & $5.6\,10^{-6}$ & $9.6\,10^{-6}$ & $1.6\,10^{-5}$& $2.5\,10^{-5}$\\
$0.7$  & $1.2\,10^{-4}$ & $9.2\,10^{-7}$ & $1.6\,10^{-4}$& $7.8\,10^{-5}$\\
$0.9$  & $3.5\,10^{-3}$ & $1.1\,10^{-2}$ & $4.1\,10^{-3}$& $7.8\,10^{-3}$\\
\hline
\end{tabular}
\end{center}
\caption{\small Percentage difference between FastKernel perturbative
  evolution of  polarized PDFs and  the Les Houches benchmark 
tables~\cite{Dittmar:2005ed}
for different
polarized PDF combinations at NLO in the ZM-VFNS scheme. 
\label{tab:lhacc}}
\vskip-0.1cm
\end{table}

Therefore, we can conclude that the accuracy of the polarized
PDF evolution in the {\tt FastKernel} framework is satisfactory
for precision phenomenology.

\clearpage

\end{document}